\documentclass[sigplan,10pt]{acmart}
\renewcommand\footnotetextcopyrightpermission[1]{}
\usepackage{optidef} 

\usepackage{xcolor}
\usepackage{algorithm}
\usepackage[noend]{algpseudocode}
\definecolor{commentgreen}{rgb}{0, 0.5, 0}

\newcommand{\LineComment}[1]{\item[] \small \textcolor{commentgreen}{\# #1}}

\usepackage{tablefootnote}
\usepackage{multirow}
\usepackage{booktabs}
\usepackage{tabularx}

\usepackage{tikz}
\usepackage{amsmath}
\newcommand{\xmark}{%
\tikz[scale=0.23] {
    \draw[line width=0.7,line cap=round] (0,0) to [bend left=6] (1,1);
    \draw[line width=0.7,line cap=round] (0.2,0.95) to [bend right=3] (0.8,0.05);
}}
\newcommand{\cmark}{%
\tikz[scale=0.23] {
    \draw[line width=0.7,line cap=round] (0.25,0) to [bend left=10] (1,1);
    \draw[line width=0.8,line cap=round] (0,0.35) to [bend right=1] (0.23,0);
}}

\usepackage{filecontents}

\usepackage{xspace}

\usepackage{listings}
\usepackage{courier}
\DeclareCaptionFormat*{mystyle}{\noindent\hrulefill\par#1.\space#3}
\usepackage{graphicx}
\usepackage{subcaption}

\usepackage[textfont={it}, font={small,bf}]{caption}

\usepackage[inline]{enumitem}
\newenvironment{denseitemize}{
	\begin{itemize}[topsep=2pt, partopsep=0pt, leftmargin=1.5em]
		\setlength{\itemsep}{2pt}
		\setlength{\parskip}{0pt}
		\setlength{\parsep}{0pt}
	}{\end{itemize}}

\newenvironment{denseenum}{
	\begin{enumerate}[topsep=2pt, partopsep=0pt, leftmargin=1.5em]
		\setlength{\itemsep}{2pt}
		\setlength{\parskip}{0pt}
		\setlength{\parsep}{0pt}
	}{\end{enumerate}}

\def\ie{{i.e.\xspace}}
\def\eg{{e.g.\xspace}}

\def\insu#1{}
\def\mosharaf#1{}

\newcommand{\name}{Entrain\xspace}
\newcommand{\parabf}[1]{\noindent\textbf{#1}\xspace}

\begin{document}

\title{Addressing Variable Heterogeneity in Distributed Multimodal Training with \name}

\author{Insu Jang}
\orcid{0009-0007-5206-2333}
\affiliation{%
	\institution{University of Michigan}
	\country{}
}
\author{Mosharaf Chowdhury}
\orcid{0000-0003-0884-6740}
\affiliation{%
	\institution{University of Michigan}
	\country{}
}

\begin{abstract}
Multimodal LLM datasets are inherently heterogeneous, with significant data variability.
Although each modality exhibits independent variability, sample-level entanglement makes it difficult to balance workloads across both modalities and batches.
We present \name, a distributed MLLM training framework that addresses both heterogeneity and variability in multimodal training workloads.
\name challenges the intuition that dynamic data variability requires dynamic model parallelism by shifting the profiling paradigm from micro-level samples to macroscopic batches.
We prove that a single, static model-parallel configuration suffices for optimal load balancing under this paradigm.
At the microscopic scale, \name introduces a hierarchical microbatch assignment algorithm that defers excess workload within each iteration to stabilize variability across microbatches.
Evaluations show that \name reduces workload variability across microbatches by up to 10.6$\times$, improving end-to-end training throughput by up to 1.40$\times$ over existing baselines.
\end{abstract}

\maketitle
\pagestyle{plain}

\section{Introduction}
\label{sec:introduction}

The rapid evolution of Large Language Models (LLMs) into Multimodal LLMs (MLLMs)~\cite{gpt4o-arxiv24,gpt5-arxiv26,gemini-arxiv23,qwen2vl-arxiv24,qwen3omni-arxiv25,llavanext-cvpr24,llavaov-tmlr25,deekseepvl-arxiv24,deepseekvl24-arxiv24} has made efficient distributed training critical, yet balanced workload distribution remains particularly challenging due to the inherent \textit{heterogeneity} and \textit{variability} of multimodal datasets.
Heterogeneity stems from the distinct computational requirements and arithmetic intensities of different modalities.
Variability arises because modality workload proportions fluctuate drastically across samples, with each modality following an independent distribution.
In text-only LLM training, recent works address sequence length variability by dynamically adapting the model-parallel configuration on the fly~\cite{hotspa-sosp24,bytescale-sigcomm25}.
A natural intuition would be to extend this dynamic reconfiguration to MLLMs.

Unfortunately, because each modality exhibits independent variability yet remains tightly entangled within a single sample, per-sample dynamic reconfiguration is prohibitively expensive.
Our analysis reveals that dynamic adaptation is fundamentally unnecessary.
While multimodal workloads are highly chaotic at the microscopic (sample or microbatch) level, the aggregate workload ratio between modalities reliably converges to a stable constant at the macroscopic scale of a global batch.
This convergence implies that a single, static model-parallel configuration suffices for optimal load balancing, provided it is anchored to the global distribution.
This exposes a critical shortcoming in existing MLLM training systems~\cite{distmm-nsdi24,disttrain-sigcomm25,optimus-atc25,dip-asplos26}, which derive their static model-parallel configurations by profiling a non-representative slice (e.g., a single sample or a small microbatch).
Profiling at a microscopic scale dominated by extreme sample-level variance captures a distorted snapshot of the dataset.
Consequently, these configurations inevitably suffer from severe load imbalances and pipeline bubbles when processing the full, highly variable dataset.

To exploit this macro-level stability while managing the unavoidable micro-level pipeline execution imbalance, we propose \name, a distributed MLLM training system that addresses multimodal variability at both macroscopic and microscopic scales.
At the macro level, \name shifts the profiling paradigm from isolated single samples or microbatches to the entire global batch.
Such macro-level profiling accurately captures the converged computational ratio between the various modalities and the core LLM.
Unlike prior works that blindly extrapolate brittle configurations from arbitrary, high-variance snapshots, \name mathematically anchors its parallel configuration to the globally converged workload distribution.
Using the Law of Large Numbers, we prove that a single, static configuration derived from this macroscopic ratio suffices to guarantee optimal load balance throughout training.
This sidesteps both the brittle configurations of micro-level profiling and the prohibitive overheads of dynamic reconfiguration.

While the macro level profiling establishes a stable global baseline, partitioning the global batch into discrete microbatches for execution inevitably re-exposes localized variability.
Because each microbatch is much smaller than the global batch, the Law of Large Numbers no longer holds at this granularity, and workload variability across microbatches persists.
\name addresses this through a decoupled microbatch assignment strategy: it treats the MLLM data flow as a producer-consumer pipeline and decouples the scheduling constraints of each modality to resolve imbalance.
With encoders as producers and the LLM as the consumer, \name ensures constant production and consumption rates independently, balancing the execution time of modality pipeline stages while avoiding buffer stalls.
Our hierarchical microbatch assignment algorithm first achieves a constant production rate by distributing encoder workloads evenly across microbatches, then balances the LLM workload by deferring excess workload from highly variable samples.

We have implemented \name on top of PyTorch and Cornstarch~\cite{cornstarch-arxiv25}.
We evaluate \name on vision-language models (based on Qwen2.5Vision with Llama3-1b and 3b models) across four multimodal datasets with distinct distributions.
Compared to DistTrain~\cite{disttrain-sigcomm25} and DIP~\cite{dip-asplos26}, \name reduces workload variability across microbatches by up to 10.6$\times$, improving end-to-end training throughput by up to 1.40$\times$.

In summary, we make the following contributions:
\begin{denseitemize}
    \item We propose \name, the first distributed training framework to address both heterogeneity and variability in multimodal training workloads via batch-level profiling and decoupled microbatch-level workload balancing.
    \item We introduce a profiling approach that targets the global batch instead of micro-level samples to derive the optimal model-parallel configuration.
    \item We design a hierarchical microbatch assignment algorithm that defers excess workload within each iteration to stabilize workload variability across microbatches.
\end{denseitemize}

\section{Background and Motivation}
\label{sec:background_motivation}

We first introduce MLLM architecture and parallelism (\S\ref{sec:bg_mllm_pipeline}), then describe the dataset characteristics that make balanced workload distribution challenging (\S\ref{sec:bg_mllm_dataset_characteristics}), and finally discuss the limitations of existing approaches (\S\ref{sec:bg_existing_works}).

\subsection{MLLM Architecture and Parallelism}
\label{sec:bg_mllm_pipeline}

\begin{figure}[t]
    \centering
    \includegraphics[width=0.8\columnwidth]{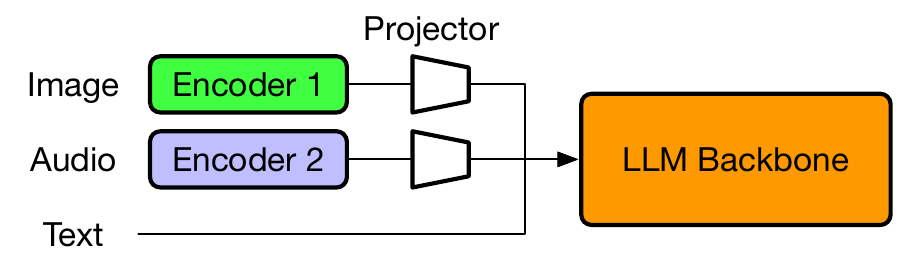}
    \caption{Multimodal LLM architecture.}
    \label{fig:bg_mllm_architecture}
\end{figure}

\begin{figure}[t]
    \centering
    \begin{subfigure}[t]{\columnwidth}
        \includegraphics[width=\columnwidth]{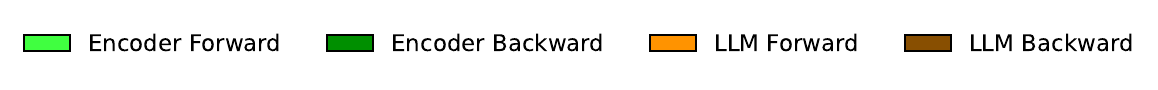}
    \end{subfigure}
    \setcounter{subfigure}{0}
    \begin{subfigure}[t]{\columnwidth}
        \includegraphics[width=\columnwidth]{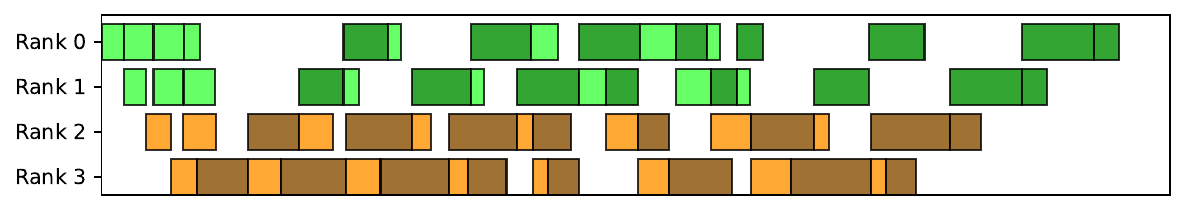}
        \caption{1F1B pipeline parallel schedule.}
        \label{fig:bg_schedule_visualization_1f1b}
    \end{subfigure}
    \begin{subfigure}[t]{\columnwidth}
        \includegraphics[width=\columnwidth]{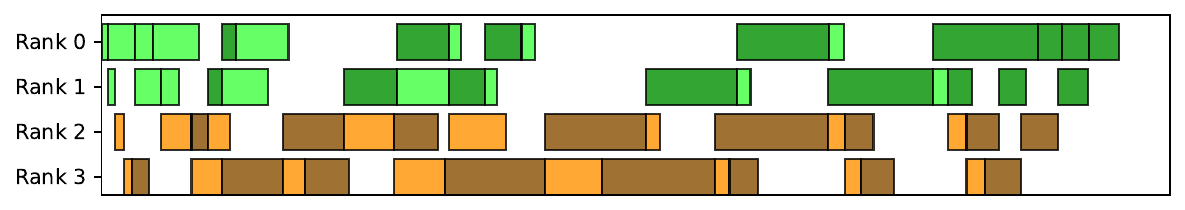}
        \caption{DistTrain~\cite{disttrain-sigcomm25} pipeline parallel schedule. Same as 1F1B but with different microbatch order.}
        \label{fig:bg_schedule_visualization_disttrain}
    \end{subfigure}
    \begin{subfigure}[t]{\columnwidth}
        \includegraphics[width=\columnwidth]{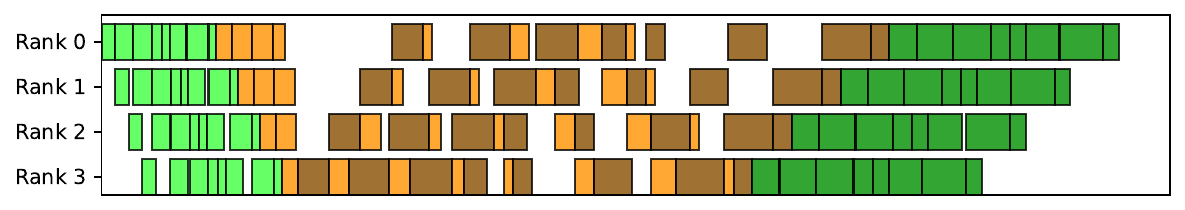}
        \caption{DIP~\cite{dip-asplos26} pipeline parallel schedule.}
        \label{fig:bg_schedule_visualization_pipeweaver}
    \end{subfigure}
    \caption{Visualization of different pipeline parallel schedules using 8 microbatches for a vision language model (VLM).}
    \label{fig:bg_schedule_visualization}
\end{figure}

To efficiently train massive MLLMs, modern distributed systems employ 4D parallelism, combining data (DP), tensor (TP), context (CP), and pipeline parallelism (PP)~\cite{sp-acl23,scalingllama-isca25}.
While these dimensions effectively scale the model across devices, the inherently heterogeneous architecture of MLLMs requires careful parallelization to balance computational workloads across the cluster.

MLLMs enforce a strict structural dependency: raw multimodal inputs must first be processed by modality-specific encoders (\eg, Vision Transformers for images, Whisper for audio), whose outputs are then projected into a unified embedding space for the core LLM backbone, as shown in Figure~\ref{fig:bg_mllm_architecture}.
Under PP, this architectural separation places encoders on earlier pipeline stages and the LLM on subsequent stages.
To maintain high throughput, the global batch is partitioned into microbatches that are injected sequentially.

This pipeline design requires strict temporal and spatial balance: execution times must remain consistent across consecutive microbatches, and all stages must have roughly equal execution times for a given microbatch.
Violations in either dimension immediately produce pipeline bubbles and stragglers.
Figure~\ref{fig:bg_schedule_visualization_1f1b} shows the standard 1F1B schedule for a vision-language model (VLM), where vision encoder stages precede the LLM stages.
DistTrain~\cite{disttrain-sigcomm25} (Figure~\ref{fig:bg_schedule_visualization_disttrain}) uses the same schedule but reorders microbatches to reduce pipeline bubbles.
DIP~\cite{dip-asplos26} (Figure~\ref{fig:bg_schedule_visualization_pipeweaver}) parallelizes modalities independently and colocates their pipeline stages.

\subsection{MLLM Dataset Characteristics}
\label{sec:bg_mllm_dataset_characteristics}

\begin{figure}[t]
    \centering
    \begin{subfigure}[t]{\columnwidth}
        \includegraphics[width=\columnwidth]{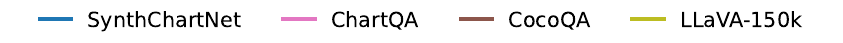}
    \end{subfigure}
    \setcounter{subfigure}{0}
    \begin{subfigure}[t]{0.49\columnwidth}
        \includegraphics[width=\columnwidth]{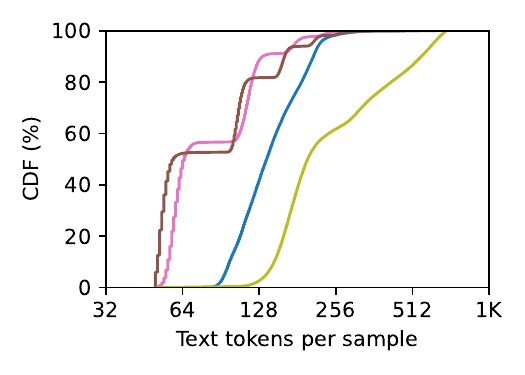}
        \caption{Number of text tokens}
        \label{fig:bg_comparison_text_tokens_cdf}
    \end{subfigure}
    \begin{subfigure}[t]{0.49\columnwidth}
        \includegraphics[width=\columnwidth]{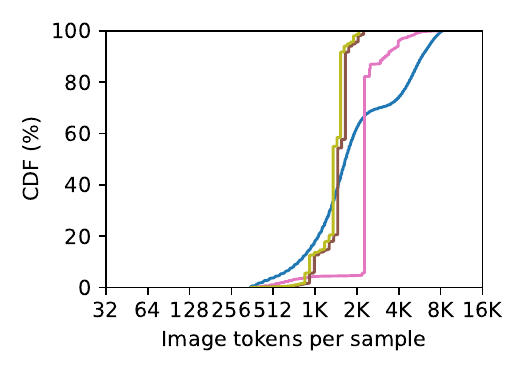}
        \caption{Number of vision tokens}
        \label{fig:bg_comparison_visual_tokens_cdf}
    \end{subfigure}
    \caption{Distributions of number of vision and text tokens in various datasets. They are independently varying.}
    \label{fig:bg_comparison_cdf}
\end{figure}

\begin{figure}[t]
    \centering
    \includegraphics[width=\columnwidth]{figures/comparison_legend.pdf}
    \includegraphics[width=\columnwidth]{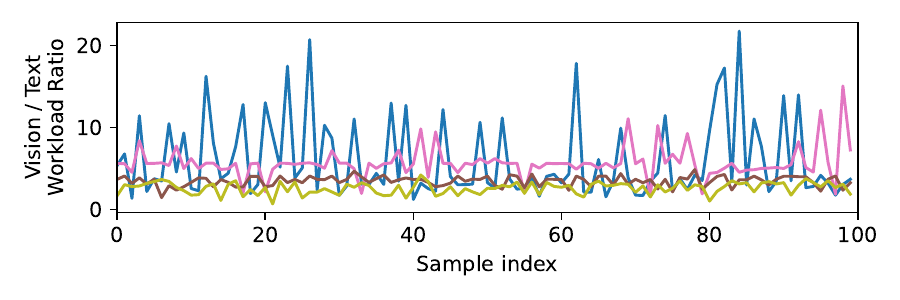}
    \caption{Workload ratio of vision encoder (Qwen2Vision) and LLM (Llama3-1B) across 100 samples in datasets.}
    \label{fig:bg_comparison_workload_ratio}
\end{figure}

While a carefully tuned parallel configuration can theoretically balance the heterogeneous architecture of an MLLM, maintaining this balance at runtime requires perfectly balanced data partitions.
However, two inherent dataset characteristics severely disrupt this: workload heterogeneity and variability.

\textit{Workload heterogeneity} refers to the systematic difference in computational characteristics across modalities.
Modality-specific encoders and the LLM have fundamentally different structures; processing a high-resolution image through a vision encoder exhibits drastically different arithmetic intensity and memory access behavior from processing text tokens through the LLM.
\textit{Workload variability} refers to the sample-to-sample fluctuation in how much of each modality appears.
Unlike unimodal text datasets, where sequence length is the dominant source of variation, each modality in an MLLM dataset follows its own workload distribution.
Figure~\ref{fig:bg_comparison_cdf} plots the CDFs of text and vision tokens across several vision-language datasets, confirming that each modality varies independently.

The challenge is further compounded by an intrinsic coupling constraint: modalities are bound within each sample and must be processed together.
The inter-modality workload ratio is therefore determined jointly per sample rather than controlled independently.
Figure~\ref{fig:bg_comparison_workload_ratio} shows that this per-sample ratio between the vision encoder and the LLM fluctuates drastically, spanning a wide range with no stable central tendency, making any fixed GPU partition inherently mismatched for the vast majority of samples it encounters.

\subsection{Limitations of Existing Works}
\label{sec:bg_existing_works}
\begin{figure}[t]
    \centering
    \includegraphics[width=\columnwidth]{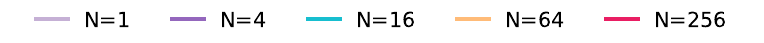}
    \includegraphics[width=\columnwidth]{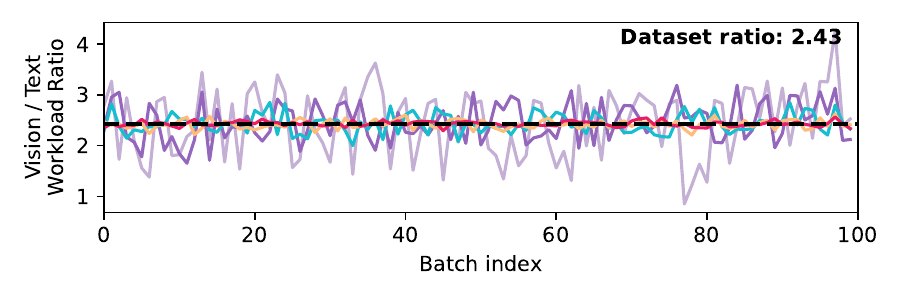}
    \caption{Workload ratio variability with different sample sizes in LLaVA-150K dataset. The mean ratio of vision-to-text workload of the entire dataset is 2.43. With larger sample size $N$, the ratio between batches becomes more stable and converges to the dataset mean.}
    \label{fig:bg_variability_ratio_vs_sample_size}
\end{figure}

Existing MLLM training systems~\cite{optimus-atc25,disttrain-sigcomm25,dip-asplos26} recognize that workload imbalances arise at two granularities -- profiling and execution -- but no single system addresses both.

\parabf{Profiling granularity.}
DistTrain~\cite{disttrain-sigcomm25} and DIP~\cite{dip-asplos26} derive parallel configurations by profiling a single sample or a small microbatch, which may not represent the entire dataset.
The high per-sample variance suggests that dynamic reconfiguration -- as applied to sequence-length variability in LLM training~\cite{hotspa-sosp24,bytescale-sigcomm25} -- is the right approach.
Counterintuitively, however, Figure~\ref{fig:bg_variability_ratio_vs_sample_size} shows that the workload ratio converges to a stable dataset mean as batch size increases, justifying a static configuration -- but only when derived from this macroscopic ratio rather than an unrepresentative micro-sample.

\begin{figure}[t]
    \centering
    \includegraphics[width=\columnwidth]{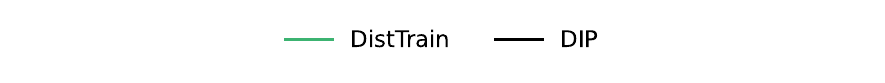}
    \includegraphics[width=\columnwidth]{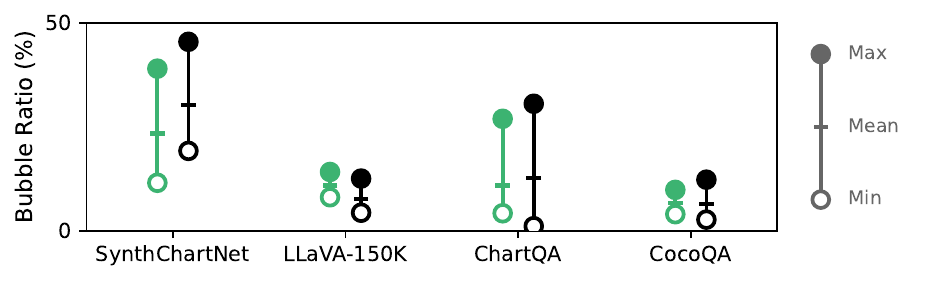}
    \caption{Pipeline bubbles of existing works vs ideal pipeline schedule with perfect workload balance.}
    \label{fig:bg_pipeline_bubble}
\end{figure}

\parabf{Execution granularity.}
Even with a correct parallel configuration, the workload ratio continues to fluctuate across microbatches.
Optimus~\cite{optimus-atc25} and DistTrain strictly couple modalities within the microbatch boundary, forcing the encoder and LLM to process an identical set of samples per microbatch.
Optimus schedules encoder stages into LLM pipeline bubbles, but with more microbatches or techniques such as zero-bubble pipeline parallelism (ZBPP)~\cite{zbpp-iclr24}, few bubbles remain to exploit, and fragmented bubbles from imbalance are not exploitable.
DistTrain relies on data reordering, which mitigates but cannot resolve workload variability across microbatches.
DIP~\cite{dip-asplos26} partitions microbatches into sub-microbatches but still confines these partitions within the rigid boundaries of the parent microbatch.
Figure~\ref{fig:bg_pipeline_bubble} shows pipeline bubbles across four datasets (100 iterations, 4-stage PP, 16 microbatches); imbalance-driven bubbles compound the irreducible data-dependency bubbles.

\section{\name Overview}
\label{sec:design}

\begin{figure}[t]
  \centering
  \includegraphics[width=\columnwidth]{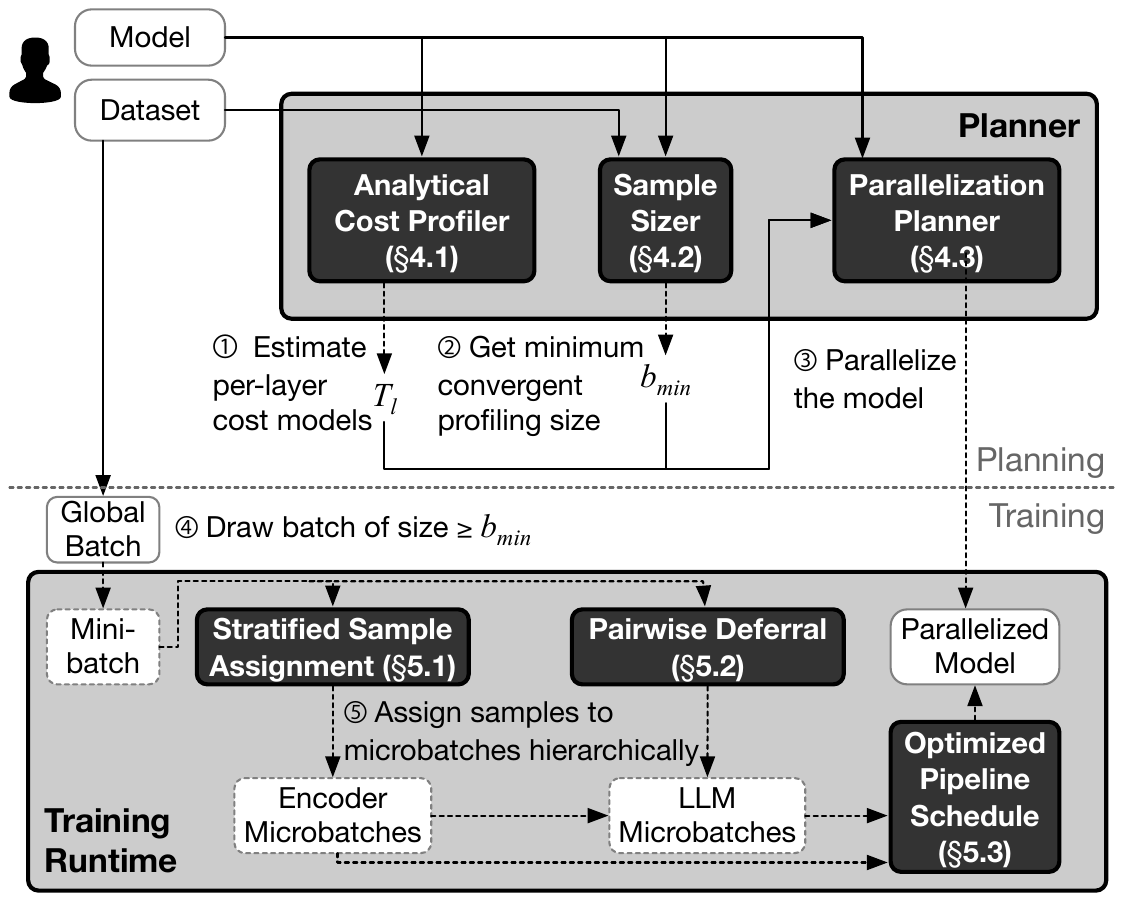}
  \caption{\name design overview.}
  \label{fig:design_overview}
\end{figure}

\name is a distributed MLLM training framework that addresses workload heterogeneity and variability in multimodal datasets.
\name maximizes training throughput by co-designing static hardware resource allocation and dynamic data scheduling for distributed multimodal training:
\begin{denseitemize}
  \item \textbf{Macroscopic profiling-based parallelization (\S\ref{sec:model_configuration})}:\\\name profiles large batch to derive a parallel configuration that provides optimal load balance across modalities throughout training.
  \item \textbf{Microscopic workload balancing via deferral (\S\ref{sec:hierarchical_microbatch_assignment})}: Inspired by the producer-consumer model, \name decouples microbatch partitioning across modalities and balances workload by deferring excess LLM computation across microbatch boundaries.
\end{denseitemize}

Figure~\ref{fig:design_overview} shows the architecture of \name, which consists of two components: the planner and the training runtime.

\parabf{Planner.}
The planner runs before training begins, analyzing the model and dataset to derive the parallel configuration to allocate hardware resources to each modality.
Instead of profiling with a single sample or a small microbatch, the planner profiles the large batch, capturing the converged workload ratio between modalities rather than a noisy, unrepresentative snapshot.
The derived parallel configuration assigns hardware resources to each modality in proportion to its computational demand.

Since profiling the full large batch before evaluating parallelization is infeasible, the planner first builds an analytical cost model by profiling with synthetic samples and fitting a polynomial equation via linear regression (\S\ref{sec:design_cost_model}).
Using this cost model, the planner then derives the minimum profiling batch size ($b_{min}$) that guarantees statistical convergence of the workload ratio (\S\ref{sec:design_math_bound}).
The user must set the global batch size to $b_{min}$ or larger; otherwise, \name does not guarantee optimal load balance across iterations.
Finally, the cost model and $b_{min}$ together enable \name to find the optimal parallel configuration that maximizes training throughput (\S\ref{sec:design_pipeline_balancing}).

\parabf{Training runtime.}
At each iteration, the runtime fetches a global batch and distributes it evenly across data-parallel replicas, each receiving a \textit{minibatch} of size $B_{global}/DP$.
To distribute samples, the runtime sorts them by encoder workload in descending order and greedily assigns each to the replica with the minimum current LLM workload.
Sorting by encoder workload descending spreads the heaviest encoder samples across replicas before smaller ones fill the gaps, balancing encoder workload.
For LLM workload, each sample is assigned to the replica that has accumulated the least LLM workload so far, preventing any single replica from accumulating disproportionately more LLM workload than the others.
Together, these heuristics balance both encoder and LLM workloads across replicas.

Each replica then partitions its minibatch into microbatches for pipeline execution.
While the planner's configuration is macroscopically optimal, this partitioning re-exposes sample-level workload variance.
Because encoder and LLM workloads vary independently across samples, balancing both simultaneously within each microbatch is intractable.
\name addresses this by decoupling microbatch boundaries across modalities: the encoder and LLM need not process the same samples in a given microbatch.
\name first strictly balances encoder microbatches (\S\ref{sec:stratified_sample_partitioning}), then \textit{defers} excess LLM workload from overloaded to underloaded microbatches, equalizing execution time across the pipeline (\S\ref{sec:global_deferral}).
The overhead of deferral due to the reversed data dependency in backward pass is handled by optimizing pipeline schedule (\S\ref{sec:backward_pass_dependency}).

\section{Macroscopic Analysis-Based Model\\ Parallelization}
\label{sec:model_configuration}

The primary objective of macroscopic analysis is to derive a single, static parallel configuration that optimally balances the heterogeneous MLLM architecture.
As established in Section~\ref{sec:bg_existing_works}, deriving this configuration via microscopic profiling fails due to the extreme sample-level variance inherent in multimodal datasets.
\name instead anchors its resource allocation to a macroscopic view of the dataset.

This section proceeds in three stages.
We first build a hardware-calibrated analytical cost model $T_{l,i}$ that estimates per-layer workloads without exhaustive profiling (\S\ref{sec:design_cost_model}).
We then use it to determine a profiling batch size $b_{min}$ large enough to yield a stable macro-level allocation (\S\ref{sec:design_math_bound}).
Finally, we derive the static parallel configuration through hierarchical pipeline balancing that first optimizes each modality component locally and then aligns them under a shared pipeline schedule (\S\ref{sec:design_pipeline_balancing}).

\subsection{Hardware-Calibrated Analytical Cost Model}
\label{sec:design_cost_model}

Determining $b_{min}$ and searching over configurations both require repeated workload estimates across many batches and candidate partitions.
Exhaustive profiling is prohibitively expensive in both cases, and pure FLOP counting is insufficient because it misses hardware-specific effects such as memory-bandwidth saturation and kernel-launch overhead.

\name therefore pre-builds a hardware-calibrated analytical cost model before any profiling or parallel configuration search begins.
It profiles every layer separately over the valid $(TP, CP)$ configurations using synthetic inputs at representative sizes (\eg, number of tokens $x \in \{64, 256, 1k, 4k, 16k\}$), then fits the measurements to a configuration-aware quadratic model $T_{l,i}(x, \text{TP}, \text{CP}) = a x^2 + b x + c$~\cite{megatron-sc21,attention-nips17} via linear regression.
We fit a model per layer because different layer types scale differently with sequence length -- \eg, attention is $O(x^2)$ while MLPs and embeddings are $O(x)$ -- and it enables estimating any pipeline stage cost by summing the costs of the individual layers it contains.

\subsection{Deriving Minimum Stable Profiling Batch Size}
\label{sec:design_math_bound}

Before beginning parallel configuration search, \name must determine a minimum profiling batch size $b_{min}$ large enough that random batches of $b_{min}$ samples consistently yield the same discrete GPU allocation under the chosen data-parallel degree $DP$.
\name establishes this guarantee in two steps.
First, it repeatedly samples batches of size $b_{min}$ and uses Bernoulli trials to test whether the resulting allocation is stable.
Second, it uses the Law of Large Numbers to show that stability at batch size $b_{min}$ implies stability for any larger global batch size $B_{global} \ge b_{min}$.

\begin{algorithm}[htbp]
    \caption{Probabilistic Macroscopic Profiling Strategy}
    \label{alg:prob_profiling}
    \footnotesize
    \begin{algorithmic}[1]
    \item[]{\textbf{Input: } Confidence level $1-\alpha$, Target error limit $p_{error}$, Initial sample size $n_0$, GPUs $N_{total}$, Data Parallel degree $DP$, Cost Models $T_{l,i}$}
    \item[]{\textbf{Output: } Stable profiling batch size $b_{min}$}
    \item[]\hrulefill
    \State $k \gets \lceil \ln(\alpha) / \ln(1 - p_{error}) \rceil$ \Comment{Compute required validation trials via Binomial bounds}
    \State $n \gets n_0$
    \Loop
    \State $\vec{P}_{ref} \gets \text{EstimateMacroscopicProportions}(n, T_{l,i})$ \label{line:prob_profiling_estimate_macroscopic_proportions}
    \State $\vec{M}_{ref} \gets \text{ProportionalAllocation}(N_{total}, DP, \vec{P}_{ref})$ \label{line:quantize_to_discrete_gpus}
    \State $IsStable \gets \text{True}$
    
    \For{$t = 1$ to $k$} \label{line:prob_profiling_loop}
        \State $\vec{P}_{test} \gets \text{EstimateMacroscopicProportions}(n, T_{l,i})$
        \State $\vec{M}_{test} \gets \text{ProportionalAllocation}(N_{total}, DP, \vec{P}_{test})$
        \If{$\vec{M}_{test} \neq \vec{M}_{ref}$}
            \State $IsStable \gets \text{False}$
            \State \textbf{break} \label{line:prob_profiling_break}
        \EndIf
    \EndFor
    
    \If{$IsStable$}
        \State \Return $n$ \Comment{Minimum stable profiling batch size $b_{min}$} \label{line:prob_profiling_return}
    \EndIf
    \State $n \gets n \times 2$ \Comment{Increase batch size if variance is too high}
    
    \EndLoop
    \end{algorithmic}
\end{algorithm}

Algorithm~\ref{alg:prob_profiling} describes the probabilistic procedure for determining the minimum stable profiling batch size $b_{min}$.
For a candidate $n$, \name first draws a reference batch and estimates the workload proportions across modality components $\vec{P}_{ref}$ (Line~\ref{line:prob_profiling_estimate_macroscopic_proportions}).
These proportions are continuous floating-point values, but GPU assignment must be discrete; thus, \name converts them into $\vec{M}_{ref}$, a vector of per-modality per-replica GPU counts, by distributing the per-replica budget of $N_{total}/DP$ GPUs proportionally across modality components and rounding to the nearest feasible integers (Line~\ref{line:quantize_to_discrete_gpus}).
It then draws $k$ additional independent validation batches (Line~\ref{line:prob_profiling_loop}), and for each one, treats the event ``the resulting discrete GPU allocation differs from the reference allocation'' as a Bernoulli failure (Line~\ref{line:prob_profiling_break}).
Because the continuous ratio is rounded to integer counts, small fluctuations between batches often fall into the same allocation (\eg, 1:0.98 and 1:1.12 both round to 1:1 when only 2 GPUs are available).

If all $k$ trials return the same allocation, standard binomial bounds imply that, with confidence $1-\alpha$, a fresh batch of size $b_{min}$ will yield a different allocation with probability at most $p_{error}$ (Line~\ref{line:prob_profiling_return}).
At batch size $b_{min}$, the Bernoulli test thus establishes that sampling noise is small enough that repeated random draws induce the same discrete allocation with high probability.
If the test fails, \name doubles the batch size and repeats.
The loop is guaranteed to terminate: by the Strong Law of Large Numbers, the sample mean converges almost surely to the population mean, so the test must eventually pass (Appendix~\ref{app:bmax_derivation}).
In our experiments, convergence occurs by batch size 256 across all evaluated datasets and GPU configurations (Appendix~\ref{app:eval-workload-ratio-with-different-batch-sizes}).

The Bernoulli test certifies stability only at the profiled size $b_{min}$; the Law of Large Numbers extends the guarantee to all larger batches.
$\vec{P}_{ref}$ and $\vec{P}_{test}$ are sample means over per-sample workloads.
By the Law of Large Numbers, they converge to the population mean as the batch size grows; so the estimator at any $B_{global} \ge b_{min}$ is at least as concentrated as the one at size $b_{min}$.
Hence, if random size-$b_{min}$ batches already induce the same allocation with high probability, any global batch of size $B_{global} \ge b_{min}$ does so with at least as high probability.
Appendix~\ref{app:statistical_bound_for_convergence} provides the full derivation.
Overall, this probabilistic procedure allows \name to use a small profiling batch while still recovering the correct macro-level hardware allocation.

\subsection{Heterogeneous Pipeline Balancing}
\label{sec:design_pipeline_balancing}

Given a stable profiling size $b_{min}$ and calibrated layer costs $T_{l,i}$, the remaining task is to derive a static parallel configuration.
An MLLM forms a heterogeneous pipeline: modality-specific encoders feed the LLM, and end-to-end throughput is bounded by $\beta_{max}$, the slowest stage across all modality components.
Rather than searching over all parallel dimensions of the full model, \name exploits the modularity of MLLMs -- each modality encoder and the LLM are individual components -- and decomposes the search into two tiers: it first optimizes each modality component locally, then evaluates the resulting per-component bottlenecks under a shared pipeline schedule.
Algorithm~\ref{alg:config_search} summarizes this search.

\begin{algorithm}[htbp]
    \caption{Heterogeneous Model Parallel Configuration Search}
    \label{alg:config_search}
    \footnotesize
    \begin{algorithmic}[1]
    \item[]{\textbf{Input: } Stable profiling batch size $b_{min}$, Global Batch $B_{global}$, Microbatch size $\mu$, GPUs $N_{total}$, Modality components $\mathcal{C}$, Cost Models $T_{l,i}$}
    \item[]{\textbf{Output: } Optimal static parallel configuration $C^*$}
    \item[]\hrulefill
    \State $\vec{P}_{pop} \gets \text{EstimateMacroscopicProportions}(b_{min}, T_{l,i})$ \label{line:estimate_macroscopic_proportions}
    \State $MaxThroughput \gets 0$
    
    \For{\textbf{each} valid hardware topology $C_{hw} = \{DP, \vec{TP}, \vec{CP}, \vec{PP}\}$}
    \State $\vec{M} \gets \text{ProportionalAllocation}(N_{total}, DP, \vec{P}_{pop})$
    \If{$C_{hw}$ violates VRAM limits or $B_{global} \not\equiv 0 \pmod{DP \cdot \mu}$}
    \State \textbf{continue}
    \EndIf
    \State $K \gets B_{global} / (DP \cdot \mu)$
    
    \LineComment{\textit{Tier 1: Intra-module balancing via Dynamic Programming}}
    \For{each modality component $i \in \mathcal{C}$}
        \State $\{\tau_{i,p}\}_{1 \le p \le PP_i} \gets \text{IntraModuleBalance}(i, PP_i, TP_i, CP_i, T_{l,i})$
        \State $\beta_i \gets \max_{1 \leq p \leq PP_i} \tau_{i,p}$
    \EndFor
    
    \LineComment{\textit{Tier 2: Inter-module balancing via Slowest-Stage Evaluation}}
    \State $\beta_{max} \gets \max_{i \in \mathcal{C}}(\beta_i)$
    \State $\mathcal{P}_{reshard} \gets \text{ComputeReshardCost}(\vec{TP}, \vec{CP}, K, B_{global}, DP)$
    \State $T_{iter} \gets \mathcal{T}_{\mathcal{S}}\Big(K, \{\tau_{i,p}\}, \beta_{max}\Big) + \mathcal{P}_{reshard}$
    \State $Throughput \gets (DP \cdot K) / T_{iter}$
    
    \If{$Throughput > MaxThroughput$}
        \State $MaxThroughput \gets Throughput$
        \State $C^* \gets C_{hw}$
    \EndIf
    
    \EndFor
    \State \Return $C^*$
    \end{algorithmic}
\end{algorithm}

\parabf{Intra-Module Balancing.}
Intra-module balancing balances the workload of pipeline stages within a modality component.
Each modality encoder and the LLM backbone are internally homogeneous, consisting of repeated layers with similar structure.
\name therefore treats each modality component $i \in \mathcal{C}$ as an independent unimodal partitioning problem, which is well-studied~\cite{alpa-osdi22,oobleck-sosp23,unity-osdi22}.
A 1D dynamic program uses the layer-wise costs evaluated under the chosen $TP_i$ and $CP_i$ to partition the modality component into $PP_i$ stages, minimizing the maximum stage latency~\cite{alpa-osdi22}.

Let $F_i(\ell, p)$ denote the minimum bottleneck latency when the first $\ell$ layers of modality component $i$ are partitioned into $p$ stages.
The recurrence is:
\begin{equation}
F_i(\ell, p) = \min_{0 \leq \ell' < \ell} \max \left( F_i(\ell', p-1), \sum_{j=\ell'+1}^{\ell} T_{j,i} \right),
\end{equation}
with base case $F_i(\ell, 1) = \sum_{j=1}^{\ell} T_{j,i}$.
Let $\tau_{i,p}$ denote the latency of stage $p$ in the optimal partition of modality component $i$; the per-component bottleneck is $\beta_i = \max_{1 \leq p \leq PP_i} \tau_{i,p} = F_i(L_i, PP_i)$.
Backtracking recovers both the stage latencies $\tau_{i,p}$ and the layer-to-stage mapping for execution.

\parabf{Inter-Module Balancing.} 
These per-component optima are only local.
When independently optimized modality components are combined into a heterogeneous pipeline, discrepancies in latency of pipeline stages across components directly translate to pipeline bubbles.
Given the shared microbatch count $K = B_{global} / (DP \cdot \mu)$, where $\mu$ is the microbatch size, \name evaluates each valid factorization using the stage-latency set $\{\tau_{i,p}\}$ and the the slowest stage latency across all components $\beta_{max} = \max_{i \in \mathcal{C}} \beta_i$.
Here, $\mathcal{T}_{\mathcal{S}}$ denotes the analytical iteration time of pipeline schedule $\mathcal{S}$~\cite{alpa-osdi22}:
\begin{equation}
    \mathcal{T}_{\mathcal{S}}\big(K,\{\tau_{i,p}\}, \beta_{max}\big)
    =
    \sum_{i \in \mathcal{C}} \sum_{p=1}^{PP_i} \tau_{i,p}
    + (K - 1)\beta_{max}.
\end{equation}
When adjacent modality components use different tensor- or context-parallel degrees, \name models the resulting communication overhead at their boundaries as a resharding cost $\mathcal{P}_{reshard}$.
The total iteration time thus includes $\mathcal{P}_{reshard}$.
The selected configuration maximizes the resulting end-to-end training throughput.

This decomposition keeps the search tractable.
For each component $i$, the search enumerates only valid spatial factorizations satisfying $TP_i \times CP_i \times PP_i = M_i$, where $M_i$ is the per-replica GPU budget for component $i$, rather than brute-forcing arbitrary parallel dimensions.
\section{Hierarchical Microbatch Assignment}
\label{sec:hierarchical_microbatch_assignment}

\begin{figure}[t]
    \centering
    \begin{subfigure}[t]{\columnwidth}
        \includegraphics[width=\columnwidth]{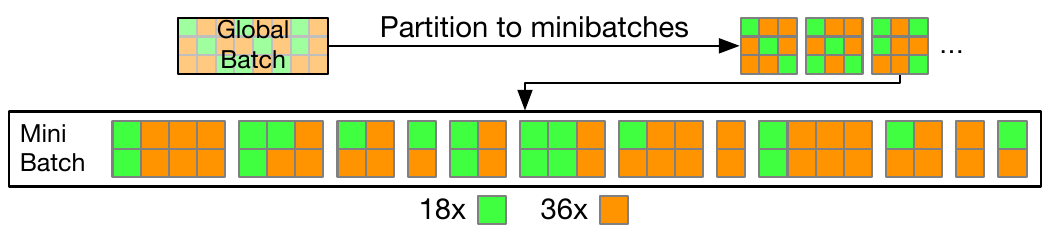}
        \caption{An example of a minibatch after the global batch is partitioned.}
        \label{fig:mandu_hierarchical_mb_1}
    \end{subfigure}
    \begin{subfigure}[t]{\columnwidth}
        \includegraphics[width=\columnwidth]{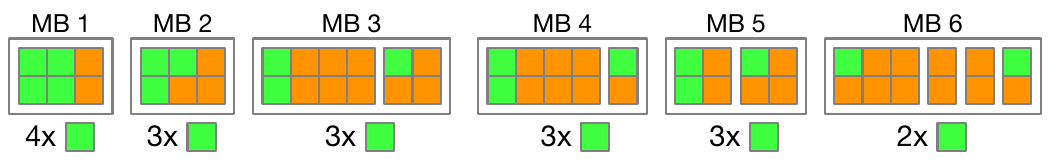}
        \caption{Stratified sample assignment to microbatches focusing on balancing encoder workload (\S\ref{sec:stratified_sample_partitioning}).}
        \label{fig:mandu_hierarchical_mb_2}
    \end{subfigure}
    \begin{subfigure}[t]{\columnwidth}
        \includegraphics[width=\columnwidth]{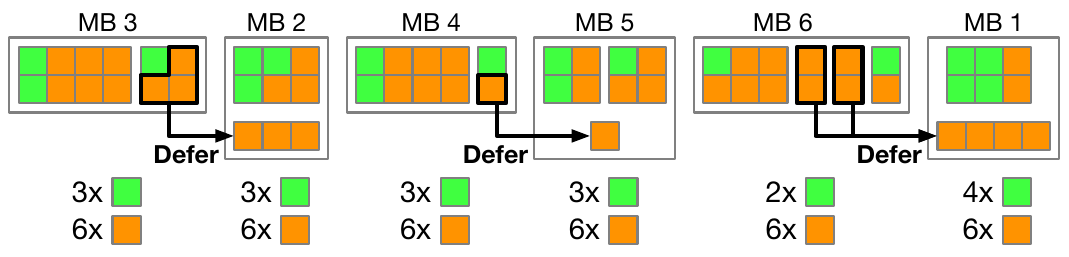}
        \caption{Pairwise deferral optimization to balance LLM workload. Microbatches are reordered (\S\ref{sec:global_deferral}).}
        \label{fig:mandu_hierarchical_mb_3}
    \end{subfigure}
    \caption{The hierarchical microbatch assignment algorithm. The number of boxes represents the amount of workload of each sample, where green boxes are for encoder workload and orange boxes are for LLM workload.}
    \label{fig:mandu_hierarchical_mb_assignment}
\end{figure}

Section~\ref{sec:design} described how \name distributes the global batch across replicas.
However, the harder challenge arises when each replica's minibatch is partitioned into microbatches for pipeline parallelism.
The small number of samples per microbatch amplifies individual workload variance, and encoder and LLM stages exhibit independent workload distributions that cannot be balanced simultaneously.
Figure~\ref{fig:mandu_hierarchical_mb_assignment} shows that encoder and LLM workloads are highly variable between microbatches and how hierarchical microbatch assignment balances the workload.
See Appendix~\ref{app:pipeline_schedule_with_pairwise_deferral} how these microbatches are executed in 1F1B pipeline schedule.

\name resolves this by adopting a \textit{producer-consumer strategy}: it exploits the pipeline buffer between the encoders (producers) and the LLM (consumer) as a workload variability absorber.
Instead of always enforcing equal sample counts per microbatch across both stages, \name decouples their schedules 
through a two-step hierarchical assignment, as depicted in Figure~\ref{fig:mandu_hierarchical_mb_assignment}.
In the first step, stratified sample assignment balances encoder execution time while laying the groundwork for the subsequent deferral (\S\ref{sec:stratified_sample_partitioning}).
In the second step, pairwise deferral optimization shifts residual LLM workload across microbatches to balance the LLM execution time without disturbing the encoder schedule (\S\ref{sec:global_deferral}).
Algorithm~\ref{alg:hierarchical_microbatch_assignment} summarizes the hierarchical microbatch assignment algorithm.

The backward pass, however, inherently reverses the data dependency; the encoder backward pass requires gradients from the LLM, thus deferral widens this latency gap for deferred samples.
\name handles this through split-backward processing and eager forward scheduling (\S\ref{sec:backward_pass_dependency}).

If the model has multiple modality encoders (\eg, Omni-modal models~\cite{qwen3omni-arxiv25,mingomni-arxiv25,olaomni-arxiv25}), merging the modality encoders into a single unified encoder module allows hierarchical microbatch assignment to be applied.

\begin{algorithm}[t]
\caption{Hierarchical microbatch assignment.}
\label{alg:hierarchical_microbatch_assignment}
\begin{algorithmic}[1]
    \item[]{\textbf{Input: } Global batch $B_{global}$, Data Parallel degree $DP$, number of microbatches $K$}
    \item[]{\textbf{Output: } Encoder microbatches $\mathcal{M}^{\text{enc}}$, LLM microbatches $\mathcal{M}^{\text{LLM}}$}
    \item[]\hrulefill
    
    \LineComment{Section~\ref{sec:design}: DP-level sample assignment to replicas}
    \State Sort $B_{global}$ by $w_{\text{encoder},i}$ in descending order \label{line:global_sort}
    \State Assign samples to each replica; let $S$ denote samples assigned to this replica \label{line:global_roundrobin}

    \LineComment{Section~\ref{sec:stratified_sample_partitioning}: stratified sample assignment to microbatches}
    \State $K_{\text{eff}} \leftarrow \min\!\left(K,\, \left\lfloor \sum_i w_{\text{encoder},i} / w_{\text{encoder},\text{max}} \right\rfloor\right)$ \label{line:keff}
    \State Partition $S$ into $S_{\text{c}}$ and $S_{\text{f}}$ by LLM workload \label{line:stratified_partitioning}
    \State Sort $S_{\text{c}}$ and $S_{\text{f}}$ by $w_{\text{encoder},i}$ in descending order \label{line:stratified_sorting}
    \State Assign $w_{\text{encoder}}$ of $S_{\text{c}}$ then $S_{\text{f}}$ to $\mathcal{M}^{\text{enc}}$ via Min-Max greedy \label{line:stratified_assign_minmax}
    
    \LineComment{Section~\ref{sec:global_deferral}: pairwise deferral optimization}
    \State Sort microbatches in $\mathcal{M}^{\text{enc}}$ by $w_{\text{LLM}}$ in descending order
    \State Partition into $S_{\text{ol}}$ (top $K/2$) and $S_{\text{ul}}$ (bottom $K/2$) \label{line:pairwise_partitioning}
    \For{each pair $(i, j)$ where $i \in S_{\text{ol}}$ and $j \in S_{\text{ul}}$}
    \State $S_{\text{deferred}} \leftarrow \textsc{SubsetSumDynamicProgramming}(i, j)$ \label{line:pairwise_compute_deferred_set}
    \State $V_{i, j} \leftarrow \textsc{CalcBottleneck}(S_{\text{deferred}}, i, j)$ \Comment{Equation~\ref{eq:bottleneck_cost_formulation}} \label{line:pairwise_compute_bottleneck_cost}
    \EndFor
    \State $L \leftarrow [w_{\text{LLM}, i} \text{ for } i \in S_{\text{ol}}]$ \label{line:pairwise_compute_standalone_cost}
    \State $T^*, \mathcal{P} \leftarrow \textsc{BottleneckMatch}(V, L, S_{\text{ol}}, S_{\text{ul}})$ \label{line:pairwise_maxcardinality}
    \State Reorder microbatches in $\mathcal{M}^{\text{enc}}$ into interleaved pairs $(ol_0, ul_0, ol_1, ul_1, \ldots)$ per $\mathcal{P}$
    \State Assign LLM workload of corresponding samples in $\mathcal{M}^{\text{enc}}$ to $\mathcal{M}^{\text{LLM}}$, deferring samples specified by $\mathcal{P}$ \label{line:pairwise_assign_deferrals}
    \State \Return $\mathcal{M}^{\text{enc}}, \mathcal{M}^{\text{LLM}}$
\end{algorithmic}
\end{algorithm}

\subsection{Stratified Sample Assignment to Microbatches}
\label{sec:stratified_sample_partitioning}

\name assigns samples to microbatches using a stratified strategy that balances encoder execution time while ensuring each microbatch retains sufficient fine-grained samples for the subsequent deferral phase.

Before assignment, \name determines the effective number of microbatches $K_{\text{eff}}$ to use (line~\ref{line:keff}).
As each sample's computation is indivisible across microbatches, a single sample with the maximum encoder workload $w_{\text{encoder},\text{max}}$ sets a lower bound on any microbatch's encoder workload.
If $K$ is too large, this microbatch is dominated by $w_{\text{encoder},\text{max}}$ while the remaining $K-1$ microbatches would be significantly lighter.
In this case, even though $K$ is given by the user, \name reduces $K$ to $K_{\text{eff}}$ so that each microbatch has workload close to $w_{\text{encoder},\text{max}}$.
If other samples' workload sum is large enough to create $K-1$ microbatches with similar or larger workload, \name uses the user's $K$ as $K_{\text{eff}}$.

After determining the effective number of microbatches, \name assigns samples to $K_{\text{eff}}$ microbatches.
Samples are first partitioned into a coarse-grained set $S_c$ with high LLM workload and a fine-grained set $S_f$ with low LLM workload (line~\ref{line:stratified_partitioning}), then assigned to microbatches sequentially.
Within each set, samples are sorted by $w_{\text{encoder},i}$ in descending order (line~\ref{line:stratified_sorting}) and assigned to microbatches via a Min-Max greedy heuristic on encoder workload (line~\ref{line:stratified_assign_minmax}).
Figure~\ref{fig:mandu_hierarchical_mb_2} illustrates the result of assigning samples in Figure~\ref{fig:mandu_hierarchical_mb_1} to microbatches by applying the stratified sample assignment.
The total encoder workload is 18 units, almost evenly distributed across 6 microbatches with 3 units each on average.
We do not consider the LLM workload balancing in this phase, so the microbatches are balanced only on the encoder workload.
Unlike traditional pipeline parallelism, \name allows microbatch sample counts to vary, prioritizing encoder balance over count uniformity.

The partition into $S_c$ and $S_f$ is to facilitate the deferral phase.
Without this partition, the greedy -- which is blind to LLM workload -- may leave some microbatches with only high-LLM-workload samples, starving them of the low-LLM-workload units needed for deferral and preventing deferral from balancing LLM workload.
By separating samples into $S_c$ and $S_f$ upfront, every microbatch is guaranteed to receive $S_f$ samples, ensuring pairwise deferral always has units available to shift and balance LLM workload delicately.
This sequential two-subset assignment does not break the encoder balancing guarantee.
The greedy always assigns to the least-loaded microbatch and each subset is independently sorted in decreasing order of encoder workload, thus the combined assignment is a valid longest processing time (LPT) list scheduling run, which by Graham's theorem yields a makespan within $(2 - 1/K)$ times optimal, where $K$ is the number of microbatches $B_{global}/(DP \cdot \mu)$~\cite{graham-scheduling-siam69}.

\subsection{Pairwise Deferral Optimization}
\label{sec:global_deferral}

\begin{figure}[t]
    \centering
    \begin{subfigure}[t]{\columnwidth}
        \includegraphics[width=\columnwidth]{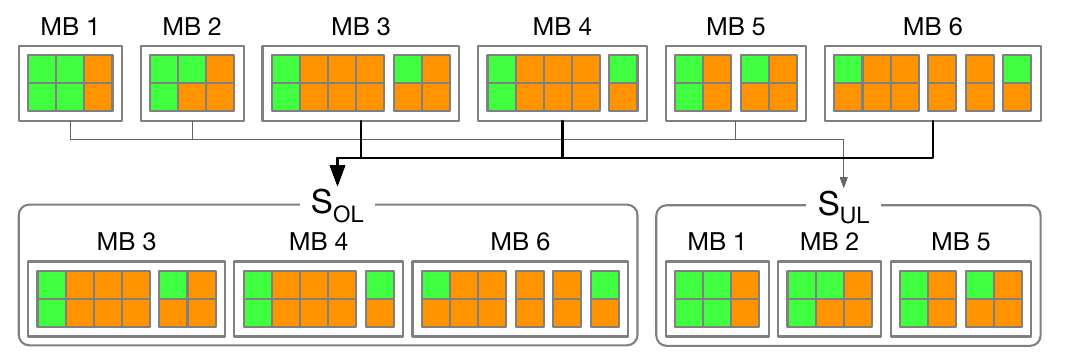}
        \caption{Dividing a set of microbatches to an overloaded set ($S_{ol}$) and an underloaded set ($S_{ul}$).}
        \label{fig:deferral_bipartite_graph_0}
    \end{subfigure}
    \begin{subfigure}[t]{\columnwidth}
        \centering
        \includegraphics[width=0.8\columnwidth]{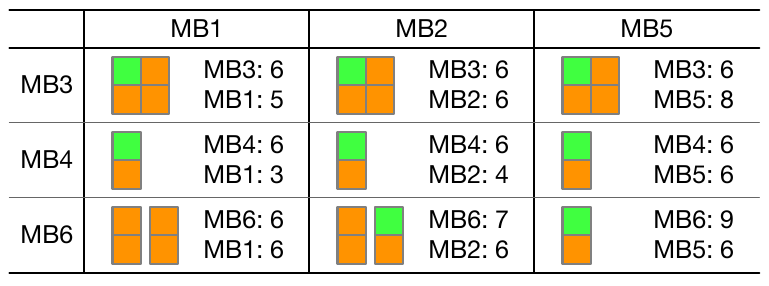}
        \caption{Computed deferral set $S_{\text{deferred}}$ for microbatch $i$ and $j$ and LLM workload $w_{\text{LLM}, i}$ and $w_{\text{LLM}, j}$ after deferral.}
        \label{fig:deferral_bipartite_graph_1}
    \end{subfigure}
    \begin{subfigure}[t]{\columnwidth}
        \centering
        \includegraphics[width=0.8\columnwidth]{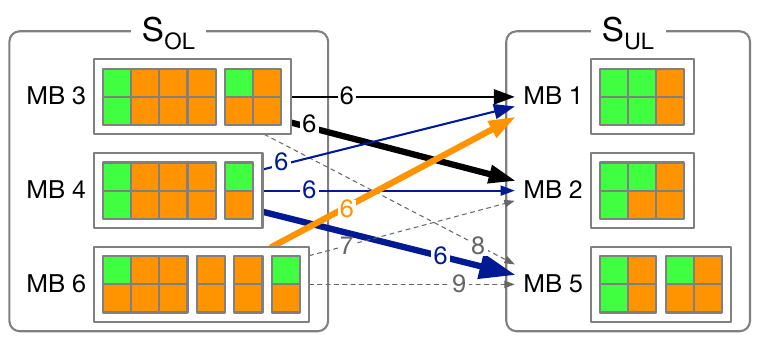}
        \caption{A bipartite graph $\mathcal{G}(S_{\text{ol}}, S_{\text{ul}}, V)$. Edges with $V_{i,j} >T$ are represented as grey dotted lines. Bold lines represent the matching $\mathcal{P}$.}
        \label{fig:deferral_bipartite_graph_2}
    \end{subfigure}
    \caption{A visualization of pairwise deferral optimization with microbatches in Figure~\ref{fig:mandu_hierarchical_mb_2}.}
    \label{fig:pairwise_deferral_optimization}
\end{figure}

After stratified sample assignment, \name balances LLM execution time across microbatches by \textit{deferring} selected sample's LLM computation from overloaded microbatches to underloaded ones, leaving the encoder schedule intact.
After deferral, microbatches are paired and reordered so that each overloaded microbatch is immediately followed by its underloaded partner, and selected samples' LLM workload is shifted between each pair.
This \textit{pairwise reordering} minimizes peak activation buffer memory.
Deferred samples must retain their encoder activations until the LLM processes them, and placing the underloaded partner immediately after its overloaded counterpart bounds the buffering duration to a single microbatch interval.
Figure~\ref{fig:mandu_hierarchical_mb_3} demonstrates the result of applying the pairwise deferral optimization to the microbatches in Figure~\ref{fig:mandu_hierarchical_mb_2}.
The encoder workload is not affected by the deferral, while the LLM workload is evenly distributed across microbatches with 6 units each.

The core challenge is that the two decisions required for deferral are tightly coupled: which microbatches to pair, and which specific samples to transfer within each pair.
\name therefore formulates deferral as a \textit{bottleneck assignment problem} that co-optimizes both decisions to minimize the maximum LLM execution time across all microbatches.
The solution proceeds in three steps: finding the optimal subset of samples to defer for every candidate pair (Optimal deferral set calculation), aggregating these into a bottleneck cost matrix (Bottleneck cost formulation), and determining the globally optimal pairing (Bottleneck matching optimization).
Figure~\ref{fig:pairwise_deferral_optimization} visualizes the deferral optimization.

\parabf{Optimal deferral set calculation.}
\name sorts the $K$ microbatches by LLM workload and partitions them into an overloaded set $S_{\text{ol}}$ (top $K/2$) and an underloaded set $S_{\text{ul}}$ (bottom $K/2$), as shown in Figure~\ref{fig:deferral_bipartite_graph_0} (line~\ref{line:pairwise_partitioning}).
For each candidate pair $(i, j)$ where $i \in S_{\text{ol}}$ and $j \in S_{\text{ul}}$, \name identifies the subset $S_{\text{deferred}}$ of samples in microbatch $i$ as an optimal deferral set between the two microbatches (line~\ref{line:pairwise_compute_deferred_set}).
The total LLM workload of $S_{\text{deferred}}$ is closest to the target transfer amount $\delta = (w_{\text{LLM},i} - w_{\text{LLM},j})/2$.
Figure~\ref{fig:deferral_bipartite_graph_1} tabulates the resulting $S_{\text{deferred}}$s and post-deferral LLM workloads.
Finding this subset is an instance of the subset sum problem, which \name solves with a discretized dynamic programming: a table $D$ of size $w'$ (the total rounded LLM workload of the overloaded microbatch) is queried for the sum $k^*$ that minimizes the residual $|\delta - k^*|$, identifying the optimal $S_{\text{deferred}}$.
This runs in pseudo-polynomial time $O(N_{\text{ol}} \times w')$, where $N_{\text{ol}}$ is the number of samples in the overloaded microbatch.

\parabf{Bottleneck cost formulation.}
With $S_{\text{deferred}}$ determined for every candidate pair, \name builds a bottleneck cost matrix $V$ and a standalone cost vector $L$.
$V_{i, j}$ is the minimum bottleneck LLM execution time achievable when microbatch $i \in S_{\text{ol}}$ defers $S_{\text{deferred}}$ to microbatch $j \in S_{\text{ul}}$ (line~\ref{line:pairwise_compute_bottleneck_cost}):
\begin{equation} \label{eq:bottleneck_cost_formulation}
V_{i, j} = \max{(w_{\text{LLM}, i} - w_{\text{deferred}, i}, w_{\text{LLM}, j} + w_{\text{deferred}, i})}
\end{equation}
$L_i$ is the cost of microbatch $i \in S_{\text{ol}}$ if it remains unpaired, \ie, $L_i = w_{\text{LLM}, i}$ (line~\ref{line:pairwise_compute_standalone_cost}).
$L$ is used in the bottleneck matching to determine whether a microbatch's LLM workload is high enough to require deferral, or it if can execute within budget on its own.

\parabf{Bottleneck matching optimization.}
Given $V$ and $L$, \name finds the minimum feasible threshold $T^*$ via \textsc{BottleneckMatch} (line~\ref{line:pairwise_maxcardinality}).
$T^*$ is the smallest value in $V \cup L$ such that every microbatch can complete LLM execution within $T^*$ either by deferring LLM workload to an underloaded partner, or by executing alone without deferral.
Since feasibility is monotone (if $T$ is feasible, any larger $T$ is also feasible), binary search over the at most $O(K^2)$ candidates in $V \cup L$ efficiently finds $T^*$.
For each candidate $T$, \name constructs a restricted bipartite graph $\mathcal{G}(S_{\text{ol}}, S_{\text{ul}}, V)$ with edges only where $V_{i,j} \leq T$, and verifies feasibility via DFS-based bipartite matching on critical microbatches.
Figure~\ref{fig:deferral_bipartite_graph_2} illustrates the bipartite graph.
Non-critical microbatches ($L_i \leq T^*$) are arbitrarily assigned to remaining $S_{\text{ul}}$ members with no deferral.
Each feasibility check costs $O(E\sqrt{K})$ where $E$ is the edge count of $\mathcal{G}$; given the small $K$, the total overhead is negligible.
The algorithm returns $T^*$ and the complete matching $\mathcal{P}$.
Microbatches are then arranged into an interleaved execution sequence $(ol_0, ul_0, ol_1, ul_1, \ldots)$ per $\mathcal{P}$ (line~\ref{line:pairwise_assign_deferrals}), with matched pairs transferring selected samples' LLM workload from the overloaded to the immediately following underloaded microbatch.

\begin{figure}[t]
\centering
\begin{subfigure}[t]{\columnwidth}
    \includegraphics[width=\columnwidth]{figures/pipeline_visualization/schedule_legend.pdf}
\end{subfigure}
\setcounter{subfigure}{0}
\begin{subfigure}[t]{\columnwidth}
    \includegraphics[width=\columnwidth]{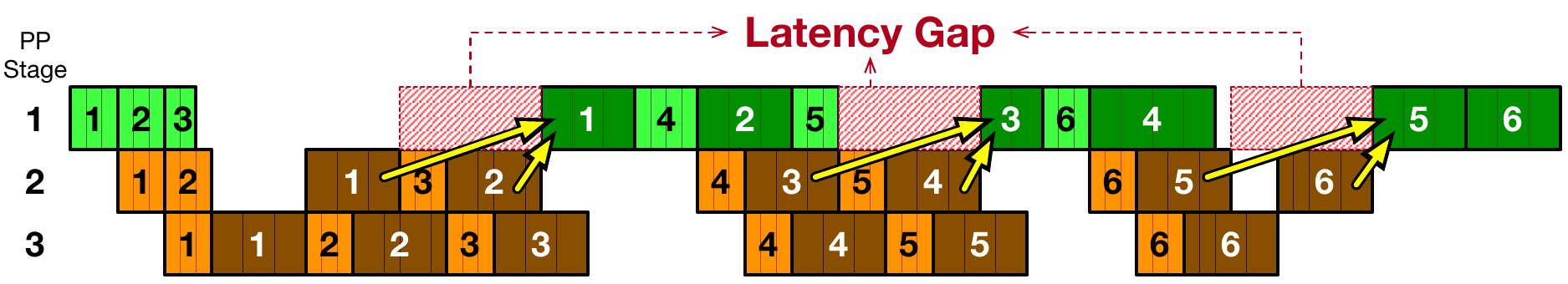}
    \caption{Backward passes of the encoder stage for microbatches deferring LLM workload are delayed until the gradients from the LLM backward pass are available.}
    \label{fig:pipeline_visual_nobpsplit}
\end{subfigure}
\begin{subfigure}[t]{\columnwidth}
    \includegraphics[width=\columnwidth]{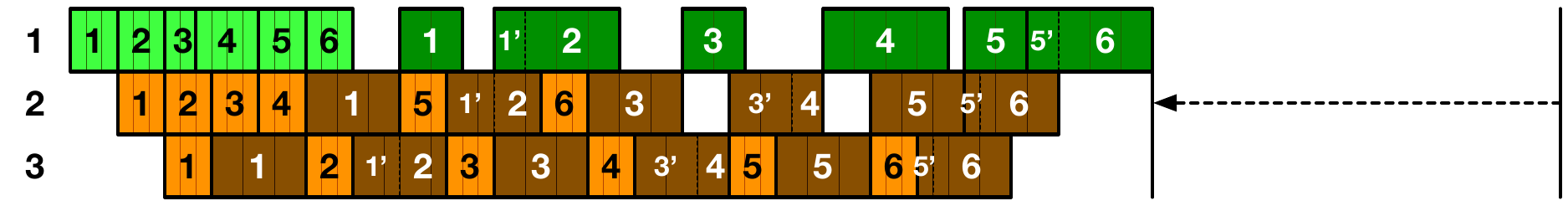}
    \caption{Split-backward processing allows the encoder backward pass to be executed immediately after getting the gradients from the LLM backward pass. $n'$ refers to deferred samples in backward pass.}
    \label{fig:pipeline_visual_bpsplit}
\end{subfigure}
\caption{Visualization of pipeline parallelism schedule of microbatches in Figure~\ref{fig:mandu_hierarchical_mb_3} with and without backward dependency optimization.}
\label{fig:backward_pass_dependency}
\end{figure}

\subsection{Optimizing Backward Pass Dependency}
\label{sec:backward_pass_dependency}
Our scheduling strategy focuses on optimizing the forward pass.
However, in pipeline parallel training, the backward pass introduces strict reverse dependencies: the gradients for the encoder cannot be computed until the corresponding gradients from the LLM are available.
Deferring the LLM workload inherently delays the completion of their forward pass, and as a result, the generation of gradients for these deferred samples is also delayed.
This creates a \textit{latency gap} for the encoder backward pass, as shown in Figure~\ref{fig:pipeline_visual_nobpsplit}.
After completing its forward pass for microbatch $i$, the encoder stage expects gradients for a whole microbatch $i$ to return.
If some LLM workload in the microbatch is deferred to microbatch $i+1$, the encoder faces a potential pipeline stall waiting for the delayed gradients to arrive.

\parabf{The split-backward processing.} To prevent this delay from stalling the encoder backward pass, we modify the execution logic to support split-backward processing.
Since the scheduler is deterministic, the encoder is aware of which specific samples within the microbatch $i$ will be deferred during the LLM forward pass.
When the backward pass for microbatch $i$ is scheduled, the encoder does not wait for the complete set of gradients.
Instead, it immediately executes the backward pass for the non-deferred samples.
The backward pass for the deferred samples is also deferred until their gradients arrive from the LLM backward pass.
Figure~\ref{fig:pipeline_visual_bpsplit} shows the pipeline parallelism schedule with split-backward processing.

\parabf{Eager forward pass execution.}
Even with split-backward processing, the backward pass workload increases later in the iteration because of the deferred samples.
To accommodate this back-loaded computation, we must reserve compute capacity in the later microbatches.
We achieve this by scheduling forward passes as eagerly as possible.
While the standard 1F1B schedule executes only $s-i+1$ forward passes in the initial warm-up phase (where $s$ is the number of stages and $i$ is the stage index starting from 1) as in Figure~\ref{fig:pipeline_visual_nobpsplit} before entering the steady phase of interleaving one forward and one backward pass, we execute as many forward passes as memory constraints allow before switching to the steady phase.
This eager execution reduces the number of forward passes required in the middle of the iteration, effectively yielding time slots that compute nodes can use to process the backward pass of deferred samples.

This makes \name more friendly to more advanced pipeline parallelism techniques, such as zero-bubble pipeline parallelism (ZBPP)~\cite{zbpp-iclr24}, which recommends executing the forward passes as eagerly as possible.

\parabf{Numerical correctness.}
With the split-backward processing, the number of backward passes -- and naturally the number of gradient accumulations -- increases.
In low-precision training (\eg, BF16, FP8, or NVFP4~\cite{hfp8-neurips19, fp4-neurips25, fp8lm-arxiv23}), such an increased number of gradient accumulations theoretically heightens the risk of numerical instability and swamping.
However, this amplified accumulation depth is structurally unavoidable in large-scale distributed training.
Many research papers have proposed algorithmic safeguards to solve this issue and thus it is no longer a destabilizing factor.
The proposed algorithms, such as stochastic rounding (SR)~\cite{sr-rsos22}, Kahan Summation~\cite{elmo-icml25}, or mixed precision with higher precision for gradients~\cite{fp8lm-arxiv23}, effectively neutralize the negative impact of increased number of gradient accumulations, ensuring convergence.

\section{Implementation}
\label{sec:implementation}

\name is implemented on top of PyTorch, Cornstarch~\cite{cornstarch-arxiv25}, and HuggingFace Transformers~\cite{transformers-emnlp20} in Python.
\name adds batch sizer and macroscopic profiler to derive the static model-parallel configuration and uses Cornstarch's multimodal 4D parallelization capability to partition the model and distribute it to GPUs accordingly.
\name exploits Cornstarch's multimodal 4D parallel execution capability to run distributed training, with the following modifications.

\parabf{Microbatch scheduler.}
We replace DistributedSampler with \name sampler that sorts samples and assigns them to the microbatches.
Deferral optimization is also executed at this point with estimated workload cost.
Executing deferral optimization at the scheduler level allows \name to be compatible with sequence packing.
Sequence packing is effective to reduce padding overhead, thus widely used~\cite{packing-arxiv22,prepacking-aistats25,longalign-emnlp24,wlbllm-osdi25,hierarchicalpacking-neurips25}.
By running deferral optimization before packing sequences and sending deferral information together with the packed microbatches, the pipeline execution engine can track the deferred samples within the packed microbatches.

\parabf{Pipeline execution engine.}
We modified the existing 1F1B~\cite{pipedream-sosp19} and Zero-Bubble Pipeline Parallelism~\cite{zbpp-iclr24} pipeline schedule to support deferral optimization.
The execution engine receives deferral information from the scheduler along with the microbatches before executing each iteration, and executes the pipeline schedule.
During the execution, the engine temporarily stores deferred activations and tracks the deferred sub-microbatches to ensure all deferred samples are properly processed in the backward pass.

\section{Evaluation}
\label{sec:evaluation}

We evaluate \name on a variety of datasets and models and compare it against DistTrain and DIP.
We summarize the results as follows:
\begin{denseitemize}
    \item \name outperforms the baselines in end-to-end training throughput by up to 1.40$\times$ (\S\ref{sec:eval-e2e-performance}).
    \item \name's proposed macroscopic profiling paradigm effectively estimates the workload ratio between modalities and provides a stable parallel configuration (\S\ref{sec:eval-analysis-of-macroscopic-profiling}).
    \item \name's hierarchical microbatch assignment and deferral optimization stabilize the workload variability across microbatches by up to 10.6$\times$ (\S\ref{sec:eval-effect-of-deferral-optimization-to-variability}).
\end{denseitemize}

\subsection{Experimental Setup}
\label{sec:experimental-setup}

\begin{table}[t]
    \centering
    \caption{Parallel configuration of \name and the baselines and execution setup. For all frameworks, TP=2, CP=1, and DP=4. E.PP and L.PP represent encoder pipeline parallel degree and LLM pipeline parallel degree, respectively. * indicates DIP colocates vision and LLM stages to the same pipeline stage.}
    \label{tab:eval-parallel-configurations}
    \small
    \begin{tabular}{c|c|ccc} 
    \toprule
    SynthChartNet                                                                   &                                                         & DistTrain & DIP & \name   \\ 
    \midrule
    \multirow{3}{*}{\begin{tabular}[c]{@{}c@{}}Execution\\Setup\end{tabular}}       & \begin{tabular}[c]{@{}c@{}}Profiling\\Size\end{tabular} & 1         & 4   & 256     \\ 
    \cmidrule{2-5}
                                                                                    & \begin{tabular}[c]{@{}c@{}}Global\\Batch\end{tabular}   & 512       & 512 & 512     \\ 
    \midrule
    \multirow{2}{*}{\begin{tabular}[c]{@{}c@{}}Qwen2Vision\\Llama3-1b\end{tabular}} & E.PP                                                    & 4         & 8*  & 5       \\ 
    \cmidrule{2-5}
                                                                                    & L.PP                                                    & 4         & 8*  & 3       \\ 
    \midrule
    \multirow{2}{*}{\begin{tabular}[c]{@{}c@{}}Qwen2Vision\\Llama3-3b\end{tabular}} & E.PP                                                    & 5         & 8*  & 4       \\ 
    \cmidrule{2-5}
                                                                                    & L.PP                                                    & 3         & 8*  & 4       \\
    \bottomrule
    \end{tabular}
\end{table}

\parabf{Cluster setup.}
We evaluate \name using 16 NVIDIA A40-48GB GPUs on 4 nodes.
Each node has 4 NVIDIA A40 GPUs and a NVIDIA Mellanox ConnectX-6 200Gbps Infiniband adaptor.
The four GPUs in a node are connected in pairs using NVLink and connected to the node via PCIe 4.0.

To emulate a larger cluster, we exploit the fact that data-parallel replicas synchronize only at the end of each iteration.
We execute one DP replica at a time using TP, PP, and CP on real hardware (16 GPUs) and execute 4 replicas sequentially.
The iteration time is taken as the maximum across the 4 executions, mirroring the all-reduce barrier to synchronize gradients.
All experiments use this emulated 64-GPU setup.


\parabf{Datasets and models.}
We evaluate \name using HuggingFace FineVision dataset~\cite{finevision-arxiv25}.
It is a multimodal dataset that contains more than 20 million samples of image and text pairs, where samples are collected from over 200 datasets.
We pick 4 subdatasets from the FineVision dataset -- SynthChartNet~\cite{synthchartnet-iccv25}, ChartQA~\cite{chartqa-acl22}, CocoQA~\cite{cocoqa-nips15}, and LLaVA-150k~\cite{llava-nips23} -- that have distinct data distributions.
We show SynthChartNet data in the main text because it is the most variable, and other datasets are shown in Appendices.

For models, we run various sizes of vision-language models (VLMs) using Qwen2.5Vision vision transformer~\cite{qwen2vl-arxiv24} to process various resolutions of images natively and Llama3~\cite{llama3-arxiv24} (1b and 3b parameters) for text processing.

\parabf{Baselines.}
We compare \name to the following baselines:
\begin{denseenum}
    \item \textit{DistTrain~\cite{disttrain-sigcomm25}}: DistTrain reorders samples to mitigate the data heterogeneity problem.
    \item \textit{DIP~\cite{dip-asplos26}}: DIP introduces decoupled modality scheduling to address heterogeneity and variability in multimodal training workloads.
\end{denseenum}
We run 4D parallelism with different parallel configurations between \name and the baselines.
Table~\ref{tab:eval-parallel-configurations} shows the parallel configurations (See Appendix~\ref{app:eval-parallel-configurations} for parallel configurations on other datasets).
DistTrain profiles 1 sample to derive a parallel configuration, while DIP profiles 1 microbatch (4 samples) to derive its parallel configuration.
We use microbatch size 4 and global batch size 512.
Although \name supports both 1F1B and ZBPP schedules, we use 1F1B for \name in evaluation.

\subsection{End-to-End Training Performance}
\label{sec:eval-e2e-performance}

\begin{figure}[t]
    \centering
    \begin{subfigure}[t]{\columnwidth}
        \includegraphics[width=\columnwidth]{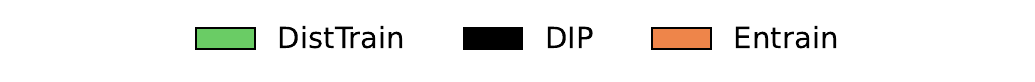}
    \end{subfigure}
    \setcounter{subfigure}{0}
    \begin{subfigure}[t]{\columnwidth}
        \includegraphics[width=\columnwidth]{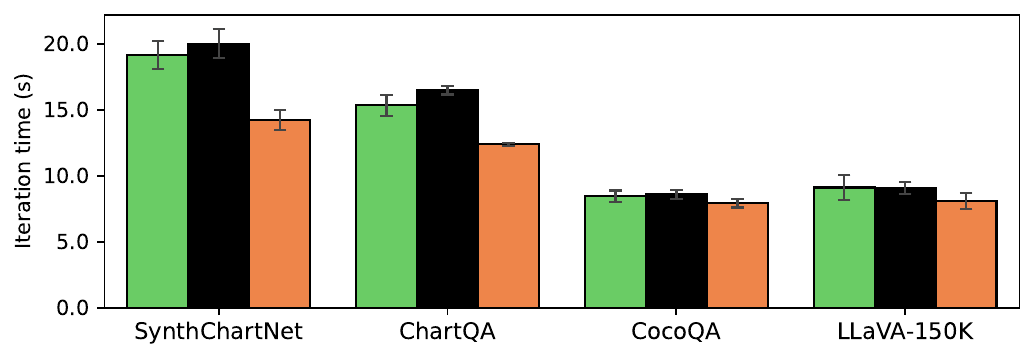}
        \caption{Qwen2.5Vision+Llama3-1b}
        \label{fig:e2e-performance-llama1b}
    \end{subfigure}
    \begin{subfigure}[t]{\columnwidth}
        \includegraphics[width=\columnwidth]{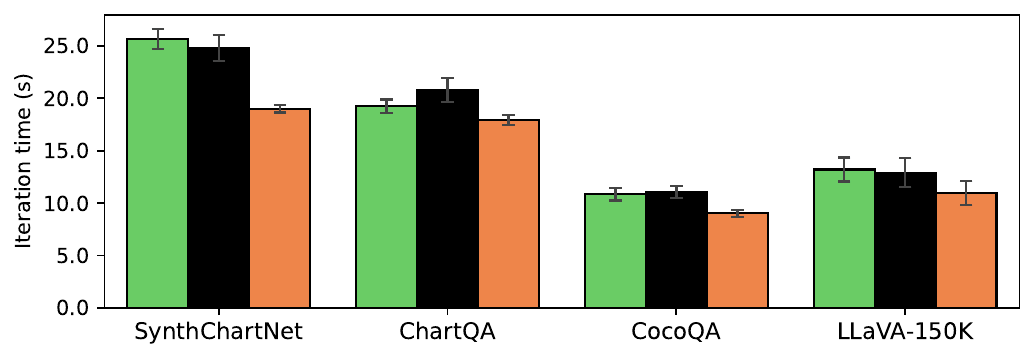}
        \caption{Qwen2.5Vision+Llama3-3b}
        \label{fig:e2e-performance-llama3b}
    \end{subfigure}
    \caption{End-to-end training performance of \name and the baselines.}
    \label{fig:e2e-performance}
\end{figure}

\begin{figure}[t]
    \centering
    \begin{subfigure}[t]{\columnwidth}
        \includegraphics[width=\columnwidth]{figures/pipeline_visualization/schedule_legend.pdf}
    \end{subfigure}
    \setcounter{subfigure}{0}
    \begin{subfigure}[t]{\columnwidth}
        \includegraphics[width=\columnwidth]{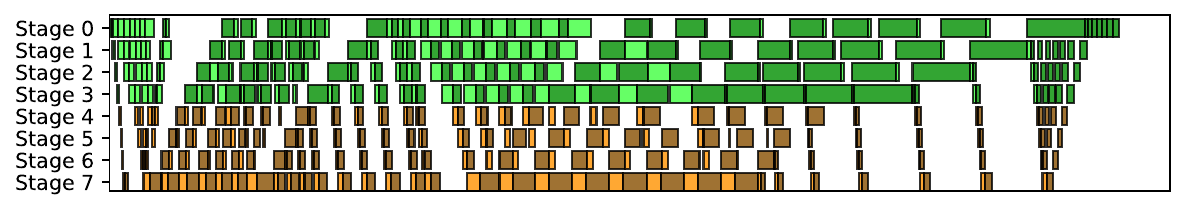}
        \caption{DistTrain pipeline schedule visualization.}
        \label{fig:eval-deferral-variability-time-llama1b-synthchartnet-disttrain}
    \end{subfigure}
    \begin{subfigure}[t]{\columnwidth}
    
        \includegraphics[width=\columnwidth]{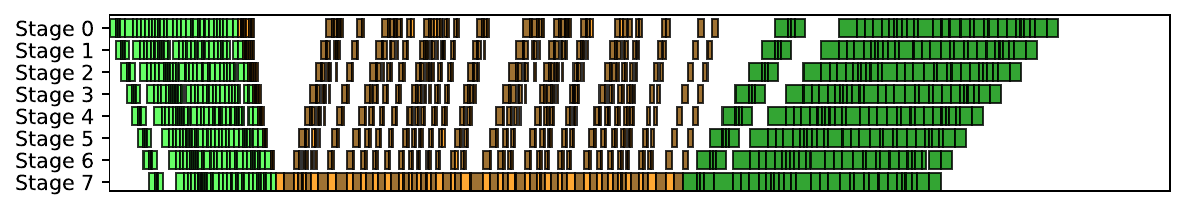}
        \caption{DIP pipeline schedule visualization.}
        \label{fig:eval-deferral-variability-time-llama1b-synthchartnet-1f1b}
    \end{subfigure}
    \begin{subfigure}[t]{\columnwidth}
        \includegraphics[width=\columnwidth]{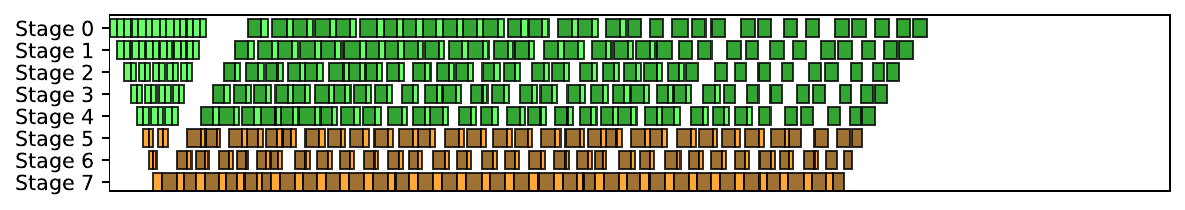}
        \caption{\name pipeline schedule visualization.}
        \label{fig:eval-deferral-variability-time-llama1b-synthchartnet-pipeweaver}
    \end{subfigure}
    \caption{Pipeline schedule visualization of \name and the baselines on SynthChartNet dataset on Qwen2.5Vision+Llama3-3b VLM.}
    \label{fig:eval-deferral-variability-time-pipeline-visualization}
\end{figure}

\parabf{Training throughput.}
Figure~\ref{fig:e2e-performance} shows end-to-end training performance of \name and the baselines using various datasets and models.
The main performance improvement of \name over the baselines comes from two factors.
First, \name's macroscopic profiling based parallel configuration is close to the optimal parallel, removing fragmented pipeline bubbles across modality pipeline stages.
Second, \name's hierarchical microbatch assignment balances the workload between microbatches, which is not possible with static microbatch partitioning.
Combined, \name achieves up to 1.40$\times$ faster end-to-end training throughput.
We also visualize one iteration of the pipeline schedule of \name and the baselines in Figure~\ref{fig:eval-deferral-variability-time-pipeline-visualization}.
\name shows balanced workload distribution between modalities and microbatches.


\begin{figure}[t]
    \centering
    \begin{subfigure}[t]{\columnwidth}
        \includegraphics[width=\columnwidth]{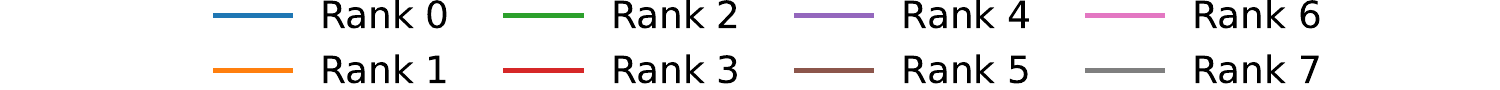}
    \end{subfigure}
    \setcounter{subfigure}{0}
    \begin{subfigure}[t]{\columnwidth}
        \includegraphics[width=\columnwidth]{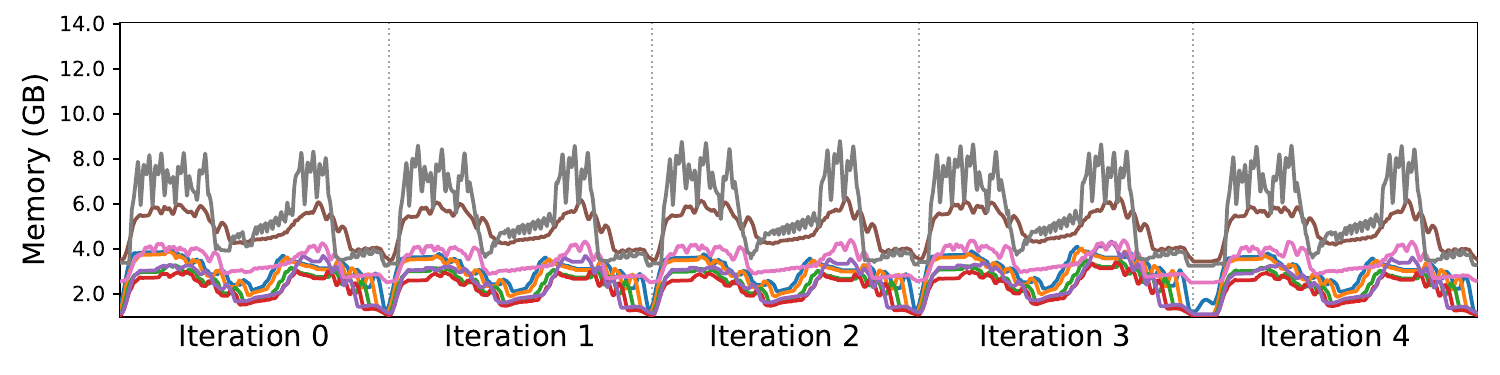}
        \caption{Memory consumption of DistTrain pipeline schedule.}
        \label{fig:e2e-memory-consumption-llama3b-synthchartnet-disttrain}
    \end{subfigure}
    \begin{subfigure}[t]{\columnwidth}
        \includegraphics[width=\columnwidth]{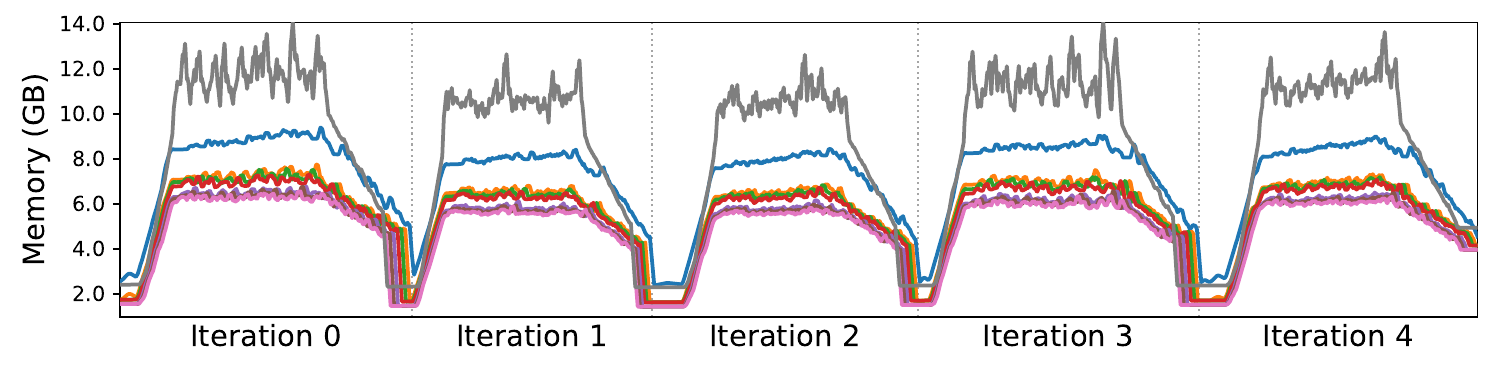}
        \caption{Memory consumption of DIP pipeline schedule.}
        \label{fig:e2e-memory-consumption-llama3b-synthchartnet-pipeweaver}
    \end{subfigure}
    \begin{subfigure}[t]{\columnwidth}
        \includegraphics[width=\columnwidth]{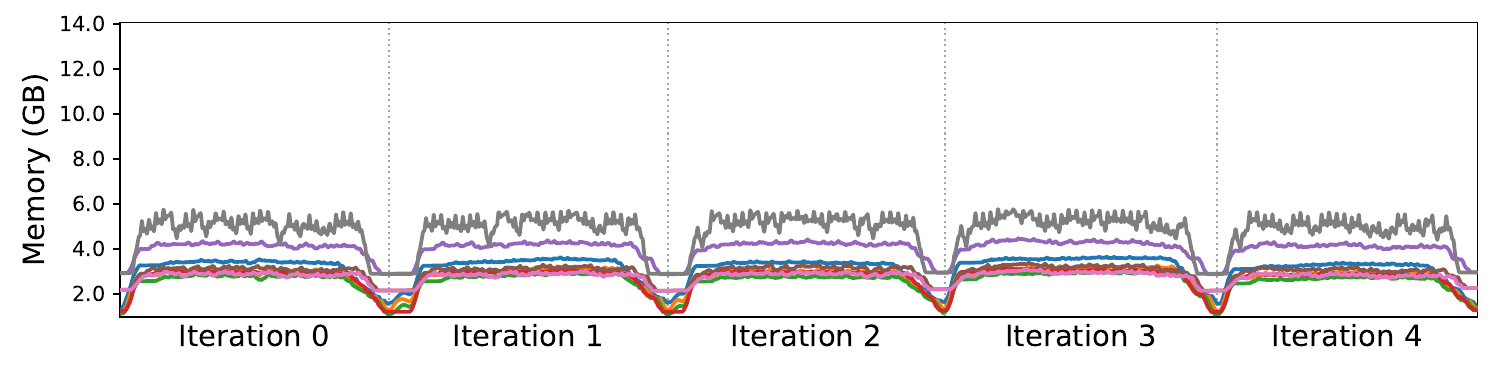}
        \caption{Memory consumption of \name pipeline schedule.}
        \label{fig:e2e-memory-consumption-llama3b-synthchartnet-mandu1f1b}
    \end{subfigure}
    \caption{Memory consumption of different schedules for SynthChartNet dataset on Qwen2.5Vision+Llama3-3b VLM.}
    \label{fig:e2e-memory-consumption}
\end{figure}

\parabf{Memory consumption.}
Figure~\ref{fig:e2e-memory-consumption} shows the memory consumption of the baselines and \name for SynthChartNet dataset on Qwen2.5Vision+Llama3-3b VLM.
\name's deferral optimization temporarily stores encoder activations in the pipeline buffer and thus requires more memory than non-deferral pipeline schedules.
However, increased memory consumption is minuscule because pairwise schedule only holds encoder activations for up to a single microbatch interval, and activation memory is very small compared to the total memory consumption.
Rather, more balanced workload distribution between modalities and microbatches makes \name's memory consumption more stable than the baselines, as shown in Figure~\ref{fig:e2e-memory-consumption-llama3b-synthchartnet-mandu1f1b}.
DistTrain (Figure~\ref{fig:e2e-memory-consumption-llama3b-synthchartnet-disttrain}), even though it follows 1F1B as \name, shows much more memory variability than \name due to imbalanced workload distribution, ending up with much higher peak memory consumption.
Meanwhile, DIP's scheduling processes all modality encoder forward passes before executing the LLM, and starts modality encoder backward pass after all LLM execution is done, as shown in Figure~\ref{fig:bg_schedule_visualization_pipeweaver}.
This requires much more memory ($\sim$12GB in rank 7) than DistTrain and \name to maintain encoder activations until they are freed during encoder backward pass, as shown in Figure~\ref{fig:e2e-memory-consumption-llama3b-synthchartnet-pipeweaver}.
See Appendix~\ref{app:e2e-memory-consumption} for memory consumption of other configurations.

\subsection{Analysis of Macroscopic Profiling}
\label{sec:eval-analysis-of-macroscopic-profiling}

\begin{table}[t]
    \centering
    \caption{Workload ratios in Bernoulli trials of SynthChartNet dataset in Qwen2.5Vision+Llama3-3b VLM using 16 GPUs.}
    \label{tab:eval-workload-ratio-with-different-batch-sizes-3b}
    \begin{tabular}{cc|c} 
        \toprule
        Batch Size & Trial Pass & Ratios Shown (Vision:LLM)                \\ 
        \midrule
        1                                                   & \xmark                                                   & 10:6, 9:7, 8:8, 7:9, 6:10  \\
        4                                                   & \xmark                                                   & 9:7, 8:8, 7:9                  \\
        16                                                  & \xmark                                                   & 9:7, 8:8, 7:9                  \\
        64                                                  & \cmark                                                   & 8:8                        \\
        256                                                 & \cmark                                                   & 8:8                        \\
        \bottomrule
    \end{tabular}
\end{table}


We analyze how the macroscopic profiling affects derivation of the workload ratio and parallel configuration.
We draw $k=59$ batches -- providing 95\% confidence level with $p_{error}=5\%$ -- with different batch sizes from the datasets and compute the workload ratio between modalities for each batch using the cost model and the metadata of the samples.
Table~\ref{tab:eval-workload-ratio-with-different-batch-sizes-3b} shows the workload ratios shown in Bernoulli trials for SynthChartNet dataset on Qwen2.5Vision+Llama3-3b VLM.
Across all datasets, smaller batch sizes show more variability in the workload ratio, which leads to more variability in GPU allocation to the modalities.
Depending on which samples are drawn, smaller batch sizes may allocate GPUs to the encoder and the LLM in different proportions (\eg, 10:6 or 6:10), while the true dataset workload ratio mean is 1.27:1.
Different trial data are shown in Appendix~\ref{app:eval-workload-ratio-with-different-batch-sizes}.

\begin{figure}[t]
    \centering
    \includegraphics[width=\columnwidth]{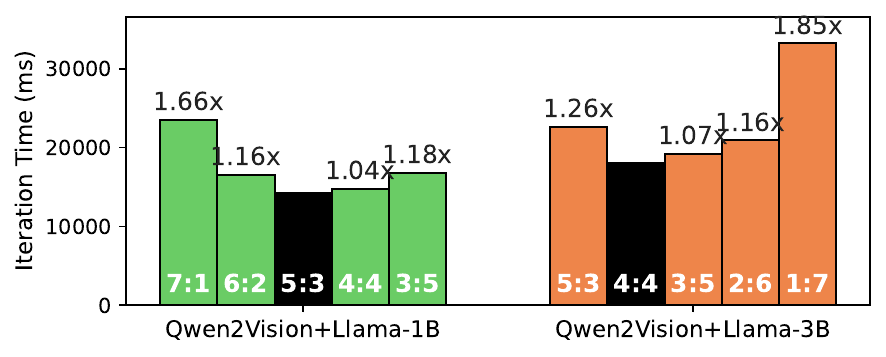}
    \caption{Sensitivity analysis on SynthChartNet dataset. Ratios are encoder-to-LLM workload ratios.}
    \label{fig:eval-sensitivity-analysis-synthchartnet}
\end{figure}

\parabf{Sensitivity analysis to macroscopic profiling quality.}
To demonstrate the impact of parallel configuration quality on performance, we profile \name under varying parallel configurations, with results shown in Figure~\ref{fig:eval-sensitivity-analysis-synthchartnet} (Appendix~\ref{app:eval-sensitivity-analysis} for other datasets).
The throughput of \name drops significantly by 85\% if a bad parallel configuration is used, while the parallel configuration derived from the macroscopic profiling with profiling size 256 -- the black bars -- is the best.

\begin{table}[t]
    \centering
    \footnotesize
    \caption{Standard deviation (std) of forward time of modalities in different pipeline schedules and datasets.}
    \label{tab:eval-deferral-variability-time-table}
    \begin{tabular}{cc|rrrr} 
    \toprule
                                &           & \multicolumn{1}{c}{\begin{tabular}[c]{@{}c@{}}Synth\\ChartNet\end{tabular}} & \multicolumn{1}{c}{ChartQA} & \multicolumn{1}{c}{CocoQA} & \multicolumn{1}{c}{Llava-150k}  \\ 
    \midrule
    \multirow{3}{*}{Vision}   & DistTrain & 208.07                                                                      & 61.34                       & 22.82                      & 17.98                           \\
                                & DIP       & 102.13                                                                      & 73.42                       & 28.79                      & 14.82                           \\
                                & \name     & \textbf{19.60}                                                              & \textbf{15.23}              & \textbf{22.11}             & \textbf{8.11}                   \\ 
    \midrule
    \multirow{3}{*}{Llama 1b} & DistTrain & 77.92                                                                       & 61.34                       & 22.82                      & 17.98                           \\
                                & DIP       & 37.45                                                                       & 73.42                       & 28.79                      & 14.82                           \\
                                & \name     & \textbf{18.79}                                                              & \textbf{15.23}              & \textbf{22.11}             & \textbf{8.11}                   \\ 
    \midrule
    \multirow{3}{*}{LLama 3b} & DistTrain & 164.55                                                                      & 35.44                       & \textbf{5.37}              & 32.92                           \\
                                & DIP       & 84.36                                                                       & 40.56                       & 7.77                       & 33.27                           \\
                                & \name     & \textbf{40.24}                                                              & \textbf{25.72}              & 5.52                       & \textbf{31.66}                  \\
    \bottomrule
    \end{tabular}
\end{table}

\subsection{Effect of Hierarchical Microbatch Assignment to Variability}
\label{sec:eval-effect-of-deferral-optimization-to-variability}

\begin{figure}[t]
    \centering
    \begin{subfigure}[t]{\columnwidth}
    \includegraphics[width=\columnwidth]{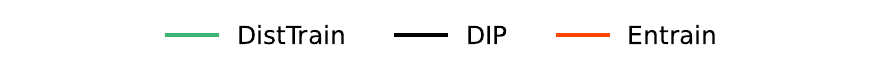}
    \end{subfigure}
    \setcounter{subfigure}{0}
    \begin{subfigure}[t]{\columnwidth}
        \includegraphics[width=\columnwidth]{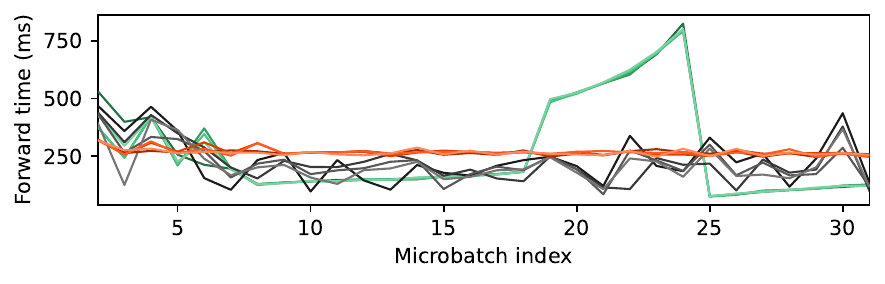}
        \caption{Qwen2.5Vision forward time variability across microbatches.}
        \label{fig:eval-deferral-variability-time-llama1b-synthchartnet-vision}
    \end{subfigure}
    \begin{subfigure}[t]{\columnwidth}
        \includegraphics[width=\columnwidth]{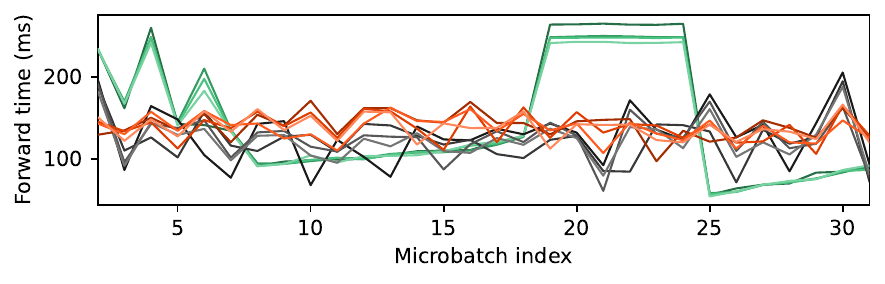}
        \caption{Llama3-1b forward time variability across microbatches.}
        \label{fig:eval-deferral-variability-time-llama1b-synthchartnet-llm}
    \end{subfigure}
    \begin{subfigure}[t]{\columnwidth}
        \includegraphics[width=\columnwidth]{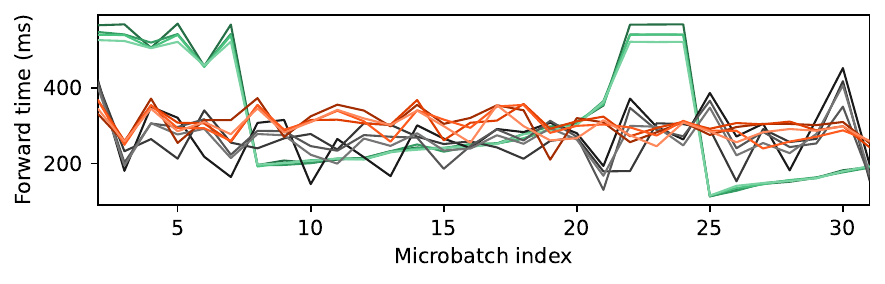}
        \caption{Llama3-3b forward time variability across microbatches.}
        \label{fig:eval-deferral-variability-time-llama3b-synthchartnet-llm}
    \end{subfigure}
    \caption{Variability of modality forward time across microbatches in SynthChartNet dataset on VLMs. Each line represents a DP replica.}
    \label{fig:eval-deferral-variability-time-synthchartnet}
\end{figure}

We evaluate how modality workload fluctuates across microbatches due to data variability, and how much the deferral optimization mitigates the imbalance.
Figure~\ref{fig:eval-deferral-variability-time-synthchartnet} shows the variability of forward time of each modality in VLMs (Qwen2.5Vision, LLama3-1b, and LLama3-3b) across microbatches in SynthChartNet dataset (Appendix~\ref{app:eval-variability-of-modality-forward-time-across-microbatches} for other datasets).
Each microbatch's forward time is computed as the sum of forward time of all corresponding pipeline stages.
DistTrain's heuristic microbatch reordering algorithm focuses on mitigating pipeline bubbles from the holistic view, hence it fails to address the data variability between microbatches and shows high variability in both vision and LLM forward time.
\name shows both vision and LLM forward time much more stable than the baselines.
More details in numerical values are shown in Table~\ref{tab:eval-deferral-variability-time-table}.
Low variability in vision forward time is due to the stratified sample assignment in the first phase of deferral optimization.
We do not statically assign samples to have homogeneous number of samples per microbatch, but assign different number of samples to each microbatch to balance the encoder workload.
Encoder workload-focused assignment breaks the balance of LLM workload, but the deferral optimization mitigates the imbalance.
Because \name defers LLM workload as atomically -- not partitioning a sample into multiple chunks and deferring only a part of the sample -- the deferral does not fully address the imbalance.
Yet, the deferral optimization still mitigates the imbalance significantly.





\section{Related Work}
\label{sec:related_work}

\parabf{Large-scale distributed training.}
Driven by the scaling law~\cite{scalinglaws-arxiv20}, distributed training frameworks such as Megatron-LM~\cite{megatron-sc21}, DeepSpeed~\cite{deepspeed-kdd20}, Colossal-AI~\cite{colossalai-icpp23}, and TorchTitan~\cite{torchtitan-iclr25} have enabled 1T+-parameter model training~\cite{qwen3-arxiv25,llama4-web}.
As the focus shifts toward larger datasets and longer contexts~\cite{chinchilla-arxiv22}, context parallelism has been proposed to distribute long sequences across GPUs~\cite{sp-acl23,loongtrain-arxiv24,flexsp-asplos25,llama3-arxiv24,wlbllm-osdi25}.

\parabf{Distributed multimodal LLM training.}
DistMM~\cite{distmm-nsdi24} is among the earliest works on distributed multimodal training, targeting CLIP-like models without LLM backbones.
DistTrain~\cite{disttrain-sigcomm25}, Optimus~\cite{optimus-atc25}, Cornstarch~\cite{cornstarch-arxiv25}, and DIP~\cite{dip-asplos26} tackle the heterogeneity between modality encoders and the LLM backbone; Cornstarch additionally handles structural characteristics such as frozen parameters and non-causal attention.
None of these systems, however, fully addresses dataset variability at both the profiling and execution granularities.

\parabf{Dataset heterogeneity and variability.}
HotSPa~\cite{hotspa-sosp24} and ByteScale~\cite{bytescale-sigcomm25} address sequence length variability in unimodal LLM training via dynamic parallelism reconfiguration, but assume a single axis of variability.
In MLLM training, each modality follows its own independent distribution yet is coupled within each sample, making the inter-modality workload ratio chaotic at fine granularities.
DistTrain~\cite{disttrain-sigcomm25} reorders samples to reduce pipeline bubbles but remains oblivious to intra-sample modality coupling; DIP~\cite{dip-asplos26} interleaves encoder and LLM pipeline stages but incurs high memory overhead from retained activations.
These works overlook that the modality workload ratio converges to a stable mean at global batch scale, a property that neither demands dynamic reconfiguration nor can be captured by micro-level profiling.

\section{Conclusion}
\label{sec:conclusion}

In this paper, we presented \name, a distributed MLLM training framework that addresses both heterogeneity and variability in multimodal training workloads.
\name counters the intuition that dynamic model-parallel configuration is necessary for MLLMs with data variability by demonstrating that at the macroscopic scale, the aggregate workload ratio between modalities reliably converges to a stable constant.
\name also addresses the fundamental limitation of variability at the microscopic scale by proposing a hierarchical microbatch assignment algorithm that defers excess workload within each iteration, stabilizing execution time across microbatches.
\name outperforms the baselines in end-to-end training throughput by up to 1.40$\times$, reducing workload variability across microbatches by up to 10.6$\times$.

\label{EndOfPaper}

\newpage

\bibliographystyle{ACM-Reference-Format}
\bibliography{ref}

@misc{gemini-arxiv23,
      title={Gemini: A Family of Highly Capable Multimodal Models}, 
      author={Google},
      year={2023},
      eprint={2312.11805},
      archivePrefix={arXiv},
      primaryClass={cs.CL},
      url={https://arxiv.org/abs/2312.11805}, 
}

@misc{qwen2vl-arxiv24,
      title={Qwen2-VL: Enhancing Vision-Language Model's Perception of the World at Any Resolution}, 
      author={Peng Wang and Shuai Bai and Sinan Tan and Shijie Wang and Zhihao Fan and Jinze Bai and Keqin Chen and Xuejing Liu and Jialin Wang and Wenbin Ge and Yang Fan and Kai Dang and Mengfei Du and Xuancheng Ren and Rui Men and Dayiheng Liu and Chang Zhou and Jingren Zhou and Junyang Lin},
      year={2024},
      eprint={2409.12191},
      archivePrefix={arXiv},
      primaryClass={cs.CV},
      url={https://arxiv.org/abs/2409.12191}, 
}

@misc{qwen3omni-arxiv25,
      title={Qwen3-Omni Technical Report}, 
      author={Jin Xu and Zhifang Guo and Hangrui Hu and Yunfei Chu and Xiong Wang and Jinzheng He and Yuxuan Wang and Xian Shi and Ting He and Xinfa Zhu and Yuanjun Lv and Yongqi Wang and Dake Guo and He Wang and Linhan Ma and Pei Zhang and Xinyu Zhang and Hongkun Hao and Zishan Guo and Baosong Yang and Bin Zhang and Ziyang Ma and Xipin Wei and Shuai Bai and Keqin Chen and Xuejing Liu and Peng Wang and Mingkun Yang and Dayiheng Liu and Xingzhang Ren and Bo Zheng and Rui Men and Fan Zhou and Bowen Yu and Jianxin Yang and Le Yu and Jingren Zhou and Junyang Lin},
      year={2025},
      eprint={2509.17765},
      archivePrefix={arXiv},
      primaryClass={cs.CL},
      url={https://arxiv.org/abs/2509.17765}, 
}

@misc{mingomni-arxiv25,
      title={Ming-Omni: A Unified Multimodal Model for Perception and Generation}, 
      author={Inclusion AI and Ant Group},
      year={2025},
      eprint={2506.09344},
      archivePrefix={arXiv},
      primaryClass={cs.AI},
      url={https://arxiv.org/abs/2506.09344}, 
}

@misc{olaomni-arxiv25,
      title={Ola: Pushing the Frontiers of Omni-Modal Language Model}, 
      author={Zuyan Liu and Yuhao Dong and Jiahui Wang and Ziwei Liu and Winston Hu and Jiwen Lu and Yongming Rao},
      year={2025},
      eprint={2502.04328},
      archivePrefix={arXiv},
      primaryClass={cs.CV},
      url={https://arxiv.org/abs/2502.04328}, 
}

@InProceedings{llavanext-cvpr24,
    author    = {Liu, Haotian and Li, Chunyuan and Li, Yuheng and Lee, Yong Jae},
    title     = {Improved Baselines with Visual Instruction Tuning},
    booktitle = {CVPR 24},
    month     = {June},
    year      = {2024},
    pages     = {26296-26306}
}

@article{llavaov-tmlr25,
title={{LL}a{VA}-OneVision: Easy Visual Task Transfer},
author={Bo Li and Yuanhan Zhang and Dong Guo and Renrui Zhang and Feng Li and Hao Zhang and Kaichen Zhang and Peiyuan Zhang and Yanwei Li and Ziwei Liu and Chunyuan Li},
journal={TMLR 25},
issn={2835-8856},
year={2025},
url={https://openreview.net/forum?id=zKv8qULV6n},
note={}
}

@misc{deekseepvl-arxiv24,
      title={{DeepSeek-VL}: Towards Real-World Vision-Language Understanding}, 
      author={Haoyu Lu and Wen Liu and Bo Zhang and Bingxuan Wang and Kai Dong and Bo Liu and Jingxiang Sun and Tongzheng Ren and Zhuoshu Li and Hao Yang and Yaofeng Sun and Chengqi Deng and Hanwei Xu and Zhenda Xie and Chong Ruan},
      year={2024},
      eprint={2403.05525},
      archivePrefix={arXiv},
      primaryClass={cs.AI},
      url={https://arxiv.org/abs/2403.05525}, 
}

@misc{deepseekvl24-arxiv24,
      title={{DeepSeek-VL2}: Mixture-of-Experts Vision-Language Models for Advanced Multimodal Understanding}, 
      author={Zhiyu Wu and Xiaokang Chen and Zizheng Pan and Xingchao Liu and Wen Liu and Damai Dai and Huazuo Gao and Yiyang Ma and Chengyue Wu and Bingxuan Wang and Zhenda Xie and Yu Wu and Kai Hu and Jiawei Wang and Yaofeng Sun and Yukun Li and Yishi Piao and Kang Guan and Aixin Liu and Xin Xie and Yuxiang You and Kai Dong and Xingkai Yu and Haowei Zhang and Liang Zhao and Yisong Wang and Chong Ruan},
      year={2024},
      eprint={2412.10302},
      archivePrefix={arXiv},
      primaryClass={cs.CV},
      url={https://arxiv.org/abs/2412.10302}, 
}

@misc{gpt4o-arxiv24,
      title={GPT-4o System Card}, 
      author={OpenAI},
      year={2024},
      eprint={2410.21276},
      archivePrefix={arXiv},
      primaryClass={cs.CL},
      url={https://arxiv.org/abs/2410.21276}, 
}

@misc{gpt5-arxiv26,
      title={OpenAI GPT-5 System Card}, 
      author={OpenAI},
      year={2026},
      eprint={2601.03267},
      archivePrefix={arXiv},
      primaryClass={cs.CL},
      url={https://arxiv.org/abs/2601.03267}, 
}

@misc{llama4-web,
	author = {Meta AI},
	title = {{The Llama 4 herd}: The beginning of a new era of natively multimodal {AI} innovation},
	howpublished = {\url{https://ai.meta.com/blog/llama-4-multimodal-intelligence/}},
	year = {2025},
	note = {[Accessed Feb 08, 2026]},
}

@inproceedings{scalingllama-isca25,
author = {Chu, Weiwei and Xie, Xinfeng and Yu, Jiecao and Wang, Jie and Phanishayee, Amar and Tang, Chunqiang and Hao, Yuchen and Huang, Jianyu and Ozdal, Mustafa and Wang, Jun and Goswami, Vedanuj and Goyal, Naman and Kadian, Abhishek and Gu, Andrew and Cai, Chris and Tian, Feng and Wang, Xiaodong and Si, Min and Balaji, Pavan and Chu, Ching-Hsiang and Park, Jongsoo},
title = {Scaling {Llama 3} Training with Efficient Parallelism Strategies},
year = {2025},
booktitle = {ISCA 25},
url = {https://doi.org/10.1145/3695053.3731410},
doi = {10.1145/3695053.3731410},
}

@inproceedings{sp-acl23,
    title = "Sequence Parallelism: Long Sequence Training from System Perspective",
    author = "Li, Shenggui  and Xue, Fuzhao  and Baranwal, Chaitanya  and Li, Yongbin  and You, Yang",
    booktitle = "ACL 23",
    year = "2023",
    url = "https://aclanthology.org/2023.acl-long.134/",
    doi = "10.18653/v1/2023.acl-long.134",
}

@inproceedings{alpa-osdi22,
author = {Zheng, Lianmin and Li, Zhuohan and Zhang, Hao and Zhuang, Yonghao and Chen, Zhifeng and Huang, Yanping and Wang, Yida and Xu, Yuanzhong and Zhuo, Danyang and Xing, Eric P. and Gonzalez, Joseph E. and Stoica, Ion},
title = {Alpa: Automating Inter- and {Intra-Operator} Parallelism for Distributed Deep Learning},
booktitle = {OSDI 22},
year = {2022},
url = {https://www.usenix.org/conference/osdi22/presentation/zheng-lianmin},
}

@inproceedings {unity-osdi22,
author = {Colin Unger and Zhihao Jia and Wei Wu and Sina Lin and Mandeep Baines and Carlos Efrain Quintero Narvaez and Vinay Ramakrishnaiah and Nirmal Prajapati and Pat McCormick and Jamaludin Mohd-Yusof and Xi Luo and Dheevatsa Mudigere and Jongsoo Park and Misha Smelyanskiy and Alex Aiken},
title = {Unity: Accelerating {DNN} Training Through Joint Optimization of Algebraic Transformations and Parallelization},
booktitle = {OSDI 22},
year = {2022},
isbn = {978-1-939133-28-1},
address = {Carlsbad, CA},
pages = {267--284},
url = {https://www.usenix.org/conference/osdi22/presentation/unger},
publisher = {USENIX Association},
month = jul
}

@inproceedings{oobleck-sosp23,
author = {Jang, Insu and Yang, Zhenning and Zhang, Zhen and Jin, Xin and Chowdhury, Mosharaf},
title = {Oobleck: Resilient Distributed Training of Large Models Using Pipeline Templates},
year = {2023},
doi = {10.1145/3600006.3613152},
booktitle = {SOSP 23},
url = {https://doi.org/10.1145/3600006.3613152}
}

@misc{llama3-arxiv24,
    title = {The Llama 3 Herd of Models},
    author = {Meta AI},
    year = {2024},
    url = {https://arxiv.org/abs/2407.21783}
}

@inproceedings{bytescale-sigcomm25,
author = {Ge, Hao and Feng, Junda and Huang, Qi and Fu, Fangcheng and Nie, Xiaonan and Zuo, Lei and Lin, Haibin and Cui, Bin and Liu, Xin},
title = {ByteScale: Communication-Efficient Scaling of LLM Training with a 2048K Context Length on 16384 GPUs},
year = {2025},
booktitle = {SIGCOMM 25},
url = {https://doi.org/10.1145/3718958.3754352}
}

@inproceedings{hotspa-sosp24,
author = {Ge, Hao and Fu, Fangcheng and Li, Haoyang and Wang, Xuanyu and Lin, Sheng and Wang, Yujie and Nie, Xiaonan and Zhang, Hailin and Miao, Xupeng and Cui, Bin},
title = {Enabling Parallelism Hot Switching for Efficient Training of Large Language Models},
year = {2024},
booktitle = {SOSP 24},
url = {https://doi.org/10.1145/3694715.3695969},
doi = {10.1145/3694715.3695969}
}

@inproceedings{disttrain-sigcomm25,
author = {Zhang, Zili and Zhong, Yinmin and Jiang, Yimin and Hu, Hanpeng and Sun, Jianjian and Ge, Zheng and Zhu, Yibo and Jiang, Daxin and Jin, Xin},
title = {DistTrain: Addressing Model and Data Heterogeneity with Disaggregated Training for Multimodal Large Language Models},
year = {2025},
booktitle = {SIGCOMM 25},
url = {https://doi.org/10.1145/3718958.3750472}
}

@inproceedings {distmm-nsdi24,
author = {Jun Huang and Zhen Zhang and Shuai Zheng and Feng Qin and Yida Wang},
title = {{DISTMM}: Accelerating Distributed Multimodal Model Training},
booktitle = {NSDI 24},
year = {2024},
url = {https://www.usenix.org/conference/nsdi24/presentation/huang}
}

@inproceedings{dip-asplos26,
author = {Xue, Zhenliang and Hu, Hanpeng and Chen, Xing and Jiang, Yimin and Song, Yixin and Mi, Zeyu and Zhu, Yibo and Jiang, Daxin and Xia, Yubin and Chen, Haibo},
title = {{DIP}: Efficient Large Multimodal Model Training with Dynamic Interleaved Pipeline},
year = {2026},
booktitle = {ASPLOS 26},
url = {https://doi.org/10.1145/3779212.3790154},
doi = {10.1145/3779212.3790154}
}

@misc{cornstarch-arxiv25,
      title={Efficient Distributed MLLM Training with Cornstarch}, 
      author={Insu Jang and Runyu Lu and Nikhil Bansal and Ang Chen and Mosharaf Chowdhury},
      year={2025},
      eprint={2503.11367},
      archivePrefix={arXiv},
      primaryClass={cs.DC},
      url={https://arxiv.org/abs/2503.11367}, 
}

@inproceedings{optimus-atc25,
  title={Optimus: Accelerating {Large-Scale Multi-Modal LLM} Training by Bubble Exploitation},
  author={Feng, Weiqi and Chen, Yangrui and Wang, Shaoyu and Peng, Yanghua and Lin, Haibin and Yu, Minlan},
  booktitle={ATC 25},
  year={2025}
}

@inproceedings{flexsp-asplos25,
author = {Wang, Yujie and Wang, Shiju and Zhu, Shenhan and Fu, Fangcheng and Liu, Xinyi and Xiao, Xuefeng and Li, Huixia and Li, Jiashi and Wu, Faming and Cui, Bin},
title = {{FlexSP}: Accelerating {Large Language Model} Training via Flexible Sequence Parallelism},
year = {2025},
booktitle = {ASPLOS 25},
url = {https://doi.org/10.1145/3676641.3715998},
doi = {10.1145/3676641.3715998}
}

@inproceedings{wlbllm-osdi25,
author = {Wang, Zheng and Cai, Anna and Xie, Xinfeng and Pan, Zaifeng and Guan, Yue and Chu, Weiwei and Wang, Jie and Li, Shikai and Huang, Jianyu and Cai, Chris and Hao, Yuchen and Ding, Yufei},
title = {{WLB-LLM}: workload-balanced 4D parallelism for large language model training},
year = {2025},
booktitle = {OSDI 25},
url = {https://www.usenix.org/conference/osdi25/presentation/wang-zheng},
}

@inproceedings{zbpp-iclr24,
title={Zero Bubble (Almost) Pipeline Parallelism},
author={Qi, Penghui and Wan, Xinyi and Huang, Guangxing and Lin, Min},
booktitle={ICLR 24},
year={2024},
url={https://openreview.net/forum?id=tuzTN0eIO5}
}

@inproceedings{hfp8-neurips19,
 author = {Sun, Xiao and Choi, Jungwook and Chen, Chia-Yu and Wang, Naigang and Venkataramani, Swagath and Srinivasan, Vijayalakshmi (Viji) and Cui, Xiaodong and Zhang, Wei and Gopalakrishnan, Kailash},
 booktitle = {NeurIPS 19},
 title = {Hybrid 8-bit Floating Point (HFP8) Training and Inference for Deep Neural Networks},
 url = {https://proceedings.neurips.cc/paper_files/paper/2019/file/65fc9fb4897a89789352e211ca2d398f-Paper.pdf},
 year = {2019}
}

@inproceedings{fp4-neurips25,
title={{FP}4 All the Way: Fully Quantized Training of Large Language Models},
author={Chmiel, Brian and Fishman, Maxim and Banner, Ron and Soudry, Daniel},
booktitle={NeurIPS 25},
year={2025},
url={https://neurips.cc/virtual/2025/loc/san-diego/poster/116331}
}

@inproceedings{elmo-icml25,
title={{ELMO} : Efficiency via Low-precision and Peak Memory Optimization in Large Output Spaces},
author={Zhang, Jinbin and Ullah, Nasib and Schultheis, Erik and Babbar, Rohit},
booktitle={ICML 25},
year={2025},
url={https://openreview.net/forum?id=d6CTIPrTTC}
}

@article{sr-rsos22,
    author = {Croci, Matteo and Fasi, Massimiliano and Higham, Nicholas J. and Mary, Theo and Mikaitis, Mantas},
    title = {Stochastic Rounding: implementation, error analysis and applications},
    journal = {Royal Society Open Science},
    issn = {2054-5703},
    doi = {10.1098/rsos.211631},
    url = {https://doi.org/10.1098/rsos.211631},
    eprint = {https://royalsocietypublishing.org/rsos/article-pdf/doi/10.1098/rsos.211631/998073/rsos.211631.pdf},
}

@misc{fp8lm-arxiv23,
      title={FP8-LM: Training FP8 Large Language Models}, 
      author={Peng, Houwen and Wu, Kan and Wei, Yixuan and Zhao, Guoshuai and Yang, Yuxiang and Liu, Ze and Xiong, Yifan and Yang, Ziyue and Ni, Bolin and Hu, Jingcheng and Li, Ruihang and Zhang, Miaosen and Li, Chen and Ning, Jia and Wang, Ruizhe and Zhang, Zheng and Liu, Shuguang and Chau, Joe and Hu, Han and Cheng, Peng},
      year={2023},
      eprint={2310.18313},
      archivePrefix={arXiv},
      primaryClass={cs.LG},
      url={https://arxiv.org/abs/2310.18313}, 
}

@inproceedings{attention-nips17,
 author = {Vaswani, Ashish and Shazeer, Noam and Parmar, Niki and Uszkoreit, Jakob and Jones, Llion and Gomez, Aidan N and Kaiser, \L ukasz and Polosukhin, Illia},
 booktitle = {NIPS 17},
 title = {Attention is All you Need},
 year = {2017},
}

@inproceedings{megatron-sc21,
author = {Narayanan, Deepak and Shoeybi, Mohammad and Casper, Jared and LeGresley, Patrick and Patwary, Mostofa and Korthikanti, Vijay and Vainbrand, Dmitri and Kashinkunti, Prethvi and Bernauer, Julie and Catanzaro, Bryan and Phanishayee, Amar and Zaharia, Matei},
title = {Efficient Large-Scale Language Model Training on GPU Clusters Using Megatron-LM},
year = {2021},
booktitle = {SC 21},
url = {https://doi.org/10.1145/3458817.3476209},
doi = {10.1145/3458817.3476209}
}

@misc{scalinglaws-arxiv20,
      title={Scaling Laws for Neural Language Models}, 
      author={Jared Kaplan and Sam McCandlish and Tom Henighan and Tom B. Brown and Benjamin Chess and Rewon Child and Scott Gray and Alec Radford and Jeffrey Wu and Dario Amodei},
      year={2020},
      eprint={2001.08361},
      archivePrefix={arXiv},
      primaryClass={cs.LG},
      url={https://arxiv.org/abs/2001.08361}, 
}

@inproceedings{deepspeed-kdd20,
      author = {Rasley, Jeff and Rajbhandari, Samyam and Ruwase, Olatunji and He, Yuxiong},
      title = {DeepSpeed: System Optimizations Enable Training Deep Learning Models with Over 100 Billion Parameters},
      year = {2020},
      booktitle = {KDD 20},
      url = {https://doi.org/10.1145/3394486.3406703},
      doi = {10.1145/3394486.3406703}
}

@inproceedings{colossalai-icpp23,
      author = {Li, Shenggui and Liu, Hongxin and Bian, Zhengda and Fang, Jiarui and Huang, Haichen and Liu, Yuliang and Wang, Boxiang and You, Yang},
      title = {Colossal-AI: A Unified Deep Learning System For Large-Scale Parallel Training},
      year = {2023},
      booktitle = {ICPP 23},
      url = {https://doi.org/10.1145/3605573.3605613},
      doi = {10.1145/3605573.3605613}
}

@misc{chinchilla-arxiv22,
      title={Training Compute-Optimal Large Language Models}, 
      author={Jordan Hoffmann and Sebastian Borgeaud and Arthur Mensch and Elena Buchatskaya and Trevor Cai and Eliza Rutherford and Diego de Las Casas and Lisa Anne Hendricks and Johannes Welbl and Aidan Clark and Tom Hennigan and Eric Noland and Katie Millican and George van den Driessche and Bogdan Damoc and Aurelia Guy and Simon Osindero and Karen Simonyan and Erich Elsen and Jack W. Rae and Oriol Vinyals and Laurent Sifre},
      year={2022},
      eprint={2203.15556},
      archivePrefix={arXiv},
      primaryClass={cs.CL},
      url={https://arxiv.org/abs/2203.15556}, 
}

@inproceedings{torchtitan-iclr25,
title={TorchTitan: One-stop PyTorch native solution for production ready {LLM} pretraining},
author={Wanchao Liang and Tianyu Liu and Less Wright and Will Constable and Andrew Gu and Chien-Chin Huang and Iris Zhang and Wei Feng and Howard Huang and Junjie Wang and Sanket Purandare and Gokul Nadathur and Stratos Idreos},
booktitle={ICLR 25},
year={2025},
url={https://openreview.net/forum?id=SFN6Wm7YBI}
}

@misc{loongtrain-arxiv24,
      title={LoongTrain: Efficient Training of Long-Sequence LLMs with Head-Context Parallelism}, 
      author={Gu, Diandian and Sun, Peng and Hu, Qinghao and Huang, Ting and Chen, Xun and Xiong, Yingtong and Wang, Guoteng and Chen, Qiaoling and Zhao, Shangchun and Fang, Jiarui and Wen, Yonggang and Zhang, Tianwei and Jin, Xin and Liu, Xuanzhe},
      year={2024},
      url={https://arxiv.org/abs/2406.18485}, 
}

@misc{qwen3-arxiv25,
      title={Qwen3 Technical Report}, 
      author={An Yang and Anfeng Li and Baosong Yang and Beichen Zhang and Binyuan Hui and Bo Zheng and Bowen Yu and Chang Gao and Chengen Huang and Chenxu Lv and Chujie Zheng and Dayiheng Liu and Fan Zhou and Fei Huang and Feng Hu and Hao Ge and Haoran Wei and Huan Lin and Jialong Tang and Jian Yang and Jianhong Tu and Jianwei Zhang and Jianxin Yang and Jiaxi Yang and Jing Zhou and Jingren Zhou and Junyang Lin and Kai Dang and Keqin Bao and Kexin Yang and Le Yu and Lianghao Deng and Mei Li and Mingfeng Xue and Mingze Li and Pei Zhang and Peng Wang and Qin Zhu and Rui Men and Ruize Gao and Shixuan Liu and Shuang Luo and Tianhao Li and Tianyi Tang and Wenbiao Yin and Xingzhang Ren and Xinyu Wang and Xinyu Zhang and Xuancheng Ren and Yang Fan and Yang Su and Yichang Zhang and Yinger Zhang and Yu Wan and Yuqiong Liu and Zekun Wang and Zeyu Cui and Zhenru Zhang and Zhipeng Zhou and Zihan Qiu},
      year={2025},
      eprint={2505.09388},
      archivePrefix={arXiv},
      primaryClass={cs.CL},
      url={https://arxiv.org/abs/2505.09388}, 
}

@inproceedings{transformers-emnlp20,
    title = "Transformers: State-of-the-Art Natural Language Processing",
    author = {Wolf, Thomas  and Debut, Lysandre  and Sanh, Victor  and
Chaumond, Julien  and Delangue, Clement  and Moi, Anthony  and Cistac, Pierric  and Rault, Tim  and Louf, Remi  and Funtowicz, Morgan  and Davison, Joe  and Shleifer, Sam  and von Platen, Patrick  and Ma, Clara  and Jernite, Yacine  and Plu, Julien  and Xu, Canwen  and Le Scao, Teven  and Gugger, Sylvain  and Drame, Mariama  and Lhoest, Quentin  and Rush, Alexander},
    booktitle = "EMNLP 20",
    year = "2020",
}

@article{graham-scheduling-siam69,
    title = {Bounds on Multiprocessing Timing Anomalies},
    author = {Graham, Ronald Lewis},
    journal = {SIAM Journal on Applied Mathematics},
    volume = {17},
    number = {2},
    pages = {416-429},
    year = {1969},
    url = {https://doi.org/10.1137/0117039}
}

@inproceedings{pipedream-sosp19,
author = {Narayanan, Deepak and Harlap, Aaron and Phanishayee, Amar and Seshadri, Vivek and Devanur, Nikhil R. and Ganger, Gregory R. and Gibbons, Phillip B. and Zaharia, Matei},
title = {PipeDream: generalized pipeline parallelism for DNN training},
year = {2019},
booktitle = {SOSP 19},
url = {https://doi.org/10.1145/3341301.3359646},
doi = {10.1145/3341301.3359646}
}

@misc{packing-arxiv22,
      title={Efficient Sequence Packing without Cross-contamination: Accelerating Large Language Models without Impacting Performance}, 
      author={Kosec, Matej and Krell, Mario Michael and Perez, Sergio P. and Fitzgibbon, Andrew},
      year={2022},
      url={https://arxiv.org/abs/2107.02027}, 
}

@inproceedings{
      prepacking-aistats25,
      title={Prepacking: A Simple Method for Fast Prefilling and Increased Throughput in Large Language Models},
      author={Zhao, Siyan and Israel, Daniel Mingyi and Van den Broeck, Guy and Grover, Aditya},
      booktitle={AISTATS 25},
      year={2025},
      url={https://openreview.net/forum?id=LBVD4krAq2}
}

@inproceedings{longalign-emnlp24,
    title = "{L}ong{A}lign: A Recipe for Long Context Alignment of Large Language Models",
    author = {Bai, Yushi and Lv, Xin and Zhang, Jiajie and He, Yuze and Qi, Ji and Hou, Lei and Tang, Jie and Dong, Yuxiao and Li, Juanzi},
    booktitle = {EMNLP 24},
    year = {2024},
    url = {https://aclanthology.org/2024.findings-emnlp.74/}
}

@inproceedings{
hierarchicalpacking-neurips25,
title={Hierachical Balance Packing: Towards Efficient Supervised Fine-tuning for Long-Context {LLM}},
author={Yao, Yongqiang and Tan, Jingru and Liang, Kaihuan and Zhang, Feizhao and Hu, Jiahao and Wu, Shuo and Niu, Yazhe and Gong, Ruihao and Lin, Dahua and Xu, Ningyi},
booktitle={NeurIPS 25},
year={2025},
url={https://openreview.net/forum?id=6BHDre6WQW}
}

@misc{finevision-arxiv25,
      title={FineVision: Open Data Is All You Need}, 
      author={Wiedmann, Luis and Zohar, Orr and Mahla, Amir and Wang, Xiaohan and Li, Rui and Frere, Thibaud and von Werra, Leandro and Gosthipaty, Aritra Roy and Marafioti, Andrés},
      year={2025},
      url={https://arxiv.org/abs/2510.17269}, 
}

@InProceedings{synthchartnet-iccv25,
    author    = {Nassar, Ahmed and Omenetti, Matteo and Lysak, Maksym and Livathinos, Nikolaos and Auer, Christoph and Morin, Lucas and de Lima, Rafael Teixeira and Kim, Yusik and Gurbuz, A. Said and Dolfi, Michele and Staar, Peter W. J.},
    title     = {SmolDocling: An ultra-compact vision-language model for end-to-end multi-modal document conversion},
    booktitle = {ICCV 25},
    month     = {October},
    year      = {2025},
    pages     = {21972-21983}
}

@inproceedings{chartqa-acl22,
    title = "{C}hart{QA}: A Benchmark for Question Answering about Charts with Visual and Logical Reasoning",
    author = {Masry, Ahmed and Long, Do Xuan and Tan, Jia Qing and Joty, Shafiq and Hoque, Enamul},
    booktitle = {ACL 22},
    year = {2022},
    url = {https://aclanthology.org/2022.findings-acl.177/}
}

@inproceedings{cocoqa-nips15,
 author = {Ren, Mengye and Kiros, Ryan and Zemel, Richard},
 booktitle = {NIPS 15},
 title = {Exploring Models and Data for Image Question Answering},
 year = {2015},
 url = {https://proceedings.neurips.cc/paper_files/paper/2015/file/831c2f88a604a07ca94314b56a4921b8-Paper.pdf}
}

@inproceedings{llava-nips23,
 author = {Liu, Haotian and Li, Chunyuan and Wu, Qingyang and Lee, Yong Jae},
 booktitle = {NIPS 23},
 title = {Visual Instruction Tuning},
 year = {2023},
 url = {https://proceedings.neurips.cc/paper_files/paper/2023/file/6dcf277ea32ce3288914faf369fe6de0-Paper-Conference.pdf}
}
\clearpage

\appendix

\section{Termination Proof for the Loop}
\label{app:bmax_derivation}

This appendix proves that the loop in Algorithm~\ref{alg:prob_profiling} terminates in finitely many iterations.

\parabf{Setup.}
Let each training sample $i$ contribute a workload vector $\mathbf{w}_i = (w_{\text{encoder},i},\, w_{\text{LLM},i})$, where $w_{\text{encoder},i}$ and $w_{\text{LLM},i}$ are the workloads of the vision encoder and LLM for that sample (following the notation of Section~\ref{sec:hierarchical_microbatch_assignment}).
Project $\mathbf{w}_i$ onto the scalar workload ratio:
\begin{equation}
    r_i \;=\; \frac{w_{\text{encoder},i}}{w_{\text{encoder},i} + w_{\text{LLM},i}} \;\in\; [0, 1]
\end{equation}
which governs the proportional GPU allocation.
Let $\nu_W = \mathbb{E}[r_i]$ denote the population mean.
Each iteration of the algorithm draws $n$ independent and identically distributed samples and computes the sample mean $\bar{r}_n$.

\parabf{Termination proof.}
Since $r_i \in [0,1]$ are i.i.d.\ bounded random variables, the Strong Law of Large Numbers guarantees that the sample mean converges to the true mean with probability~1:
\begin{equation}
    \bar{r}_n \;\longrightarrow\; \nu_W \quad\text{as } n \to \infty.
    \label{eq:lln_convergence}
\end{equation}
The proportional GPU allocation function $Q(\cdot)$ is piecewise constant: it rounds the continuous ratio to the nearest integer split, producing a finite number of breakpoints in $[0,1]$.
Let $d > 0$ denote the distance from $\nu_W$ to the nearest breakpoint.
This is positive as long as $\nu_W$ does not fall exactly on a breakpoint, which holds for any continuously-valued workload distribution.

By Equation~\ref{eq:lln_convergence}, there exists a finite threshold $n^*$ such that for all $n \ge n^*$, every independently drawn batch of size $n$ satisfies $|\bar{r}_n - \nu_W| < d$, and hence produces the same allocation $Q(\bar{r}_n) = Q(\nu_W)$.
Therefore, once $n$ reaches $n^*$, all $k$ Bernoulli validation draws agree with the reference batch, $\mathit{IsStable} = \mathrm{True}$, and the loop returns.
The loop terminates in finitely many doublings.

\parabf{Convergence rate.}
The above argument does not specify the value of $n^*$.
Under the Central Limit Theorem approximation, one can show $n^* \le (6\,\sigma_{pop}/d)^{2}$, at which point the probability of any single batch producing a different allocation falls to $\approx 0.0000002\%$.
In practice, the Bernoulli test passes at batch sizes far smaller than this bound, as shown empirically in Appendix~\ref{app:eval-workload-ratio-with-different-batch-sizes}.

\section{Probabilistic Proof of Configuration Stability}
\label{app:statistical_bound_for_convergence}

This appendix formalizes the guarantee used in Section~\ref{sec:design_math_bound}.
The argument has two steps.
Section~\ref{app:statistical_bound_for_convergence_1} bounds the probability that a random profiling batch of size $n$ yields a configuration different from the observed reference allocation.
Section~\ref{app:statistical_bound_for_convergence_2} then shows that the same decision remains valid for any larger batch size $b \ge n$, in particular the runtime global batch size $B_{global}$, because the workload-ratio estimator concentrates as batch size grows.

\subsection{Bounding the Error Rate at Batch Size $n$}
\label{app:statistical_bound_for_convergence_1}
Let $C_{ref}$ denote the discrete allocation returned by the first reference batch of size $n$ under the chosen data-parallel degree $DP$.
For each subsequent independent validation batch of size $n$, let $C_n$ denote the allocation returned by the same procedure.
Define the failure event as $C_n \neq C_{ref}$, and let
\begin{equation}
p_{error} = \Pr(C_n \neq C_{ref} \mid C_{ref})
\end{equation}
be the probability that a fresh size-$n$ batch yields a different allocation than the observed reference allocation.

After fixing the reference batch, running the procedure on $k$ additional independent validation batches gives $k$ Bernoulli trials.
The probability of observing zero failures is:
\begin{equation}
\Pr(\text{0 errors in } k \text{ trials}) = (1 - p_{error})^k
\end{equation}

For observing zero failures to constitute significant evidence against an error rate of $p_{error}$, we require $(1 - p_{error})^k \leq \alpha$.
Solving the boundary case for the minimum number of trials:
\begin{equation}
k = \left\lceil \frac{\ln(\alpha)}{\ln(1 - p_{error})} \right\rceil
\end{equation}

Therefore, observing identical configurations across $k$ independent validation trials allows the system to state with confidence $1 - \alpha$ that a random batch of size $n$ will differ from the observed reference allocation $C_{ref}$ with probability at most $p_{error}$.
For example, with $\alpha = 0.05$ and $p_{error} = 0.05$, the required number of trials is $k \approx 59$.

\subsection{Lifting the Guarantee from $n$ to Larger Batch Sizes}
\label{app:statistical_bound_for_convergence_2}
Section~\ref{app:statistical_bound_for_convergence_1} only certifies stability for random profiling batches of size $n$ relative to the observed reference allocation $C_{ref}$.
However, the configuration is used at training time for larger global batches, so we must show that the same decision remains valid at those larger scales.

Let each sample $i$ contribute a workload vector $\mathbf{w}_i = (w_{\text{encoder},i},\, w_{\text{LLM},i})$.
Let $\mu = \mathbb{E}[\mathbf{w}_i]$ and $\Sigma = \text{Cov}(\mathbf{w}_i)$.
For any batch size $b \ge n$, the estimated macroscopic workload ratio is derived from the sample mean
\begin{equation}
\bar{\mathbf{w}}_b = \frac{1}{b} \sum_{i=1}^{b} \mathbf{w}_i .
\end{equation}
Assuming independent and identically distributed sampling, its covariance scales as
\begin{equation}
\text{Cov}(\bar{\mathbf{w}}_b) = \frac{1}{b}\Sigma .
\end{equation}
Thus, the estimator becomes more concentrated as $b$ grows.

Now consider the mapping from the continuous workload ratio to a discrete configuration.
Because GPUs are indivisible, this mapping partitions the workload space into piecewise-constant decision regions.
Let $V(C_{ref})$ denote the region that maps to the observed reference allocation $C_{ref}$.

From the binomial argument in Section~\ref{app:statistical_bound_for_convergence_1}, the estimator at batch size $n$ already lands in this region with high probability:
\begin{equation}
\Pr(\bar{\mathbf{w}}_n \in V(C_{ref}) \mid C_{ref}) \ge 1 - p_{error} .
\end{equation}

Because proportional allocation maps continuous workload ratios to integer GPU counts via rounding, each decision region $V(C_{ref})$ is a convex polytope.
Moreover, the high success rate established above implies that the population mean $\mu$ lies in the interior of $V(C_{ref})$: if $\mu$ were outside or near the boundary of $V(C_{ref})$, the size-$n$ estimator would frequently land outside the region, contradicting the Bernoulli test.

For any $b \ge n$, the estimator $\bar{\mathbf{w}}_b$ shares the same mean $\mu$ but has strictly smaller covariance $\Sigma/b \preceq \Sigma/n$.
Because $\mu$ lies in the interior of the convex region $V(C_{ref})$, concentrating more tightly around $\mu$ can only increase the probability of remaining in $V(C_{ref})$.
The allocation selected from profiling batches of size $n$ therefore remains valid for any larger batch size, including the runtime global batch size $B_{global}$ and, as a limiting case, the full dataset size $N$.

Once a modest profiling batch size $n$ is large enough that repeated random draws agree on the same discrete allocation, averaging over any larger batch only reinforces that decision.

\section{Pipeline Schedule with Pairwise Deferral Optimization}
\label{app:pipeline_schedule_with_pairwise_deferral}

\begin{figure}[t]
    \centering
    \begin{subfigure}[t]{\columnwidth}
        \includegraphics[width=\columnwidth]{figures/pipeline_visualization/schedule_legend.pdf}
    \end{subfigure}
    \setcounter{subfigure}{0}
    \begin{subfigure}[t]{\columnwidth}
        \includegraphics[width=\columnwidth]{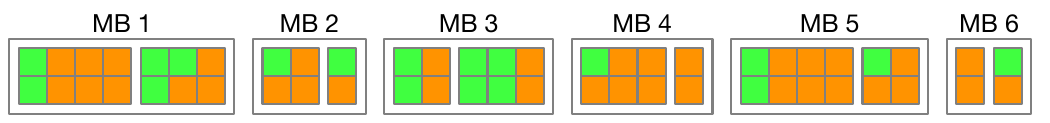}
        \includegraphics[width=\columnwidth]{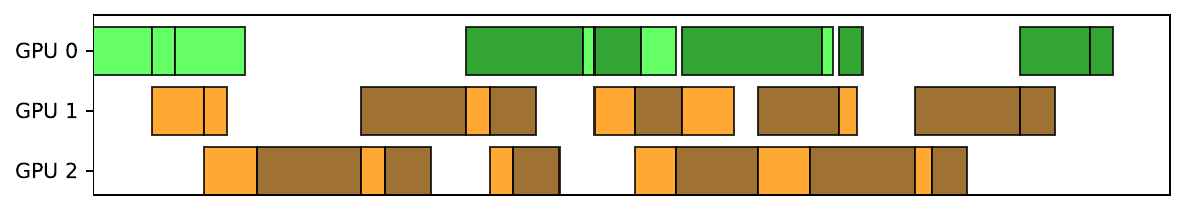}
        \caption{Pipeline schedule without pairwise deferral optimization. Microbatches have static (two) number of samples.}
    \end{subfigure}
    \begin{subfigure}[t]{\columnwidth}
        \includegraphics[width=\columnwidth]{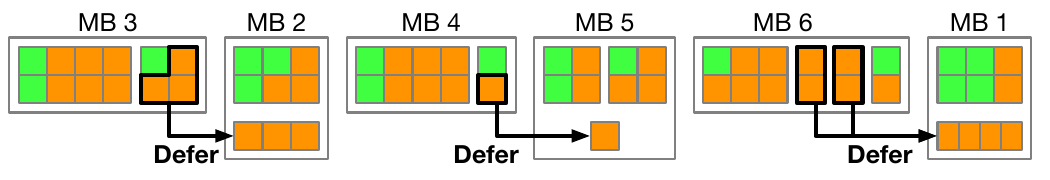}
        \includegraphics[width=\columnwidth]{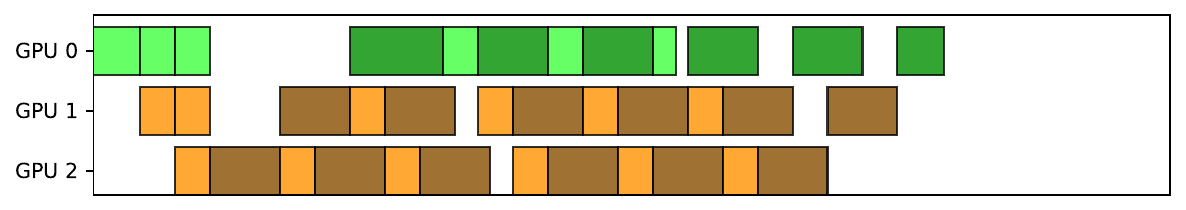}
        \caption{Pipeline schedule with pairwise deferral optimization.}
    \end{subfigure}
    \caption{Comparison of 3-stage pipeline parallel schedule without and with pairwise deferral optimization.}
    \label{fig:pipeline_schedule_with_pairwise_deferral_optimization}
\end{figure}

Figure~\ref{fig:pipeline_schedule_with_pairwise_deferral_optimization} shows the pipeline parallel schedule with 3 pipeline stages without and with pairwise deferral optimization.
One stage and two stages are assigned to vision and LLM, respectively, since total workload ratio is 1:2 (18:36).
After pairwise deferral optimization, some LLM workloads are shifted to the follow-up microbatches, resulting in a more balanced pipeline schedule.
Pairwise deferral only happens in LLM, thus the encoder schedule remains unchanged.

\section{Parallel Configurations}
\label{app:eval-parallel-configurations}

Table~\ref{tab:eval-parallel-configurations} shows the parallel configuration of \name and the baselines on Qwen2.5Vision + Llama3-1b VLM and Qwen2.5Vision + Llama3-3b VLM.
Note that for DIP, vision stages and LLM stages are colocated.

\begin{table}
    \centering
    \caption{Parallel configurations of \name and the baselines on various datasets.}
    \label{tab:eval-parallel-configurations-various-datasets}
    \begin{tabular}{c|c|c|c|c} 
    \toprule
                                                                                    &      & \multicolumn{3}{c}{DistTrain/DIP/\name}  \\ 
    \cmidrule{3-5}
                                                                                    &      & LLaVA-150k & ChartQA & CocoQA            \\ 
    \midrule
    \multirow{2}{*}{\begin{tabular}[c]{@{}c@{}}Qwen2Vision\\Llama3-1b\end{tabular}} & E.PP & 5/8/4      & 5/8/5   & 4/8/5             \\
                                                                                    & L.PP & 3/8/4      & 3/8/3   & 4/8/3             \\ 
    \midrule
    \multirow{2}{*}{\begin{tabular}[c]{@{}c@{}}Qwen2Vision\\Llama3-3b\end{tabular}} & E.PP & 4/8/3      & 4/8/4   & 4/8/3             \\
                                                                                    & L.PP & 4/8/5      & 4/8/4   & 4/8/5             \\
    \bottomrule
    \end{tabular}
    \end{table}

\section{Pipeline Schedule Memory Consumption}
\label{app:e2e-memory-consumption}

\subsection{Qwen2.5Vision + Llama3-1b}
\begin{figure}[t]
    \centering
    \begin{subfigure}[t]{\columnwidth}
        \includegraphics[width=\columnwidth]{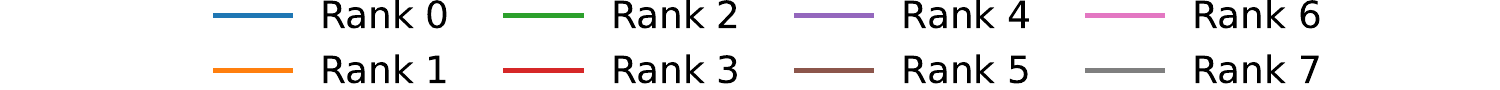}
    \end{subfigure}
    \setcounter{subfigure}{0}
    \begin{subfigure}[t]{\columnwidth}
        \includegraphics[width=\columnwidth]{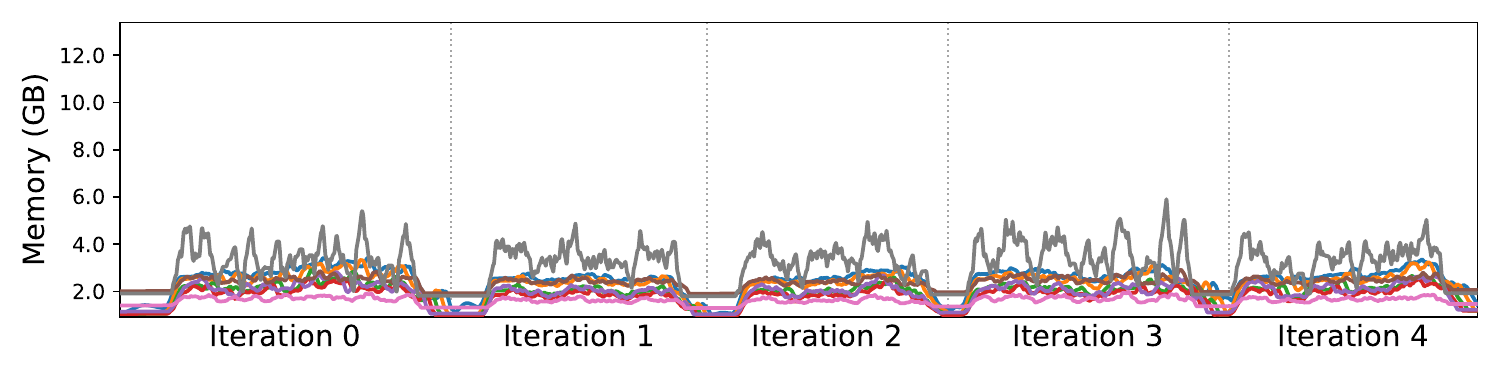}
        \caption{Memory consumption of 1F1B pipeline schedule.}
        \label{fig:e2e-memory-consumption-llama1b-synthchartnet-1f1b}
    \end{subfigure}
    \begin{subfigure}[t]{\columnwidth}
        \includegraphics[width=\columnwidth]{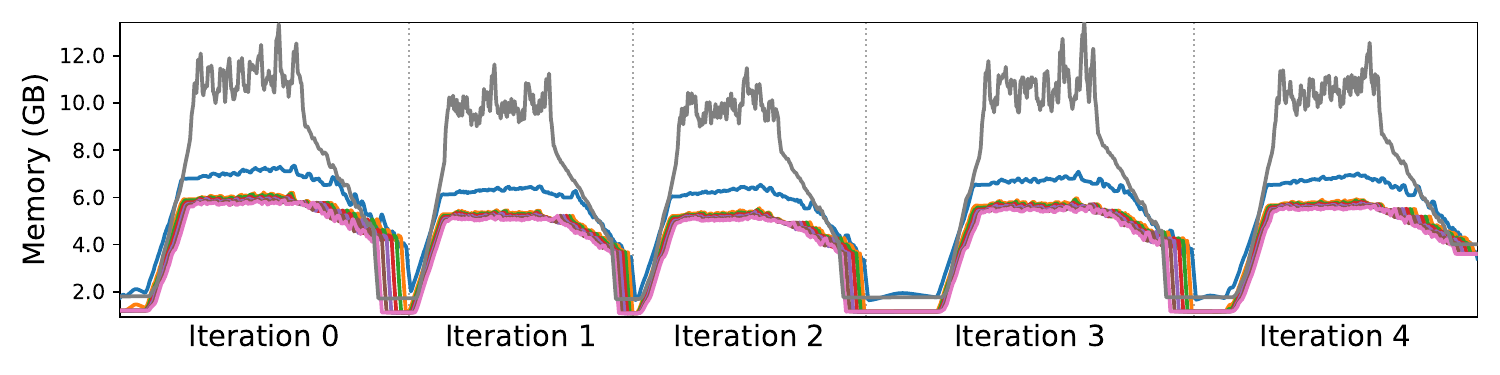}
        \caption{Memory consumption of DIP pipeline schedule.}
        \label{fig:e2e-memory-consumption-llama1b-synthchartnet-pipeweaver}
    \end{subfigure}
    \begin{subfigure}[t]{\columnwidth}
        \includegraphics[width=\columnwidth]{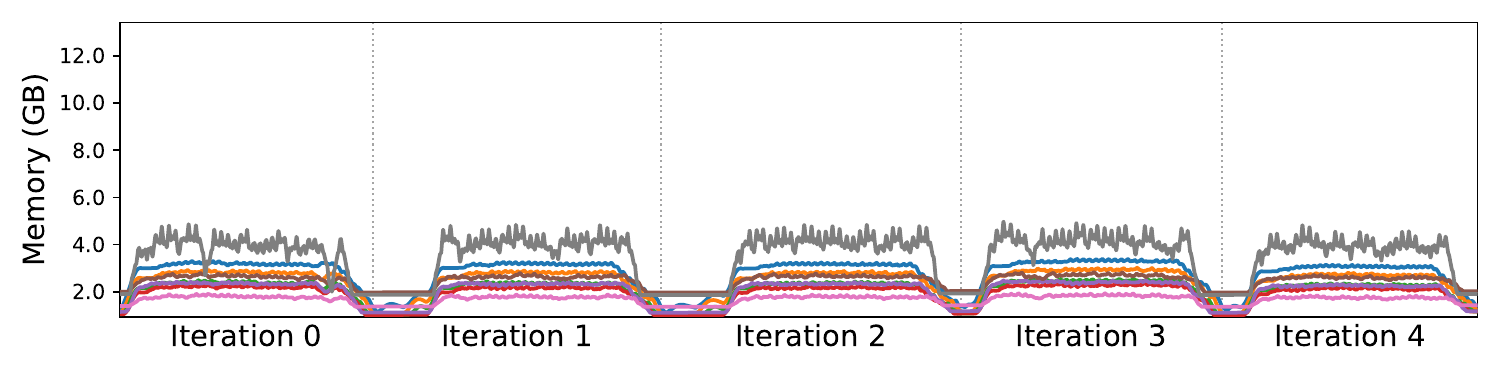}
        \caption{Memory consumption of \name-1F1B pipeline schedule.}
        \label{fig:e2e-memory-consumption-llama1b-synthchartnet-mandu1f1b}
    \end{subfigure}
    \caption{SynthChartNet on Qwen2.5Vision+Llama3-1b VLM.}
    \label{fig:e2e-memory-consumption-llama1b-synthchartnet}
\end{figure}

\begin{figure}[t]
    \centering
    \begin{subfigure}[t]{\columnwidth}
        \includegraphics[width=\columnwidth]{evaluation/e2e/memory_consumption/llama_1b/pipeline_memory_consumption_legend.pdf}
    \end{subfigure}
    \setcounter{subfigure}{0}
    \begin{subfigure}[t]{\columnwidth}
        \includegraphics[width=\columnwidth]{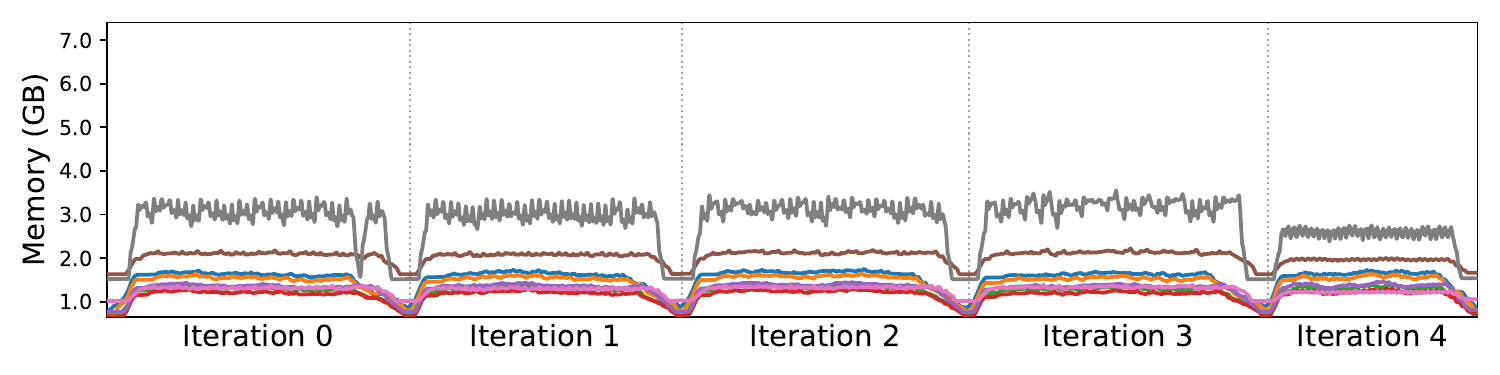}
        \caption{Memory consumption of 1F1B pipeline schedule.}
        \label{fig:e2e-memory-consumption-llama1b-llava150k-1f1b}
    \end{subfigure}
    \begin{subfigure}[t]{\columnwidth}
        \includegraphics[width=\columnwidth]{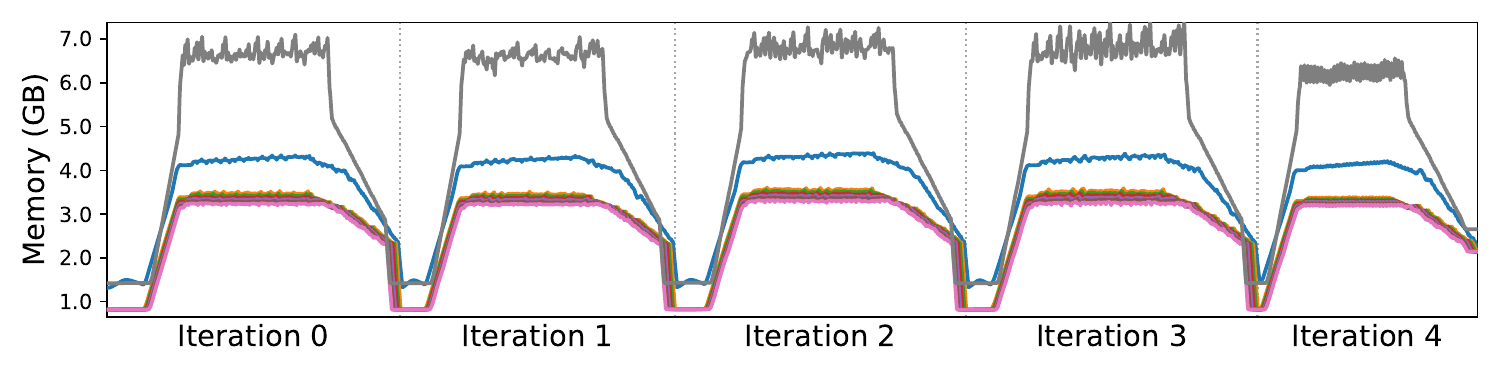}
        \caption{Memory consumption of DIP pipeline schedule.}
        \label{fig:e2e-memory-consumption-llama1b-llava150k-pipeweaver}
    \end{subfigure}
    \begin{subfigure}[t]{\columnwidth}
        \includegraphics[width=\columnwidth]{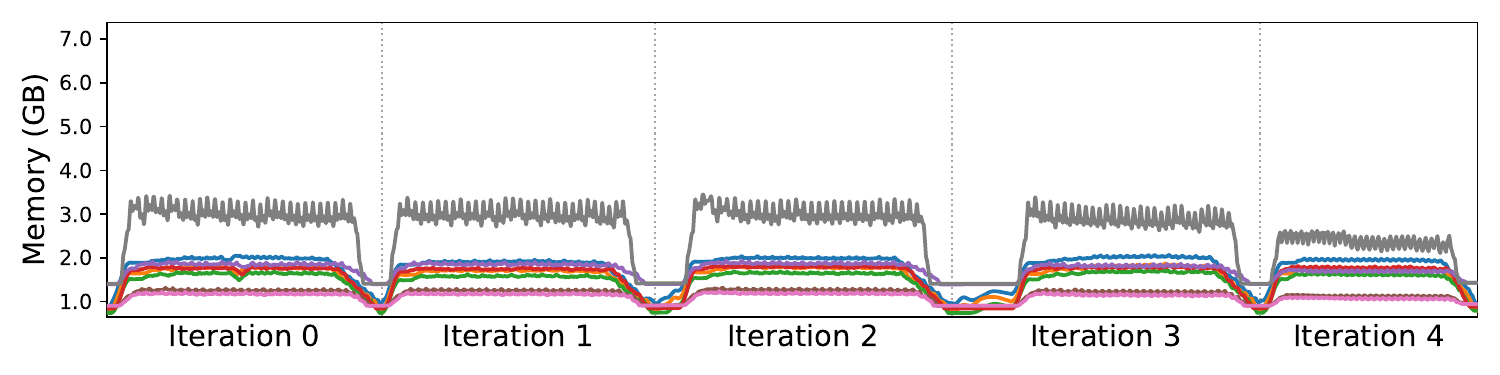}
        \caption{Memory consumption of \name pipeline schedule.}
        \label{fig:e2e-memory-consumption-llama1b-llava150k-mandu1f1b}
    \end{subfigure}
    \caption{LLaVA-150k on Qwen2.5Vision+Llama3-1b VLM.}
    \label{fig:e2e-memory-consumption-llama1b-llava150k}
\end{figure}

\begin{figure}[t]
    \centering
    \begin{subfigure}[t]{\columnwidth}
        \includegraphics[width=\columnwidth]{evaluation/e2e/memory_consumption/llama_1b/pipeline_memory_consumption_legend.pdf}
    \end{subfigure}
    \setcounter{subfigure}{0}
    \begin{subfigure}[t]{\columnwidth}
        \includegraphics[width=\columnwidth]{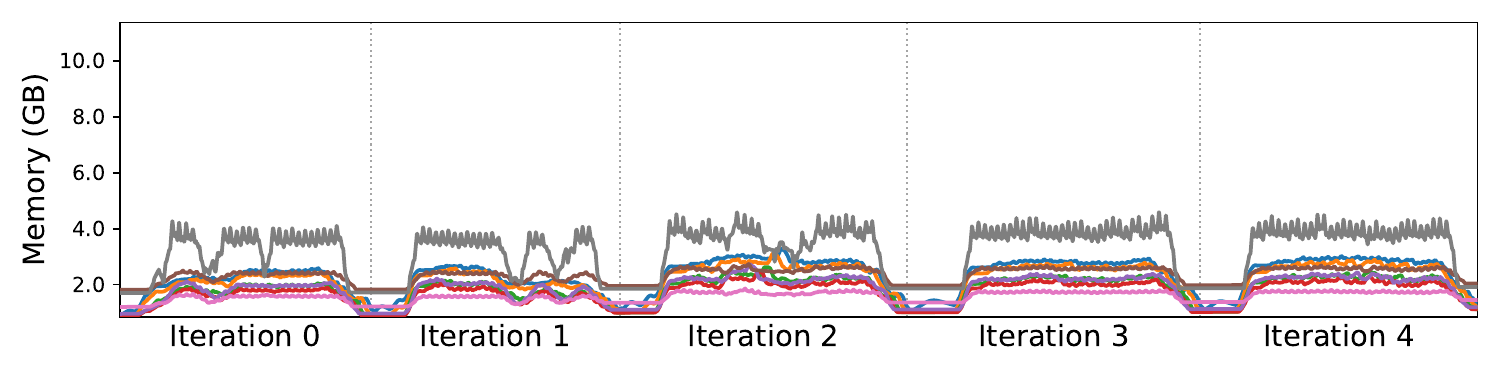}
        \caption{Memory consumption of 1F1B pipeline schedule.}
        \label{fig:e2e-memory-consumption-llama1b-chartqa-1f1b}
    \end{subfigure}
    \begin{subfigure}[t]{\columnwidth}
        \includegraphics[width=\columnwidth]{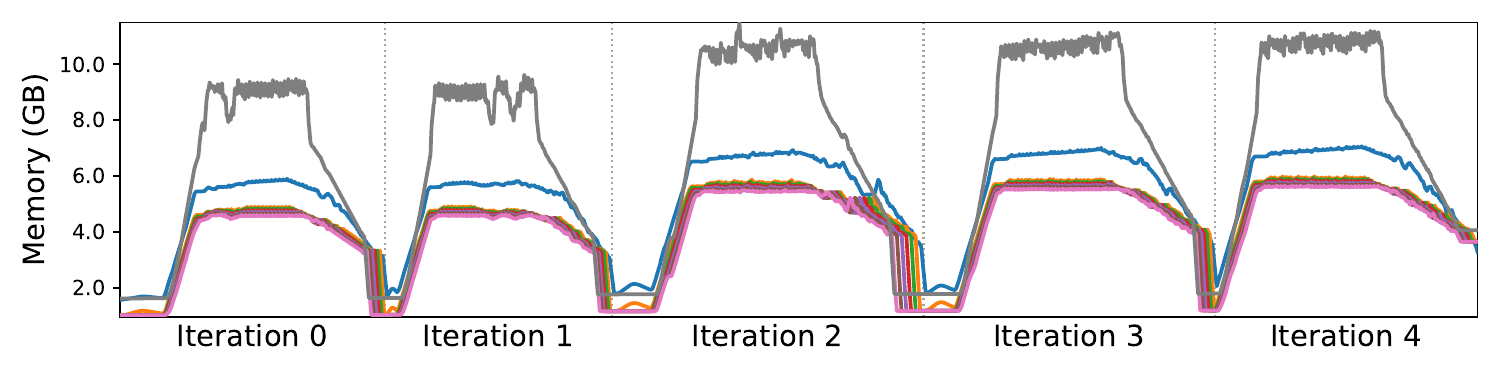}
        \caption{Memory consumption of DIP pipeline schedule.}
        \label{fig:e2e-memory-consumption-llama1b-chartqa-pipeweaver}
    \end{subfigure}
    \begin{subfigure}[t]{\columnwidth}
        \includegraphics[width=\columnwidth]{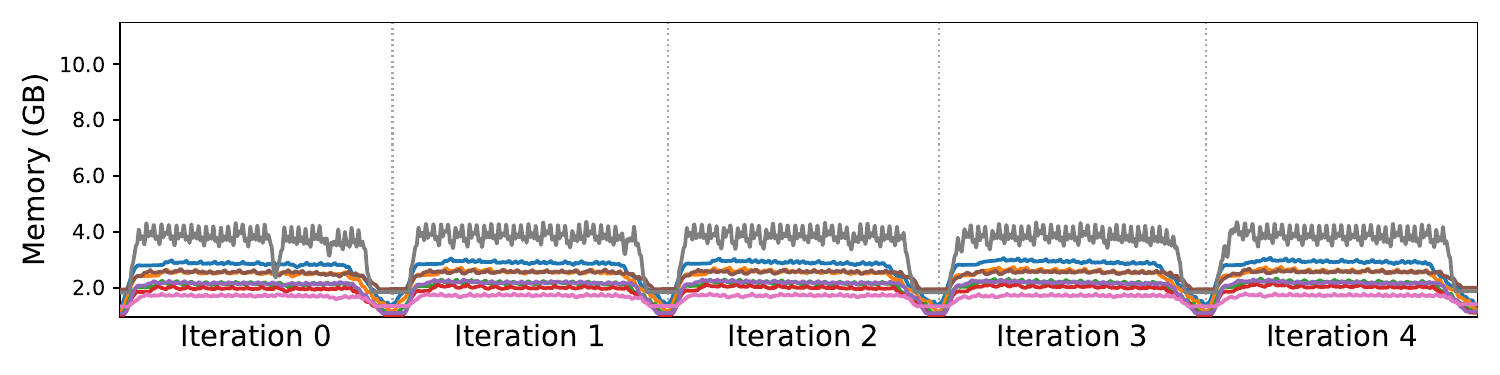}
        \caption{Memory consumption of \name pipeline schedule.}
        \label{fig:e2e-memory-consumption-llama1b-chartqa-mandu1f1b}
    \end{subfigure}
    \caption{ChartQA on Qwen2.5Vision+Llama3-1b VLM.}
    \label{fig:e2e-memory-consumption-llama1b-chartqa}
\end{figure}

\begin{figure}[t]
    \centering
    \begin{subfigure}[t]{\columnwidth}
        \includegraphics[width=\columnwidth]{evaluation/e2e/memory_consumption/llama_1b/pipeline_memory_consumption_legend.pdf}
    \end{subfigure}
    \setcounter{subfigure}{0}
    \begin{subfigure}[t]{\columnwidth}
        \includegraphics[width=\columnwidth]{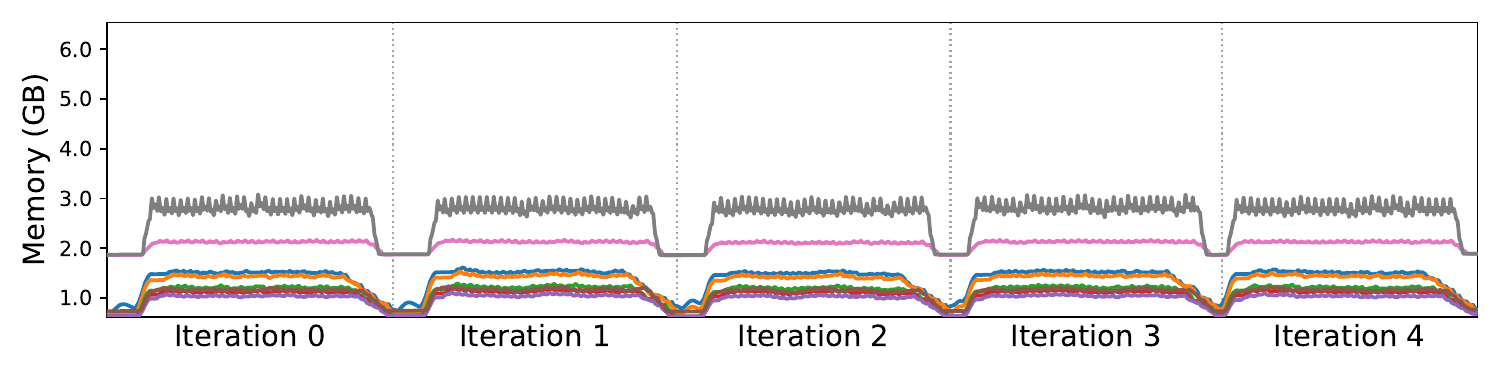}
        \caption{Memory consumption of 1F1B pipeline schedule.}
        \label{fig:e2e-memory-consumption-llama1b-cocoqa-1f1b}
    \end{subfigure}
    \begin{subfigure}[t]{\columnwidth}
        \includegraphics[width=\columnwidth]{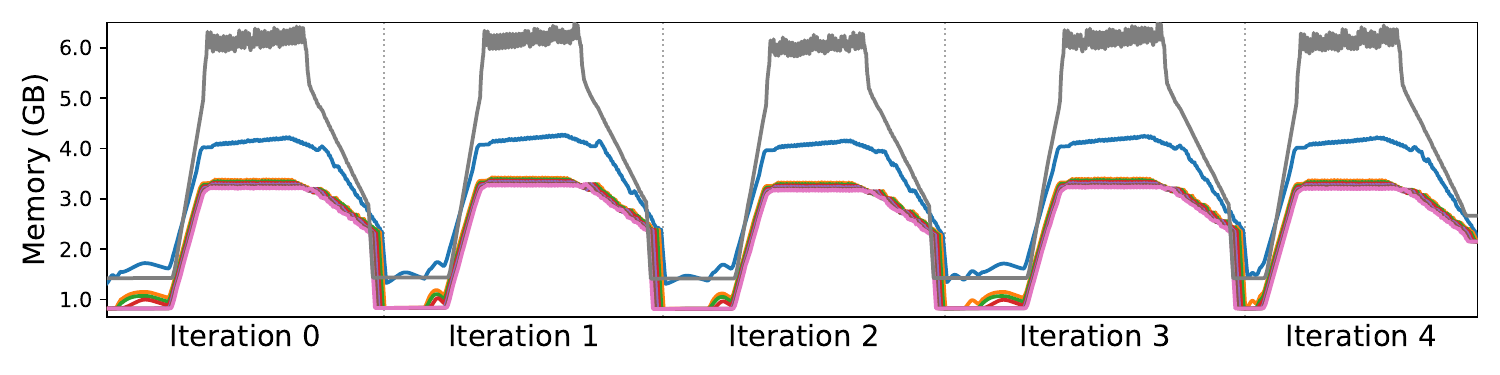}
        \caption{Memory consumption of DIP pipeline schedule.}
        \label{fig:e2e-memory-consumption-llama1b-cocoqa-pipeweaver}
    \end{subfigure}
    \begin{subfigure}[t]{\columnwidth}
        \includegraphics[width=\columnwidth]{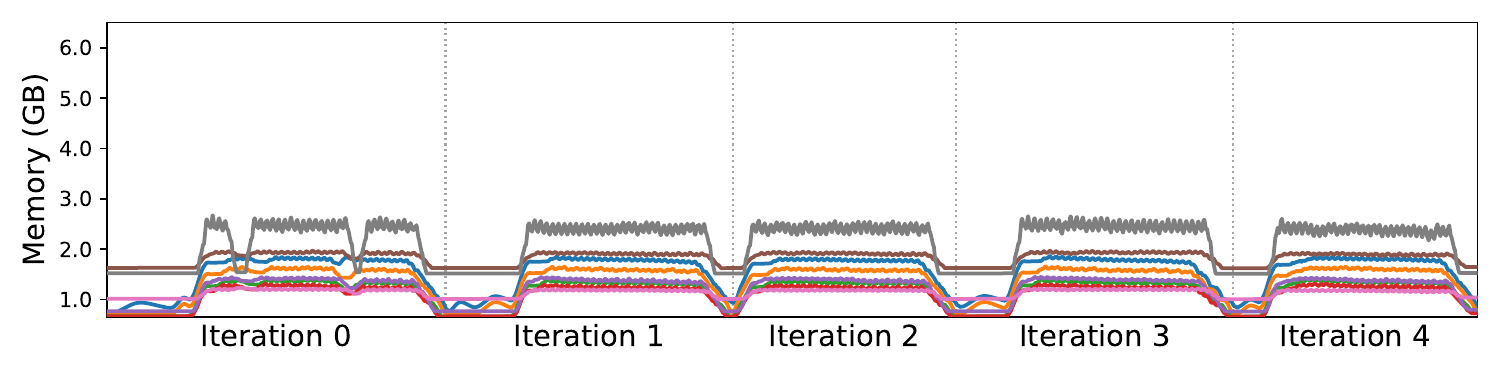}
        \caption{Memory consumption of \name pipeline schedule.}
        \label{fig:e2e-memory-consumption-llama1b-cocoqa-mandu1f1b}
    \end{subfigure}
    \caption{CocoQA on Qwen2.5Vision+Llama3-1b VLM.}
    \label{fig:e2e-memory-consumption-llama1b-cocoqa}
\end{figure}

Figure~\ref{fig:e2e-memory-consumption-llama1b-synthchartnet}, Figure~\ref{fig:e2e-memory-consumption-llama1b-llava150k}, Figure~\ref{fig:e2e-memory-consumption-llama1b-chartqa}, and Figure~\ref{fig:e2e-memory-consumption-llama1b-cocoqa} show the pipeline schedule memory consumption of SynthChartNet, LLaVA-150k, ChartQA, and CocoQA datasets on Qwen2.5Vision + Llama3-1b VLM.

\subsection{Qwen2.5Vision + Llama3-3b}
\begin{figure}[t]
    \centering
    \begin{subfigure}[t]{\columnwidth}
        \includegraphics[width=\columnwidth]{evaluation/e2e/memory_consumption/llama_3b/pipeline_memory_consumption_legend.pdf}
    \end{subfigure}
    \setcounter{subfigure}{0}
    \begin{subfigure}[t]{\columnwidth}
        \includegraphics[width=\columnwidth]{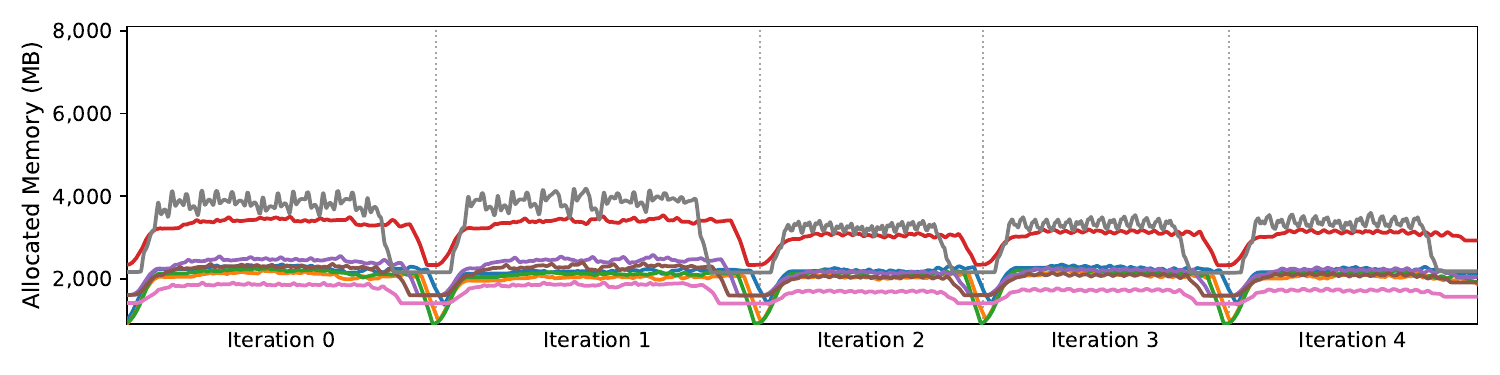}
        \caption{Memory consumption of 1F1B pipeline schedule.}
        \label{fig:e2e-memory-consumption-llama3b-llava150k-1f1b}
    \end{subfigure}
    \begin{subfigure}[t]{\columnwidth}
        \includegraphics[width=\columnwidth]{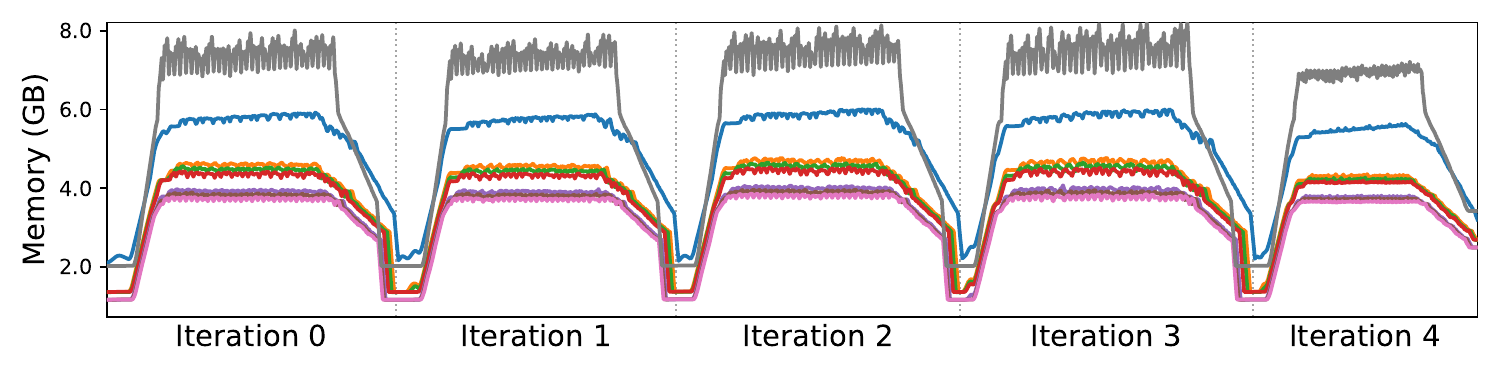}
        \caption{Memory consumption of DIP pipeline schedule.}
        \label{fig:e2e-memory-consumption-llama3b-llava150k-pipeweaver}
    \end{subfigure}
    \begin{subfigure}[t]{\columnwidth}
        \includegraphics[width=\columnwidth]{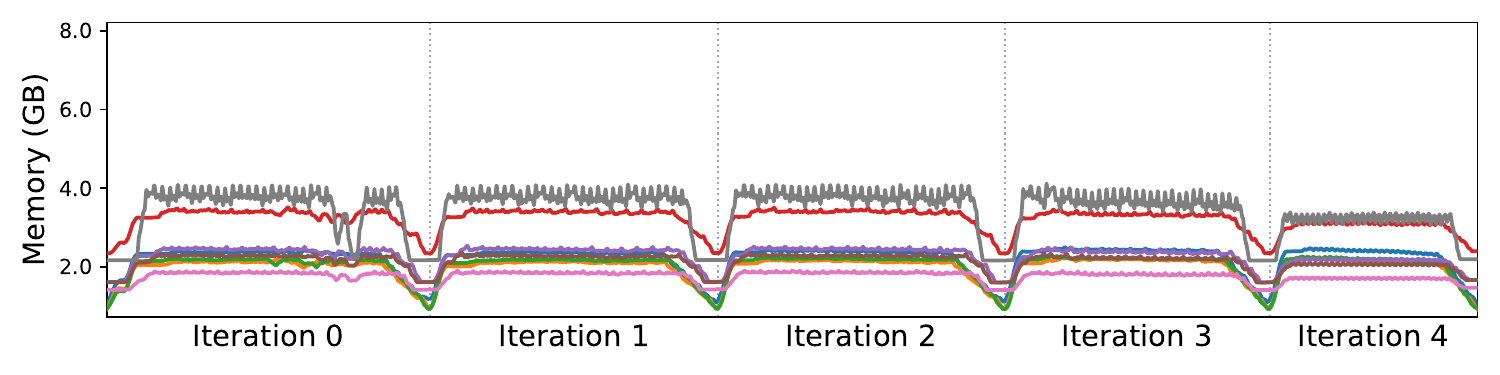}
        \caption{Memory consumption of \name pipeline schedule.}
        \label{fig:e2e-memory-consumption-llama3b-llava150k-mandu1f1b}
    \end{subfigure}
    \caption{LLaVA-150k on Qwen2.5Vision+Llama3-1b VLM.}
    \label{fig:e2e-memory-consumption-llama3b-llava150k}
\end{figure}

\begin{figure}[t]
    \centering
    \begin{subfigure}[t]{\columnwidth}
        \includegraphics[width=\columnwidth]{evaluation/e2e/memory_consumption/llama_3b/pipeline_memory_consumption_legend.pdf}
    \end{subfigure}
    \setcounter{subfigure}{0}
    \begin{subfigure}[t]{\columnwidth}
        \includegraphics[width=\columnwidth]{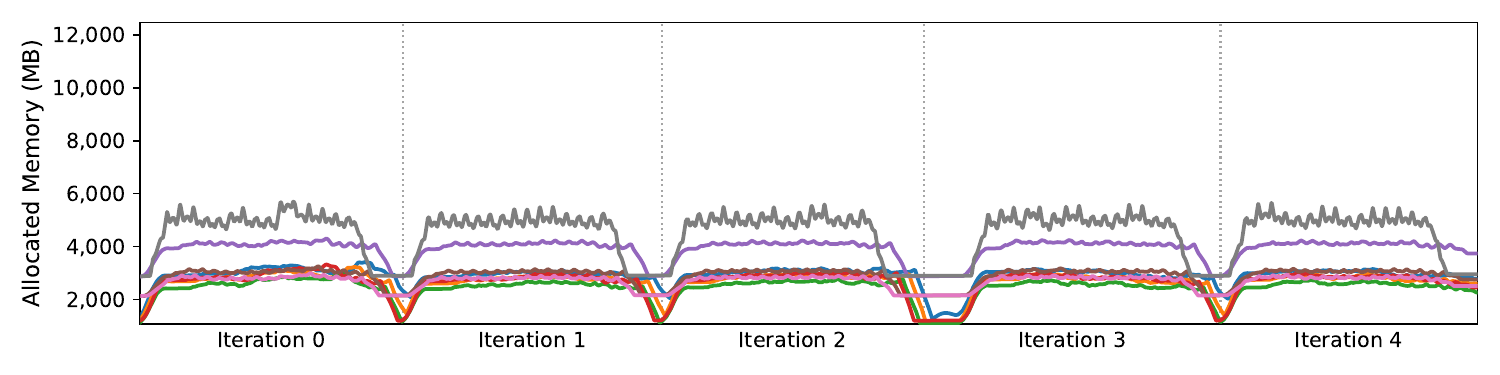}
        \caption{Memory consumption of 1F1B pipeline schedule.}
        \label{fig:e2e-memory-consumption-llama3b-chartqa-1f1b}
    \end{subfigure}
    \begin{subfigure}[t]{\columnwidth}
        \includegraphics[width=\columnwidth]{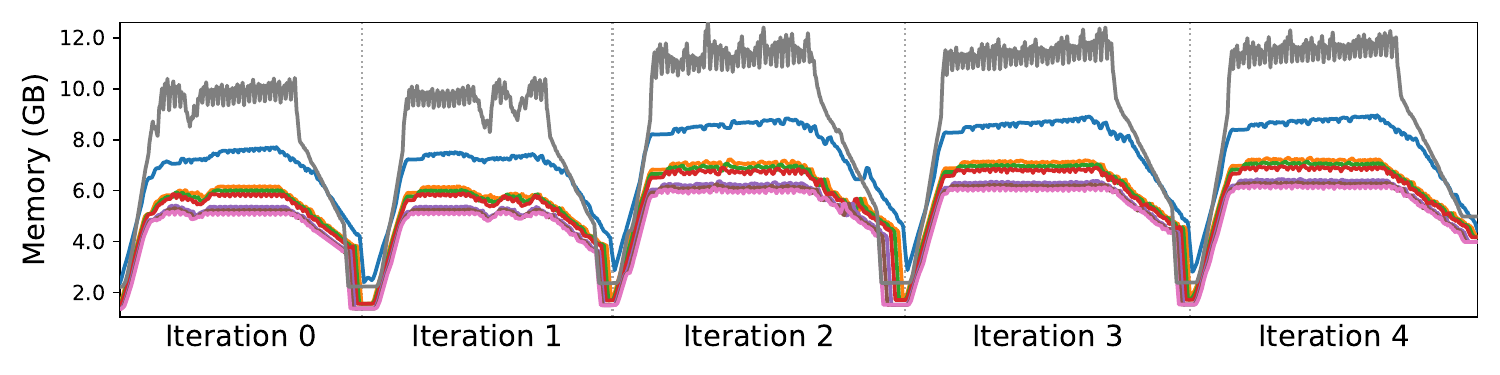}
        \caption{Memory consumption of DIP pipeline schedule.}
        \label{fig:e2e-memory-consumption-llama3b-chartqa-pipeweaver}
    \end{subfigure}
    \begin{subfigure}[t]{\columnwidth}
        \includegraphics[width=\columnwidth]{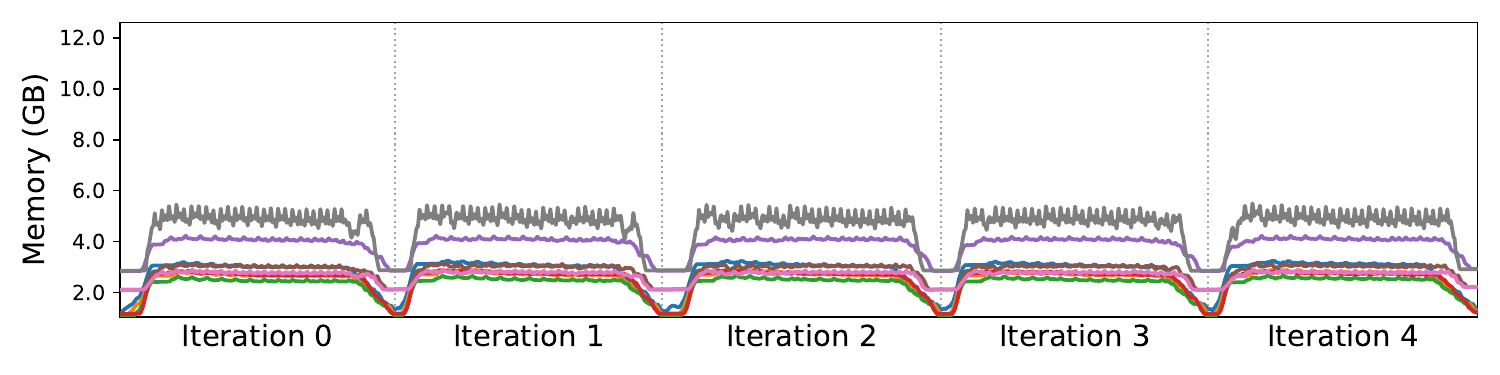}
        \caption{Memory consumption of \name pipeline schedule.}
        \label{fig:e2e-memory-consumption-llama3b-chartqa-mandu1f1b}
    \end{subfigure}
    \caption{ChartQA on Qwen2.5Vision+Llama3-1b VLM.}
    \label{fig:e2e-memory-consumption-llama3b-chartqa}
\end{figure}

\begin{figure}[t]
    \centering
    \begin{subfigure}[t]{\columnwidth}
        \includegraphics[width=\columnwidth]{evaluation/e2e/memory_consumption/llama_3b/pipeline_memory_consumption_legend.pdf}
    \end{subfigure}
    \setcounter{subfigure}{0}
    \begin{subfigure}[t]{\columnwidth}
        \includegraphics[width=\columnwidth]{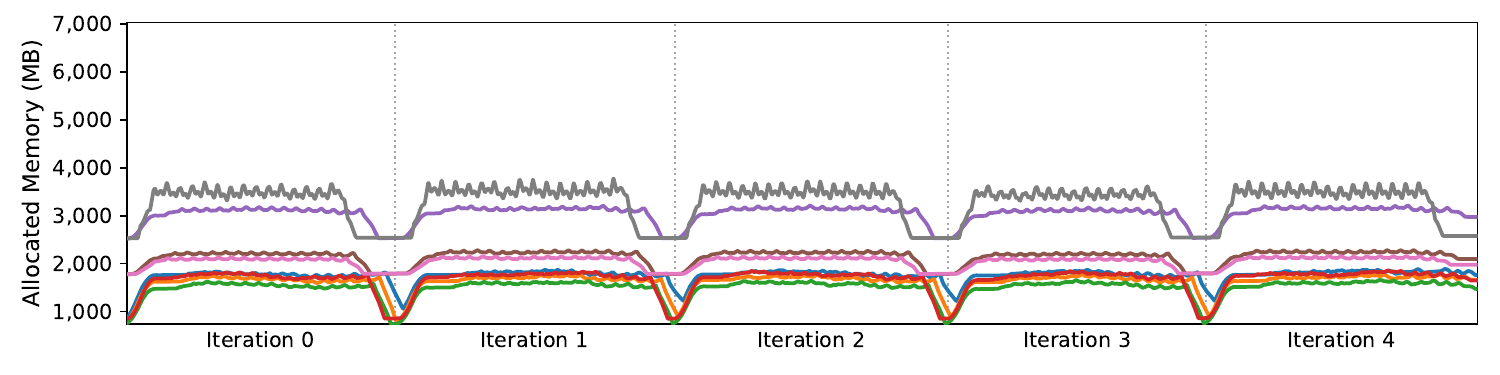}
        \caption{Memory consumption of 1F1B pipeline schedule.}
        \label{fig:e2e-memory-consumption-llama3b-cocoqa-1f1b}
    \end{subfigure}
    \begin{subfigure}[t]{\columnwidth}
        \includegraphics[width=\columnwidth]{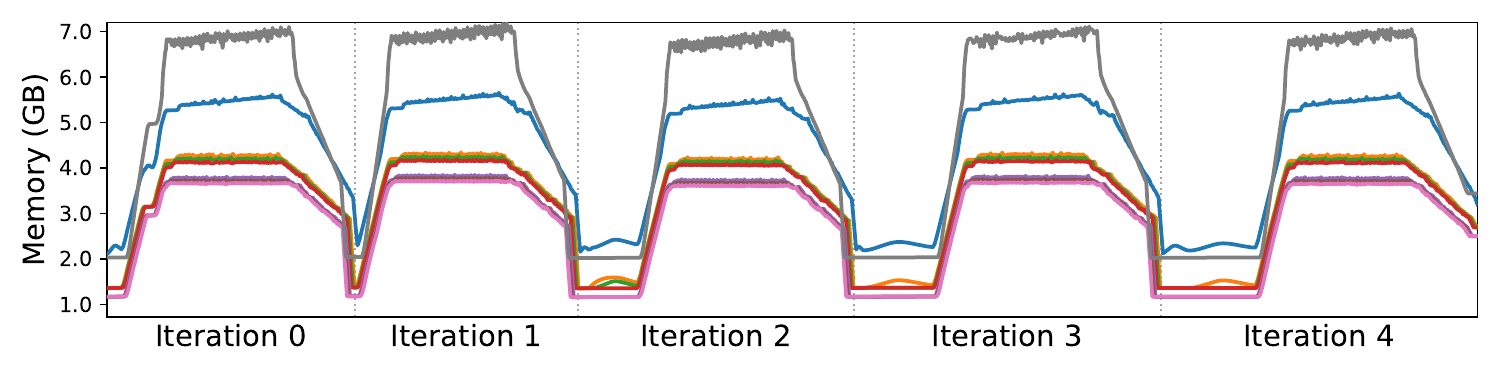}
        \caption{Memory consumption of DIP pipeline schedule.}
        \label{fig:e2e-memory-consumption-llama3b-cocoqa-pipeweaver}
    \end{subfigure}
    \begin{subfigure}[t]{\columnwidth}
        \includegraphics[width=\columnwidth]{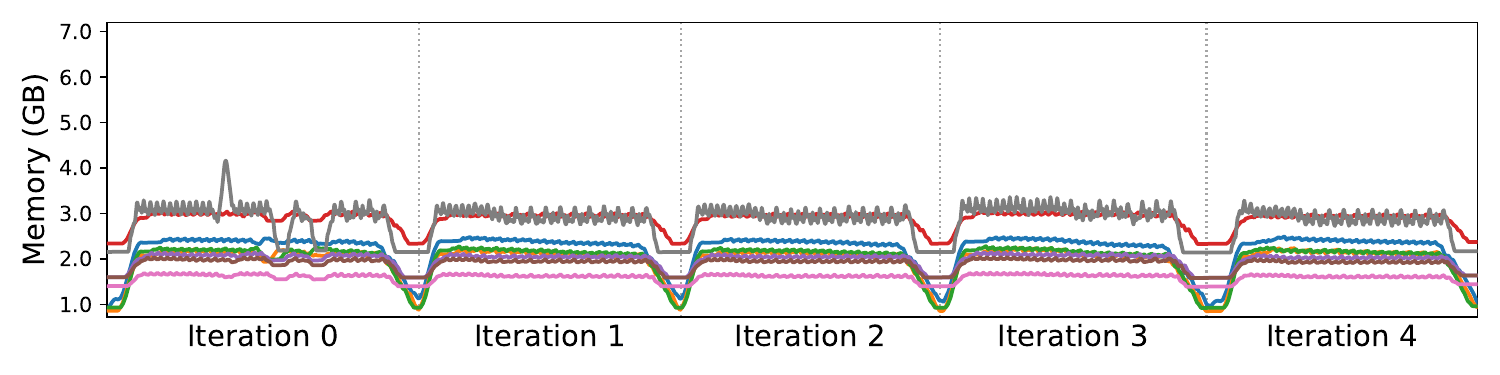}
        \caption{Memory consumption of \name pipeline schedule.}
        \label{fig:e2e-memory-consumption-llama3b-cocoqa-mandu1f1b}
    \end{subfigure}
    \caption{CocoQA on Qwen2.5Vision+Llama3-1b VLM.}
    \label{fig:e2e-memory-consumption-llama3b-cocoqa}
\end{figure}

Figure~\ref{fig:e2e-memory-consumption-llama3b-llava150k}, Figure~\ref{fig:e2e-memory-consumption-llama3b-chartqa}, Figure~\ref{fig:e2e-memory-consumption-llama3b-cocoqa} show the pipeline schedule memory consumption of LLaVA-150k, ChartQA, and CocoQA datasets on Qwen2.5Vision + Llama3-3b VLM.

\section{Workload Ratios in Bernoulli Trials}
\label{app:eval-workload-ratio-with-different-batch-sizes}

\subsection{Qwen2.5Vision + Llama3-1b}

Table~\ref{tab:app-eval-synthchartnet-1b}, Table~\ref{tab:app-eval-llava150k-1b}, Table~\ref{tab:app-eval-chartqa-1b}, and Table~\ref{tab:app-eval-cocoqa-1b} show the workload ratios in Bernoulli trials of SynthChartNet, LLaVA-150k, ChartQA, and CocoQA datasets using Qwen2.5Vision + Llama3-1b.

\begin{table}[H]
    \centering
    \caption{Workload ratios in Bernoulli trials of SynthChartNet dataset using Qwen2.5Vision + Llama3-1b.}
    \label{tab:app-eval-synthchartnet-1b}
    \begin{tabular}{cc|c} 
        \toprule
        \begin{tabular}[c]{@{}c@{}}Batch\\Size\end{tabular} & \begin{tabular}[c]{@{}c@{}}Trial\\Pass\end{tabular} & Ratios Shown (Vision:LLM)  \\ 
        \midrule
        1                                                   &    \xmark                                                 & 11:5, 10:6, 9:7            \\
        4                                                   &    \xmark                                           & 11:5, 10:6, 9:7            \\
        16                                                  &    \xmark                                           & 11:5, 10:6                 \\
        64                                                  &    \cmark                                           & 10:6                       \\
        256                                                 &    \cmark                                           & 10:6                       \\
        \bottomrule
    \end{tabular}
\end{table}

\begin{table}[H]
    \centering
    \caption{Workload ratios in Bernoulli trials of LLaVA-150k dataset using Qwen2.5Vision + Llama3-1b.}
    \label{tab:app-eval-llava150k-1b}
    \begin{tabular}{cc|c} 
        \toprule
        \begin{tabular}[c]{@{}c@{}}Batch\\Size\end{tabular} & \begin{tabular}[c]{@{}c@{}}Trial\\Pass\end{tabular} & Ratios Shown (Vision:LLM)  \\ 
        \midrule
        1                                                   &   \xmark                                          & 10:6, 9:7, 8:8             \\
        4                                                   &   \xmark                                            & 10:6, 9:7, 8:8             \\
        16                                                  &   \xmark                                            & 9:7, 8:8                   \\
        64                                                  &   \cmark                                            & 9:7                        \\
        256                                                 &   \cmark                                            & 9:7                        \\
        \bottomrule
        \end{tabular}
\end{table}

\begin{table}[H]
    \centering
    \caption{Workload ratios in Bernoulli trials of ChartQA dataset using Qwen2.5Vision + Llama3-1b.}
    \label{tab:app-eval-chartqa-1b}
    \begin{tabular}{cc|c} 
        \toprule
        \begin{tabular}[c]{@{}c@{}}Batch\\Size\end{tabular} & \begin{tabular}[c]{@{}c@{}}Trial\\Pass\end{tabular} & Ratios Shown (Vision:LLM)  \\ 
        \midrule
        1                                                   &   \xmark                                          & 10:6, 9:7                  \\
        4                                                   &   \xmark                                            & 10:6, 9:7                  \\
        16                                                  &   \cmark                                            & 10:6                       \\
        64                                                  &   \cmark                                            & 10:6                       \\
        256                                                 &   \cmark                                            & 10:6                       \\
        \bottomrule
    \end{tabular}
\end{table}

\begin{table}[H]
    \centering
    \caption{Workload ratios in Bernoulli trials of CocoQA dataset using Qwen2.5Vision + Llama3-1b.}
    \label{tab:app-eval-cocoqa-1b}
    \begin{tabular}{cc|c} 
        \toprule
        \begin{tabular}[c]{@{}c@{}}Batch\\Size\end{tabular} & \begin{tabular}[c]{@{}c@{}}Trial\\Pass\end{tabular} & Ratios Shown (Vision:LLM)  \\ 
        \midrule
        1                                                   &   \xmark                                          & 10:6, 9:7                  \\
        4                                                   &   \xmark                                            & 10:6, 9:7                  \\
        16                                                  &   \cmark                                            & 10:6                       \\
        64                                                  &   \cmark                                            & 10:6                       \\
        256                                                 &   \cmark                                            & 10:6                       \\
        \bottomrule
    \end{tabular}
\end{table}

\subsection{Qwen2.5Vision + Llama3-3b}

Table~\ref{tab:app-eval-llava150k-3b}, Table~\ref{tab:app-eval-chartqa-3b}, and Table~\ref{tab:app-eval-cocoqa-3b} show the workload ratios in Bernoulli trials of LLaVA-150k, ChartQA, and CocoQA datasets using Qwen2.5Vision + Llama3-3b.

\begin{table}[H]
    \centering
    \caption{Workload ratios in Bernoulli trials of LLaVA-150k dataset using Qwen2.5Vision + Llama3-3b.}
    \label{tab:app-eval-llava150k-3b}
    \begin{tabular}{cc|c} 
    \toprule
    \begin{tabular}[c]{@{}c@{}}Batch\\Size\end{tabular} & \begin{tabular}[c]{@{}c@{}}Trial\\Pass\end{tabular} & Ratios Shown (Vision:LLM)  \\ 
    \midrule
    1                                                   &         \xmark                                      & 7:9, 6:10, 5:11            \\
    4                                                   &         \xmark                                      & 7:9, 6:10                  \\
    16                                                  &         \xmark                                      & 7:9, 6:10                  \\
    64                                                  &         \xmark                                      & 7:9, 6:10                  \\
    256                                                 &         \cmark                                      & 7:9                        \\
    \bottomrule
    \end{tabular}
\end{table}

\begin{table}[H]
    \centering
    \caption{Workload ratios in Bernoulli trials of ChartQA dataset using Qwen2.5Vision + Llama3-3b.}
    \label{tab:app-eval-chartqa-3b}
    \begin{tabular}{cc|c} 
        \toprule
        \begin{tabular}[c]{@{}c@{}}Batch\\Size\end{tabular} & \begin{tabular}[c]{@{}c@{}}Trial\\Pass\end{tabular} & Ratios Shown (Vision:LLM)  \\ 
        \midrule
        1                                                   &      \xmark                                         & 8:8, 7:9                   \\
        4                                                   &      \cmark                                         & 8:8                        \\
        16                                                  &      \cmark                                         & 8:8                        \\
        64                                                  &      \cmark                                         & 8:8                        \\
        256                                                 &      \cmark                                         & 8:8                        \\
        \bottomrule
    \end{tabular}
\end{table}

\begin{table}[H]
    \centering
    \caption{Workload ratios in Bernoulli trials of CocoQA dataset using Qwen2.5Vision + Llama3-3b.}
    \label{tab:app-eval-cocoqa-3b}
    \begin{tabular}{cc|c} 
        \toprule
        \begin{tabular}[c]{@{}c@{}}Batch\\Size\end{tabular} & \begin{tabular}[c]{@{}c@{}}Trial\\Pass\end{tabular} & Ratios Shown (Vision:LLM)  \\ 
        \midrule
        1                                                   &      \xmark                                         & 8:8, 7:9                   \\
        4                                                   &      \cmark                                         & 8:8                        \\
        16                                                  &      \cmark                                         & 8:8                        \\
        64                                                  &      \cmark                                         & 8:8                        \\
        256                                                 &      \cmark                                         & 8:8                        \\
        \bottomrule
    \end{tabular}
\end{table}

\section{Sensitivity Analysis}
\label{app:eval-sensitivity-analysis}

\begin{figure}[H]
    \centering
    \includegraphics[width=\columnwidth]{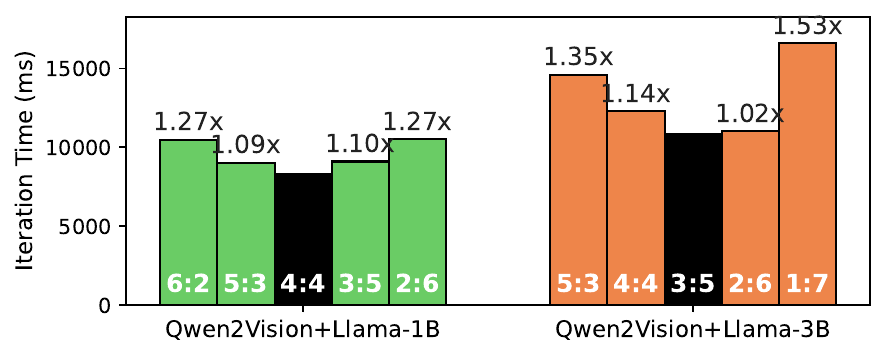}
    \caption{Sensitivity analysis of the profiling batch size on LLaVA-150k dataset.}
    \label{fig:app-eval-sensitivity-analysis-llava150k}
\end{figure}

\begin{figure}[H]
    \centering
    \includegraphics[width=\columnwidth]{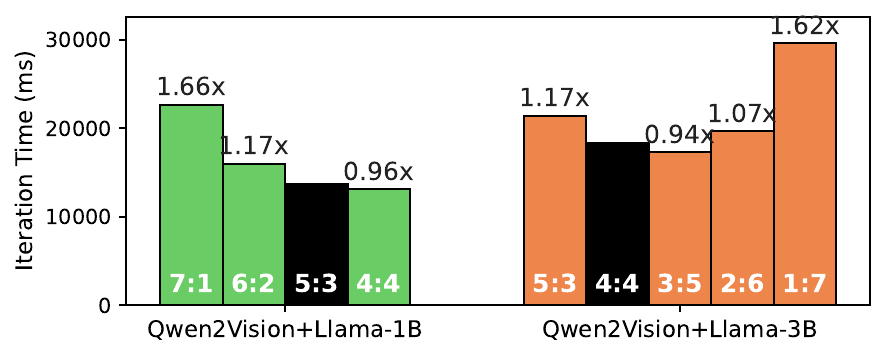}
    \caption{Sensitivity analysis of the profiling batch size on ChartQA dataset.}
    \label{fig:app-eval-sensitivity-analysis-chartqa}
\end{figure}

\begin{figure}[H]
    \centering
    \includegraphics[width=\columnwidth]{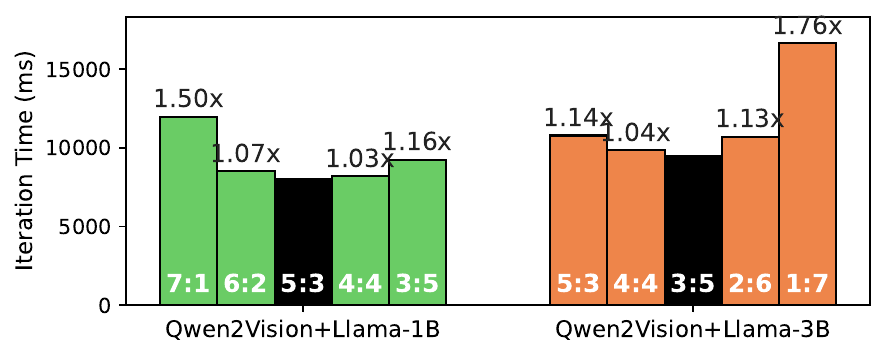}
    \caption{Sensitivity analysis of the profiling batch size on CocoQA dataset.}
    \label{fig:app-eval-sensitivity-analysis-cocoqa}
\end{figure}

Figure~\ref{fig:app-eval-sensitivity-analysis-llava150k}, Figure~\ref{fig:app-eval-sensitivity-analysis-chartqa}, and Figure~\ref{fig:app-eval-sensitivity-analysis-cocoqa} show the sensitivity analysis of the profiling batch size on LLaVA-150k, ChartQA, and CocoQA datasets.

\section{Variability of Modality Forward Time Across Microbatches}
\label{app:eval-variability-of-modality-forward-time-across-microbatches}

\begin{figure}[H]
    \centering
    \begin{subfigure}[t]{\columnwidth}
        \includegraphics[width=\columnwidth]{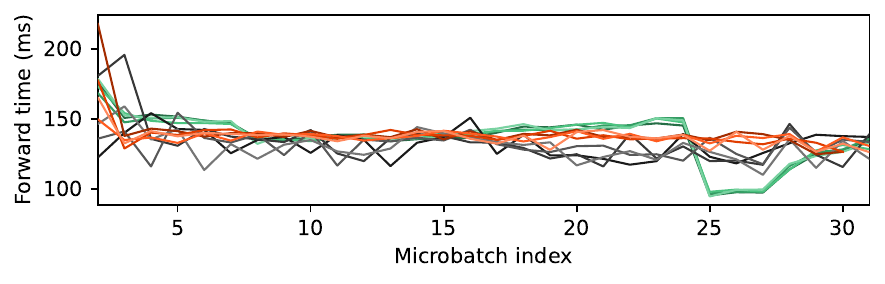}
        \caption{Qwen2.5Vision on LLaVA-150k dataset.}
    \end{subfigure}
    \begin{subfigure}[t]{\columnwidth}
        \includegraphics[width=\columnwidth]{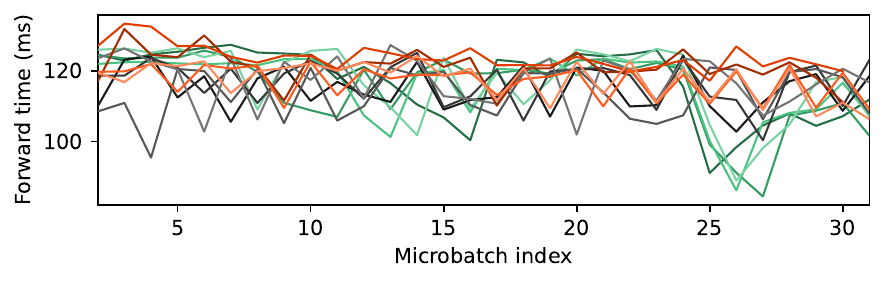}
        \caption{Llama3-1b on LLaVA-150k dataset.}
    \end{subfigure}
    \begin{subfigure}[t]{\columnwidth}
        \includegraphics[width=\columnwidth]{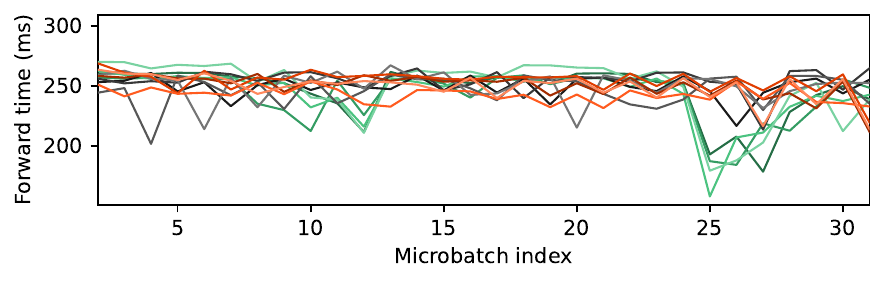}
        \caption{Llama3-3b on LLaVA-150k dataset.}
    \end{subfigure}
    \caption{Variability of modality forward time across microbatches in LLaVA-150k dataset.}
    \label{fig:app-eval-variability-of-modality-forward-time-across-microbatches-llava150k}
\end{figure}

\begin{figure}[H]
    \centering
    \begin{subfigure}[t]{\columnwidth}
        \includegraphics[width=\columnwidth]{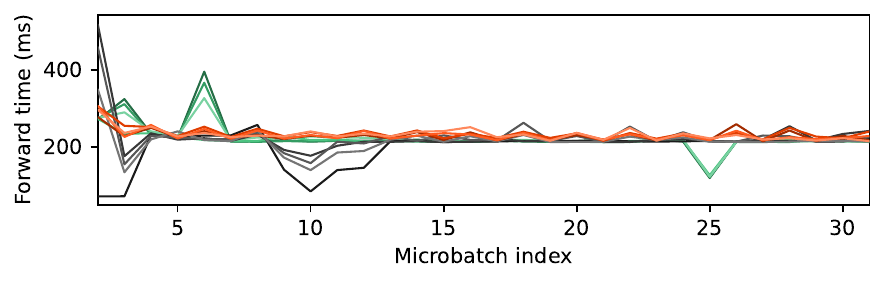}
        \caption{Qwen2.5Vision on ChartQA dataset.}
    \end{subfigure}
    \begin{subfigure}[t]{\columnwidth}
        \includegraphics[width=\columnwidth]{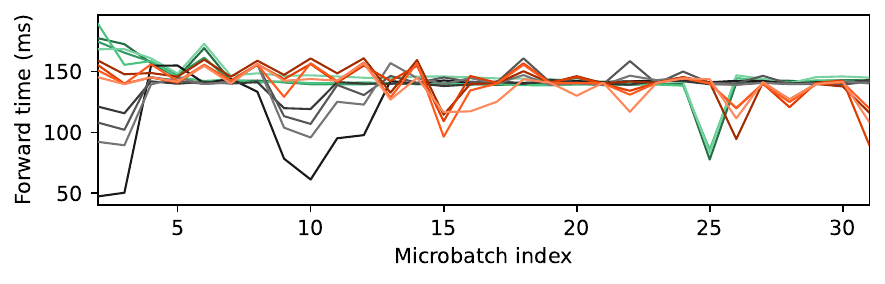}
        \caption{Llama3-1b on ChartQA dataset.}
    \end{subfigure}
    \begin{subfigure}[t]{\columnwidth}
        \includegraphics[width=\columnwidth]{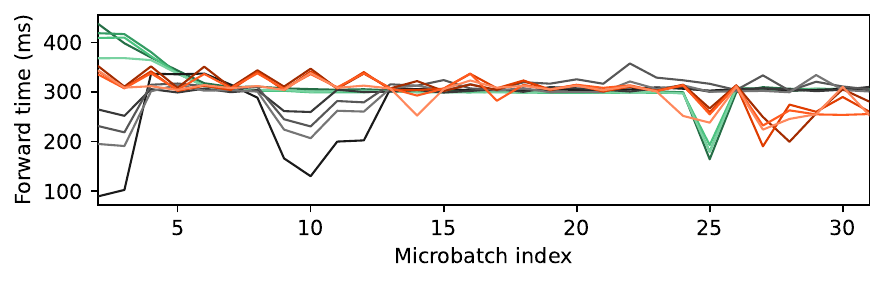}
        \caption{Llama3-3b on ChartQA dataset.}
    \end{subfigure}
    \caption{Variability of modality forward time across microbatches in ChartQA dataset.}
    \label{fig:app-eval-variability-of-modality-forward-time-across-microbatches-chartqa}
\end{figure}

\begin{figure}[H]
    \centering
    \begin{subfigure}[t]{\columnwidth}
        \includegraphics[width=\columnwidth]{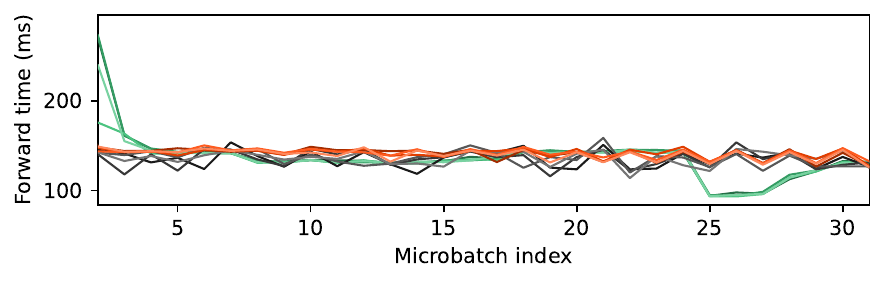}
        \caption{Qwen2.5Vision on CocoQA dataset.}
    \end{subfigure}
    \begin{subfigure}[t]{\columnwidth}
        \includegraphics[width=\columnwidth]{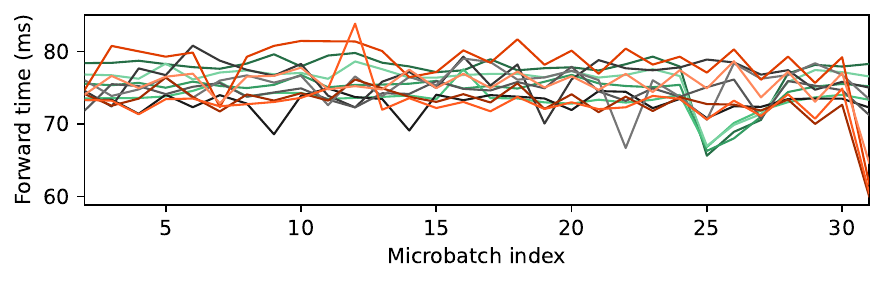}
        \caption{Llama3-1b on CocoQA dataset.}
    \end{subfigure}
    \begin{subfigure}[t]{\columnwidth}
        \includegraphics[width=\columnwidth]{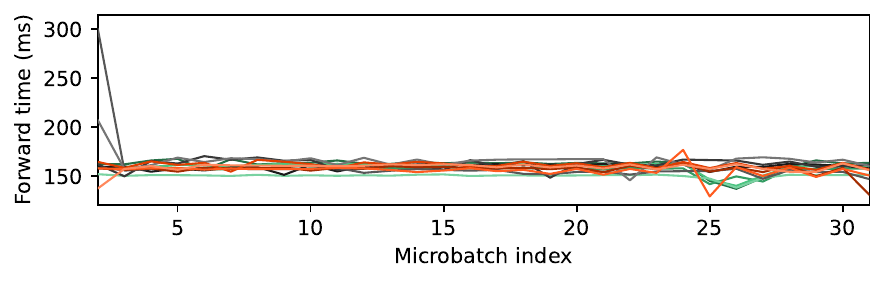}
        \caption{Llama3-3b on CocoQA dataset.}
    \end{subfigure}
    \caption{Variability of modality forward time across microbatches in CocoQA dataset.}
    \label{fig:app-eval-variability-of-modality-forward-time-across-microbatches-cocoqa}
\end{figure}

Figure~\ref{fig:app-eval-variability-of-modality-forward-time-across-microbatches-llava150k}, Figure~\ref{fig:app-eval-variability-of-modality-forward-time-across-microbatches-chartqa}, and Figure~\ref{fig:app-eval-variability-of-modality-forward-time-across-microbatches-cocoqa} show the variability of modality forward time across microbatches in LLaVA-150k, ChartQA, and CocoQA datasets.

\end{document}